\documentclass[11pt,a4paper,oneside]{article}
\usepackage[top=3cm, bottom=3cm, left=2cm, right=2cm]{geometry}
\usepackage{amsthm,amsmath,amssymb,mathrsfs,dsfont,mathtools}
\usepackage{xcolor}
\usepackage{bm}
\usepackage{enumitem}
\usepackage{graphicx}
\usepackage[colorinlistoftodos]{todonotes}
\usepackage[colorlinks=true, citecolor=blue, urlcolor=blue,
linkcolor = blue]{hyperref}
\usepackage{booktabs,subcaption}
\usepackage[all]{nowidow}
\usepackage{setspace}
\usepackage{algorithm}
\usepackage[noend]{algpseudocode}
\usepackage{tikz}

\usepackage[sf]{titlesec}
\usepackage{sectsty}
\allsectionsfont{\centering \normalfont\scshape}
\usepackage{adjustbox}
\usepackage{lineno}

\usepackage{natbib}
\usepackage[noabbrev,capitalise]{cleveref}
\creflabelformat{equation}{#2\textup{(#1)}#3}
\usepackage{indentfirst}

\usepackage{authblk}
\title{Compressive Bayesian non-negative matrix factorization for mutational signatures analysis}
\date{}
\author[1]{Alessandro Zito}
\author[1]{Jeffrey W. Miller}

\affil[1]{Department of Biostatistics, Harvard University, Boston, MA, 02115, U.S.A.}

\newtheorem{theorem}{Theorem}
\newtheorem{corollary}{Corollary}
\newtheorem{lemma}{Lemma}
\newtheorem{proposition}{Proposition}

\theoremstyle{definition}
\newtheorem{definition}{Definition}

\newcommand{\pr}{\mathds{P}}

\definecolor{revision_color}{HTML}{00008B}
\newcommand{\revision}[1]{\textcolor{black}{#1}}
\allowdisplaybreaks

\usepackage[resetlabels]{multibib}
\newcites{Supp}{References}

\begin{document}
\maketitle

\begin{abstract}
Non-negative matrix factorization (NMF) is a popular tool for dimensionality reduction, especially for count matrices. However, inferring an appropriate number of factors is challenging. Existing approaches based on information criteria or nonparametric sparsity-inducing priors tend to be computationally burdensome or highly sensitive to prior choices. Moreover, theoretical properties of the posterior distribution of Poisson NMF parameters endowed with shrinkage priors remain under-explored. This paper introduces a novel Bayesian NMF method that automatically infers the number of factors while also incorporating information on the latent factors from previous studies. This is achieved using compressive hyperpriors, which are hierarchical priors that make the sample-specific weights of unneeded factors concentrate near zero in the posterior. We provide novel distribution theory for posterior inference to elucidate this shrinkage mechanism, both in finite samples and asymptotically. We apply our method to mutational signatures analysis in cancer genomics, in simulations and on real data from breast cancer. Compared to state-of-the-art alternatives, our method is more robust to mild overdispersion and improves detection and estimation of signatures aligned with prior information.
\end{abstract}


\section{Introduction}\label{sec:intro}
Non-negative matrix factorization (NMF) is a popular dimensionality reduction method that approximates a non-negative matrix as the product of two lower-dimensional non-negative matrices of a desired rank  \citep{Lee_seung_nature}. NMF has been widely applied to both discrete and continuous data \citep{gillis2021nonnegative}, and there is a rich literature on discrete Bayesian NMF, encompassing parametric \citep{Cemgil_2009} and non-parametric models \citep{Gopalan_2014, Ayed_Caron_2021}, multi-study frameworks \citep{Grabski_2023}, and spatially dependent structures \citep{Townes_Engelhardt_2023}. See \citet{Zhou_Carin_2015} for a summary of Bayesian NMF methods for count data.

Poisson NMF methods have been highly effective in cancer genomics, where they have been used to discover a wide range of mutational signatures from mutation counts in many tumor types
\citep{Nik_zainal_2012, Alexandrov_2020}. Each signature is a latent vector of the probabilities with which different types of mutations occur, and corresponds to a specific mutagenic process, such as dysregulated DNA repair mechanisms or tobacco smoking. 
Each patient's mutation count profile is modeled via a weighted combination of signatures, where the weights are called ``exposures.''
Accurately detecting signature exposures in patients can improve the effectiveness of
therapies \citep{Aguirre_2018}. However, choosing an appropriate number of signatures to infer---that is, the number of factors---is a challenging statistical problem: using too many factors leads to overfitting, while using too few
can obscure relevant signals due to merging of signatures that should be kept distinct. This article proposes a novel 
Bayesian Poisson NMF approach to overcome this issue.

There exists a wide variety of approaches to inferring the number of factors, or equivalently, the rank of the factorization. 
Traditional likelihood-based methods solve the problem by minimizing the Bayesian information criterion or related indicators, fitting a separate solution for a range of plausible rank choices \citep{Rosales_2016,Islam_2022}.
However, such strategies rely on rather heuristic elbow-type rules, and are often computationally burdensome. Alternatively, automatic relevance determination through Bayesian hierarchical priors is a faster model-based solution, but state-of-the-art algorithms for Poisson NMF \citep{Tan_Fevotte_2013, kim2016somatic} only provide point estimates, disregarding uncertainty quantification entirely.  Moreover, the theoretical properties of Poisson NMF in such settings are still largely unexplored, which makes the selection mechanism lack a transparent interpretation.
Recent nonparametric Bayesian factor models offer an elegant alternative based on sparsity-inducing shrinkage priors that increasingly penalize model dimensionality \citep{Bhattacharya_Dunson_2011, Gopalan_2014, Legramanti_2020}. However, the rank inferred by these models is extremely sensitive to the values of the hyperparameters regulating the shrinkage (e.g., in the case of \citealp{Bhattacharya_Dunson_2011}, see the analysis of \citealp{DURANTE2017198}), especially when the likelihood is incorrectly specified relative to the true data-generating process. Additionally, crucially, the implied column exchangeability assumption limits the use of factor-specific prior information from previous investigations.

We present a novel Poisson factorization method
that circumvents the above limitations. We are motivated by mutational signatures analysis, but the method applies to any count matrix. Our model, which we refer to as compressive NMF, 
uses conjugate Dirichlet and gamma priors for the signatures and the exposures, respectively, and performs automatic rank selection via \emph{compressive hyperpriors} on the global means of the exposures. These are data-dependent hyperpriors having a prior mean close to zero, and a posterior that gives equal weight to the prior mean and the sample mean of the latent exposures.
As a result, setting the rank to a sufficiently large value that upper bounds the true rank, the exposures concentrate at negligible values for redundant signatures, effectively removing the whole factor from the decomposition in the same spirit as overfitted mixture models \citep{Rousseau_Mengersen_2011}. We present novel distribution theory for the posterior of the parameters of a Poisson NMF to investigate this shrinkage behavior, both in finite-sample settings and asymptotically. Specifically, our large sample results elucidate the role that prior hyperparameters have on the shrinkage, enabling model choices backed by a meaningful interpretation.
Posterior inference and rank selection are carried out via Gibbs sampling with multinomial data augmentation \citep{Dunson_Herring_2005}, bypassing the need to fit multiple models or employ discrete Metropolis-Hastings moves to vary the rank. We also investigate the sensitivity to prior choices and misspecification caused by mild overdispersion, finding that the model is robust in many settings.   

Additionally, our model can be straightforwardly extended to leverage factor-specific prior information. 
For example, the \emph{Catalogue Of Somatic Mutations In Cancer}  (COSMIC) database provides a curated set of mutational signatures and their putative etiologies \citep{Alexandrov_2020}. By using informative Dirichlet priors centered at the COSMIC signatures, our model obtains improved sensitivity to detect the presence of signatures and infers unambiguous matches to the original database.

The paper is organized as follows. \cref{sec:methodology} defines the model and discusses inference.
In \cref{sec:theory}, we provide theoretical results characterizing the posterior and its concentration.
\cref{sec:simulation} contains a simulation study comparing to leading methods, and \cref{sec:application} presents an application
on breast cancer.
We conclude with a brief discussion in \cref{sec:discussion}. 

\subsection{Background on mutational signatures analysis}\label{sec:mutational-signatures}

Cancer development in humans is connected to the accumulation of mutations in the DNA of somatic cells.
When considering single-base substitutions (SBS), mutations are classified according to which of the four nucleotide bases was present before and after the mutation, on the strand containing the pyrimidine before the mutation occurred. Recalling that adenine (A) and guanine (G) are \emph{purines} while cytosine (C) and thymine (T) are  \emph{pyrimidines} and that C always binds with G and T with A, there are six possible types of substitutions, namely, C$>$T, C$>$G, C$>$A, T$>$A, T$>$C and T$>$G, where C$>$T indicates a C that mutates into a T. To account for context-specific variability due to adjacent bases, mutations are further classified according to which bases (A, G, C, or T) occur on the $5$' and the $3$' sides on the strand containing the pre-substitution pyrimidine. This creates
$6\times 4 \times 4 = 96$ types of single-base substitutions, called
\emph{mutational channels}; see Section~\ref{subsec:substitutions} for further details.%

It has been observed that many mutation-causing processes consistently produce each type of mutation at a particular rate: for instance, ultraviolet radiation produces a large number of C$>$T substitutions in melanoma
\citep{Alexandrov_2020}. 
Due to this, the mutational processes acting on somatic cells can be characterized using mutational signatures \citep{Nik_zainal_2012}, which encode the latent probabilities of each mutation type. A curated set of known signatures is maintained in  the COSMIC database
\citep{Alexandrov_2020}, along with their experimentally validated etiologies.
For example, SBS7a, b, c, and d are linked to ultraviolet light exposure.
Others,
such as SBS60, require further investigation to understand whether they arise from true biological processes or are
technical artifacts.%

\section{Methodology}\label{sec:methodology}
\subsection{Poisson non-negative matrix factorization model}
Let $X_{m n}$ represent the number of mutations for channel $m$ in sample $n$, where $m = 1, \ldots, M$ and $n = 1,\ldots, N$, and let $\bm{X}\in \mathds{N}^{M\times N}$ denote the matrix with entries $X_{mn}$, where $\mathds{N} = \{0,1,2,\ldots\}$ is the set of nonnegative integers.
The goal of NMF is to find two non-negative matrices $\bm{R} \in \mathds{R}_+^{M\times K}$ and $\boldsymbol{\Theta}\in \mathds{R}_+^{K \times N}$ such that $\bm{X} \approx \bm{R} \,\boldsymbol{\Theta}$, with the \emph{rank} $K$ typically chosen so that $K \leq \min\{M, N\}$. The $k$th column of $\bm{R}$, denoted $\bm{r}_{k} = (r_{1 k}, \ldots, r_{M k})$, is referred to as the $k$th \emph{mutational signature}. The $k$th row of $\boldsymbol{\Theta}$, denoted $\boldsymbol{\theta}_k = (\theta_{k 1}, \ldots, \theta_{k N})$, is the vector of weights representing the \emph{exposure} to signature $k$ in each of the $N$ samples.

From a probabilistic perspective, it is natural to model the mutation counts as
\begin{equation}\label{eq:Poisson}
X_{m n} \sim \mathrm{Poisson}\bigg(\sum_{k = 1}^K r_{m k}\theta_{k n}\bigg)
\end{equation}
independently for $m = 1, \ldots, M$ and $n = 1, \ldots, N$, where $\mathrm{Poisson}(\lambda)$ denotes the Poisson distribution with mean and variance $\lambda$.
In Section~\ref{sec:rationale}, we show that \cref{eq:Poisson} can be derived from first principles by modeling the occurrences of mutations as continuous-time Markov processes across the genome. This provides a biological justification for the model. 

\subsection{Prior}
For the prior distribution on the signatures $\bm{r}_k$ and exposures $\boldsymbol{\theta}_k$, we take
\begin{align}\label{eq:prior_R}
\bm{r}_k = (r_{1 k},\ldots,r_{M k}) &\sim \mathrm{Dir}(\alpha, \ldots, \alpha), \\
\theta_{k 1},\ldots,\theta_{k N}\mid \mu_k \;&\sim\; \mathrm{Ga}(a,\, a/\mu_k), \label{eq:prior_theta}
\end{align}
independently for $k = 1,\ldots,K$, where $\alpha>0$, $a>0$, \revision{and $\mu_k > 0$ is given a hyperprior $\mu_k \sim \pi(\mu_k)$.} Here, $\mathrm{Ga}(a,b)$ denotes the gamma distribution with mean $a/b$ and  variance $a/b^2$.  Thus, the prior mean of the exposures is $E(\theta_{k n}\mid \mu_k) = \mu_k$ and the variance is $\mathrm{var}(\theta_{kn} \mid \mu_k) = \mu_k^2/a$, implying that $\mu_k$ is a global parameter that controls the overall contribution of signature $\bm{r}_k$ to the total mutation count generated by process $k$, through each $\theta_{kn}$, which act as local parameters. Moreover, we impose that $\sum_{m=1}^M r_{m k} = 1$, a constraint that is standard for mutational signatures and has previously been used in NMF models \citep{geiger2024connectionnonnegativematrixfactorization}. Under this constraint, we pose a Dirichlet prior on the signatures (\cref{eq:prior_R}).
This choice has the twofold advantage of avoiding scaling ambiguities in both the signature vectors $\bm{r}_{k}$ and their exposures $\boldsymbol{\theta}_k$, and facilitating construction of informative priors using known mutational signatures.

We refer to $\mu_1, \ldots, \mu_K$ as \emph{relevance weights}, following the usage of this type of prior structure in automatic relevance determination for shrinking the weights of unneeded factors to near-zero values \citep{Tan_Fevotte_2013}.
Here, we take a fully Bayesian approach and quantify uncertainty in $r_k$ and $\theta_k$ rather than just optimizing them.
An overview of existing NMF methods and differences with ours is presented in Section \ref{sec:comparison_literature}, including a comparison with global-local shrinkage priors.

\subsection{Compressive hyperprior}\label{subsec:compressive_hyperprior_prop}
The hyperprior on the relevance weights $\pi(\mu_k)$ plays a crucial role in inferring the number of factors. Several approaches have been developed to provide sparsity in Gaussian factorization models, such as spike-and-slab priors that introduce exact zeros in the exposures \citep{Veronica_2016} or induce cumulative shrinkage to near-zero values for redundant factors  \citep{Legramanti_2020}.
However, in our model, we found that spike-and-slab priors over $\mu_k$ tend to make inference difficult, likely due to the strong multimodal nature of the resulting posterior.  Instead, we propose a more direct alternative, letting
\begin{equation}\label{eq:hyperprior}
\mu_k \sim \mathrm{InvGa}(a N + 1,\, \varepsilon a N)
\end{equation}
independently for $k = 1, \ldots, K$, where $\mathrm{InvGa}(a_0,b_0)$  denotes the inverse-gamma distribution with mean $b_0/(a_0-1)$ when $a_0>1$. 
Here, $a$ is the shape parameter from the prior on $\theta_{k n}$ in \cref{eq:prior_theta}, and we set $\varepsilon>0$ to be a small constant, such as $\varepsilon = 0.001$. 
This makes $\mu_k$ small \emph{a priori}, since $E(\mu_k) = \varepsilon$ and and $\mathrm{var}(\mu_k) = \varepsilon^2/(aN -1)$  Further, the full conditional mean is 
$E(\mu_k\mid \boldsymbol{\Theta}) = \varepsilon/2  + \overline{\theta}_k/2,$ where $\overline{\theta}_k = N^{-1}\sum_{n = 1}^N \theta_{k n}$, implying that the prior mean $\varepsilon$ and the average exposure $\overline{\theta}_k$ for signature $k$ have equal influence on the posterior for $\mu_k$ at any sample size $N$. We refer to \cref{eq:hyperprior} as a \emph{compressive hyperprior}. This should not be interpreted as a meaningful representation of prior uncertainty about $\mu_k$, but rather as a design to yield a posterior with good computational and accuracy properties. 

Our prior structure has the important advantage of favoring sparse solutions: for any extra unneeded signatures, \cref{eq:hyperprior} encourages the corresponding relevance weights $\mu_k$ to be small, on the order of $\varepsilon$. However, the implied shrinkage mechanism differs from traditional Gaussian factor models, which usually assume independent standard normal priors over the factors and separate priors for the variance of the loadings, because $\mu_k$ controls both the mean and the variance of the exposures $\theta_{kn}$. While this might initially appear restrictive, in \cref{thm:concentration,thm:normalityCLT,pro:loadings_shrink} we show that it does not overly shrink $\theta_{k n}$ for factors that are needed to fit the data. Moreover, small departures from the assumed Poisson NMF likelihood do not strongly affect the number of factors used by the model.  As $N$ increases, a hyperprior on $\mu_k$ that does not depend on $N$ would be overwhelmed by the likelihood since the number of parameters $\theta_{k n}$ grows with $N$, leading to the inclusion of spurious extra signatures when the model is misspecified due to overdispersion. Meanwhile, the dependence on $N$ as in \cref{eq:hyperprior} drives out spurious extra signatures while preserving significant ones by maintaining a balance between the contribution from the exposures and from the hyperprior; see Section~\ref{subsec:comparison_fixed_comp}. 

\subsection{\revision{Choice of model settings}}\label{sec:model_settings}
We now discuss the role of the values of $K$, $\varepsilon$, $a$, and $\alpha$ in our model. 
It turns out that, while the posterior distribution is significantly affected by $a$, the precise values of $K$, $\varepsilon$, and $\alpha$ are less relevant. We recommend the following default values: $K = 20$, $\varepsilon = 0.001$, $a = 1$, and $\alpha = 0.5$, justified by the theory in \cref{sec:theory} and sensitivity analyses in Sections~\ref{sec:sensitivity_analyses} and \ref{sec:additional_application_results}.

The hyperprior mean, $\varepsilon$, is the value that the relevance weights $\mu_k$ and exposures $\theta_{k n}$ for unused signatures will be driven down to.  Thus, it is not critical to choose a particular value of $\varepsilon$, as long as it is considerably less than the smallest true nonzero exposure.

In our model, $K$ represents the maximum number of signatures that might be encountered in the data.
When the true number of signatures $K^0$ is less than $K$ and $\varepsilon$ is small, the compressive hyperprior drives the exposures $\theta_{kn}$ for the $K - K^0$ redundant factors down to~$\varepsilon$. 
In Section~\ref{subsec:hyperparams_sens}, we observe that any value of $K \geq K^0$ works equally well. Hence, this shrinkage mechanism mirrors the automatic selection of the number of active components in overfitted mixture models \citep{Rousseau_Mengersen_2011}. In our setting, we find $K = 20$ to be a good compromise between the average number of signatures found across cancer types (around 10 according to \citealp{Alexandrov_2020}), the possibility of discovering novel signatures in the data, and the additional computational cost of using a larger $K$.
More generally, if all signatures remain active during estimation, we suggest raising $K$ to a larger value and re-running the model.

The gamma shape parameter $a$ has an important and non-obvious role, which is that it serves as a threshold for inclusion of signatures in the model. More precisely, in \cref{sec:theory} we show that a signature will be included only if the average number of mutations attributed to that signature is greater than $a$ (\cref{fig:kummer}); thus, $a = 1$ is generally justified. 
Putting a prior on $a$ is not recommended, since we find that it hurts performance; see Section~\ref{subsec:Random_a_section}.

\revision{The Dirichlet concentration $\alpha$ in \cref{eq:prior_R} controls the prior entropy of the signatures. Small values of $\alpha$ lead to low-entropy signatures in which only a few channels have nonnegligible probability, whereas larger values of $\alpha$ lead to flatter, more uniform signatures.
We find $\alpha = 0.5$ to adapt well to both low- and high-entropy signatures without being excessively informative, but similar results were obtained with $\alpha = 0.25$ and $\alpha = 1$ in our simulations.}

\subsection{Posterior inference}\label{sec:inference}
Posterior inference for the hierarchical model defined by \cref{eq:Poisson,eq:prior_R,eq:prior_theta,eq:hyperprior} can be efficiently performed via Gibbs sampling. 
Since the sum of independent Poisson random variables is also Poisson distributed, we can equivalently write the NMF model in \cref{eq:Poisson} as
\begin{equation}\label{eq:hierarchical-model}
X_{m n} = \textstyle\sum_{k =1}^K Y_{m n k},\qquad  Y_{m n k}\mid \mu_k, r_k,\theta_k \; \sim\;\mathrm{Poisson}(r_{m k}\theta_{k n}).
\end{equation}
Each auxiliary variable \revision{$Y_{m n k}\in\mathds{N}$} can be interpreted as the number of mutations due to signature $k$ in channel $m$ for sample $n$.
Defining the vector $Y_{m n} = (Y_{m n 1}, \ldots, Y_{m n K})$, one can demonstrate that  $Y_{m n} \mid X_{m n},R,\boldsymbol{\Theta}$ follows a multinomial distribution
\citep{Dunson_Herring_2005}.
The rest of the sampler relies on standard conjugate updates:
\begin{itemize}[leftmargin=*, align=left]
    \item For each $m$ and $n$, update $(\bm{Y}_{m n} \mid \bm{X}, \bm{R}, \boldsymbol{\Theta}) \sim \mathrm{Multinomial}\{X_{m n}, (q_{m n 1}, \ldots, q_{m n K}) \}$ with probabilities     
    $q_{m n k} = r_{m k} \theta_{k n}/Q_{m n}$ and $Q_{m n} = \sum_{k = 1}^K r_{m k} \theta_{k n}$.
    \item For each $k$, update 
    $(\bm{r}_k \mid \bm{Y})\sim \mathrm{Dir}(\alpha + \sum_{n=1}^N Y_{1 n k},\,\ldots, \,\alpha + \sum_{n=1}^N Y_{M n k}).$
    \item For each $n$ and $k$, update $(\theta_{k n}\mid \bm{Y}, \boldsymbol{\mu}) \sim\mathrm{Ga}(a + \sum_{m=1}^M Y_{m n k},\; a/\mu_{k} + 1).$
    \item\label{alg:update-mu} For each $k$, update 
    $(\mu_k \mid \boldsymbol{\Theta})\sim\mathrm{InvGa}(2aN + 1,\; \varepsilon aN + a\sum_{n=1}^{N} \theta_{k n}).$
\end{itemize}

The model is symmetric with respect to the order of the factors, in the sense that the priors and likelihood are invariant to permutations of $k = 1,\ldots,K$. While attractive from a modeling standpoint, this symmetry could potentially lead to label switching when running the Gibbs sampler, complicating the calculation of posterior estimates.  In Section~\ref{subsec:label_switching}, we demonstrate a technique for checking whether label switching has occurred.  While we have not encountered any label switching issues, in such cases one may employ alignment algorithms similar to those proposed for Bayesian factor models \citep{Poworoznek2025}.

We do not prune inactive factors from the factorization, since the compressive hyperprior shrinks the exposures for inactive factors to a very small value, close to $\varepsilon$.
In Section~\ref{subsec:RMSE_differences}, we show that the reconstruction error between $\bm{X}$ and $\bm{R}\boldsymbol{\Theta}$ is 
nearly identical, regardless of whether we drop the inactive factors or keep them in the factorization.

\subsection{Informative priors based on known signatures}\label{subsec:infoPriors}
A favorable aspect of mutational signatures analysis is the abundance of historical data on signatures across many cancer types.
It is natural to leverage such prior information as follows. Suppose $\bm{s}_k = (s_{1 k}, \ldots, s_{M k})$, for $k = 1,\ldots,K_{\mathrm{pre}}$, are pre-defined mutational signatures known to occur in cancer, with $\sum_{m= 1}^M s_{mk} =1$. To allow for variation in signatures across studies, we let $\boldsymbol{\rho}_k = (\rho_{1 k}, \ldots, \rho_{M k})$ denote a study-specific version of $\bm{s}_k$.
We then generalize \cref{eq:Poisson} by independently modeling
\begin{equation}\label{eq:IzziModel}
X_{m n} \sim \mathrm{Poisson}\bigg(\sum_{k = 1}^{K_{\mathrm{pre}}} \rho_{m k}\omega_{k n}  + \sum_{k = 1}^{K_{\mathrm{new}}} r_{m k}\theta_{k n}\bigg),
\end{equation}
where $r_{m k}$ and $\theta_{k n}$ are given the priors in \cref{eq:prior_R,eq:prior_theta}, respectively, and
\begin{align}\label{eq:priors_IzziModel}
\boldsymbol{\rho}_k \sim\mathrm{Dir}(\beta_k s_{1 k}, \ldots, \beta_k \revision{s_{M k}}), \quad \!
\omega_{k n}\mid\psi_k \sim\mathrm{Ga}(a,\, a/\psi_k), \quad \!
\psi_k \sim \mathrm{InvGa}(a N + 1,\, \varepsilon a N).
\end{align}
Thus, the prior on $\bm{\rho}_k$ is centered at $\bm{s}_k$, while the exposures $\omega_{k n}$ and corresponding relevance weights $\psi_k$ are given the same prior and compressive hyperprior as $\theta_{k n}$ and $\mu_k$, respectively.
Gibbs sampling can be performed as in \cref{sec:inference}, with minor adjustments; see Section~\ref{sec:SamplerInform}.

Here, $\beta_k$ is the concentration of the prior for $\boldsymbol{\rho}_k$ around the point $\bm{s}_k$ in the simplex. To ensure that \cref{eq:priors_IzziModel} encodes the same prior uncertainty around each COSMIC signature, we tune $\beta_k$ so that the median cosine similarity between $\bm{s}_k$ and random draws from  $\boldsymbol{\rho}_k\sim \mathrm{Dir}(\beta_k s_{1 k}, \ldots, \beta_k s_{M k})$ is approximately 0.975; see Section~\ref{sec:SamplerInform} for details. 
The informative prior in \cref{eq:priors_IzziModel} does not ensure identifiability of the NMF parameters, which requires stricter conditions on the exposure matrix; see Ch. 4 of \citet{gillis2021nonnegative}. However, in our empirical analysis, tuning $\beta_k$ appropriately helps disentangle signatures with correlated exposures, such as those associated with similar biological processes; see Section~\ref{subsec:MergingSigs}.

The model in \cref{eq:IzziModel} is reminiscent of the recovery-discovery model discussed by \citet{Grabski_2023}, when only a single study is taken into consideration. However, unlike \citet{Grabski_2023}, we impose normalization of the signatures to avoid scaling ambiguities, and elicit the informative prior more directly. Moreover, in our framework, the prior rank $K_{\mathrm{pre}}+K_{\mathrm{new}}$ is often greater than $N$. This is at odds with the classic approach to NMF, where the factorization rank is typically smaller than the rank of $\bm{X}$. However, the compressive mechanism behind our \cref{eq:priors_IzziModel} still ensures a parsimonious representation in the posterior, such that only the active signatures have a nonnegligible relevance weight $\psi_k$. 

\section{Theory}\label{sec:theory}

\subsection{Distributional properties of the model}
We establish closed-form expressions for $\mu_k\mid \bm{Y}$ and $\theta_{k n}\mid \bm{Y}$, where $\bm{Y}$ is the tensor of latent counts $(Y_{m n k})\in \mathds{R}^{M\times N \times K}$ in \cref{eq:hierarchical-model}, and characterize their asymptotic distributions. These results help elucidate the behavior of the posterior since the distribution of $\mu_k \mid \bm{X}$ is a mixture of $\mu_k\mid \bm{Y}$ over the posterior of $\bm{Y}$ given $\bm{X}$. See Section~\ref{sec:proof} for the proofs. 

First, the density of $\mu_k\mid \bm{Y}$ can be expressed in terms of confluent hypergeometric functions. We refer to the resulting novel family of distributions as \emph{inverse Kummer}.
\begin{definition}\label{def:InvKummer}
The \emph{inverse Kummer} distribution with parameters $\lambda > 0$, $\beta > 0$, $\delta > 0$, and $\gamma \in \mathds{R}$ is a continuous distribution on $(0,\infty)$ with probability density function 
\begin{align}\label{eq:InvKumm-definition}
\pi(\mu) = \frac{\mu^{-(\lambda - \gamma) - 1}(1 + \mu/\delta)^{-\gamma}e^{-\beta/\mu}}{\delta^{\gamma - \lambda}\,\Gamma(\lambda) \, U(\lambda,\lambda + 1 - \gamma, \beta/\delta)}.
\end{align}
We write $\mu \sim \mathrm{InvKummer}(\lambda, \beta, \gamma, \delta)$ to denote that $\mu$ has the density in \cref{eq:InvKumm-definition}.
\end{definition}
Here, $U(a, b, z)$ denotes the confluent hypergeometric function of the second kind, 
$$
U(a, b, z) = \frac{1}{\Gamma(a)}\int_{0}^{\infty} t^{a-1}(1+t)^{b-a-1}e^{-zt}\mathrm{d}t,
$$
with $z>0$.
We call this an inverse Kummer distribution since if $\mu \sim \mathrm{InvKummer}(\lambda, \beta, \gamma, \delta)$ then $1/\mu$ follows a Kummer distribution, which was introduced by \citet{Armero_Bayarri_1997} when studying a $\mathrm{M/M/}\infty$ queuing problem. 
The inverse Kummer is a generalization of the inverse gamma: when $\gamma =0$, the $\delta$ parameter simplifies out of the density, and we have  $\mathrm{InvKummer}(\lambda, \beta, 0, \delta) = \mathrm{InvGa}(\lambda, \beta)$ (Proposition \ref{pro:InvGamma}).
This distribution plays a key role in our model.
Let $Y_{nk} = \sum_{m = 1}^MY_{mnk}$ be the total number of mutations in sample $n$ attributed to signature $k$ and let $\overline{Y}_k = N^{-1}\sum_{n = 1}^NY_{nk}$ be the average number of mutations attributed to $k$ across all samples $n=1,\ldots,N$.
\begin{theorem}\label{thm:InvKumPost}
Under the model in \cref{eq:hierarchical-model}  with priors in \cref{eq:prior_R,eq:prior_theta,eq:hyperprior}, 
$$(\mu_k \mid \bm{Y}) \sim \mathrm{InvKummer}\big(2 a N + 1,\, \varepsilon a N,\, N\overline{Y}_k  + a N,\, a\big).$$
\end{theorem}
Figure~\ref{fig:kummer}(A) shows the density of $\mu_k \mid \bm{Y}$ for various $\overline{Y}_k$ and $N$ values. Similarly, we show that $\theta_{k n}\mid Y$ has a closed-form density in Theorem~\ref{thm:post_theta_dist}. We report its expectation below.
\begin{theorem}\label{pro:PostTheta}
In the same setting as \cref{thm:InvKumPost},
$$
E(\theta_{kn} \mid \bm{Y}) =  (a + Y_{nk}) \,\frac{U\{2aN + 1, N(a - \overline{Y}_k) + 1, \varepsilon N\}}{U\{2aN + 1, N(a - \overline{Y}_k) + 2, \varepsilon N\}}.
$$
\end{theorem}
Despite its complicated-looking form, $E(\theta_{k n}\mid \bm{Y})$ is highly tractable and easily approximated when $N\to\infty$.  We show this in \cref{eq:compressive_vs_fixed} below.

\subsection{\revision{Asymptotic results and compressive property}}\label{sec:compressive-property}
The essence of the compressive hyperprior is that for unneeded factors, the relevance weights are shrunk to values close to $\varepsilon$, while needed factors undergo minimal shrinkage. We characterize this behavior by studying the concentration of  $\mu_k\mid \bm{Y}$ as $N\to\infty$ in the next two results, where $a_N = o(b_N)$ means that $a_N/b_N \to 0$ as $N \to \infty$.
\begin{theorem}\label{thm:concentration}
Consider the model in \cref{eq:hierarchical-model}  with priors in \cref{eq:prior_R,eq:prior_theta,eq:hyperprior}. For a given $k$, suppose that $\overline{Y}_k \to y_k$ as $N\to\infty$ for some $y_k\geq 0$.
Then, for any sequence $c_1, c_2, \ldots \in [0, \infty)$ such that $c_N \to \infty$,   $c_N N^{-1/2} \to 0$, and  $|\overline{Y}_k - y_k| = o(c_N N^{-1/2})$, we have
$$
\pr(|\mu_k - \mu_k^*| \leq \revision{c_N N^{-1/2}} \;\big\vert\; \bm{Y}) \xrightarrow[N\to\infty]{} 1,
$$
where $\mu_k^* = 2 a \varepsilon / [\{(y_k-a+\varepsilon)^2 + 8 a\varepsilon\}^{1/2} - (y_k - a + \varepsilon)]$.
\end{theorem}
This result shows that $\mu_k\mid \bm{Y}$ concentrates at a fixed point $\mu_k^*$ at a $N^{-1/2}$ rate whenever $\overline{Y}_k$ converges to $y$ at a $N^{-1/2}$ rate, which is a common assumption for a sample mean. Moreover,~the quantity $N^{1/2}(\mu_k- \mu_k^*)$ is also asymptotically normally distributed, as the next result shows.
\begin{theorem}\label{thm:normalityCLT}
In the same setting of \cref{thm:concentration}, there exist constants $D_1, D_2, \ldots \in \mathds{R}$ and $v_1, v_2, \ldots\in \mathds{R}$ such that
$$
\{N^{1/2}(\mu_k - \mu_k^*) - \Delta_N\} \mid \bm{Y} \ \xrightarrow[N\to\infty]{\mathrm{d}} \ \mathrm{N}\bigg(0, \; \frac{{\mu_k^*}^3({\mu_k^*} + a)}{2a{\mu_k^*}^2 + a^2\varepsilon}\bigg),
$$
where $\Delta_N = D_N N^{1/2} (\overline{Y}_k - y_k) + v_N$, $\lim_{N\to\infty}D_N =  {\mu_k^*}^2/\{2a\varepsilon + {\mu_k^*}(y_k - a + \varepsilon)\}$, and $v_N \to 0$. 
\end{theorem}
Here, $\stackrel{\mathrm{d}}{\to}$ denotes convergence in distribution, and $\mathrm{N}(m, s^2)$ is the normal distribution with mean $m$ and variance $s^2$.
We report the analytic form for $\Delta_N$ and the quantities $D_N$ and $v_N$ in Theorem~\ref{thm:concentration_normality_mu}, along with the proof.

\begin{figure}
\centering
    \includegraphics[width =0.95\linewidth]{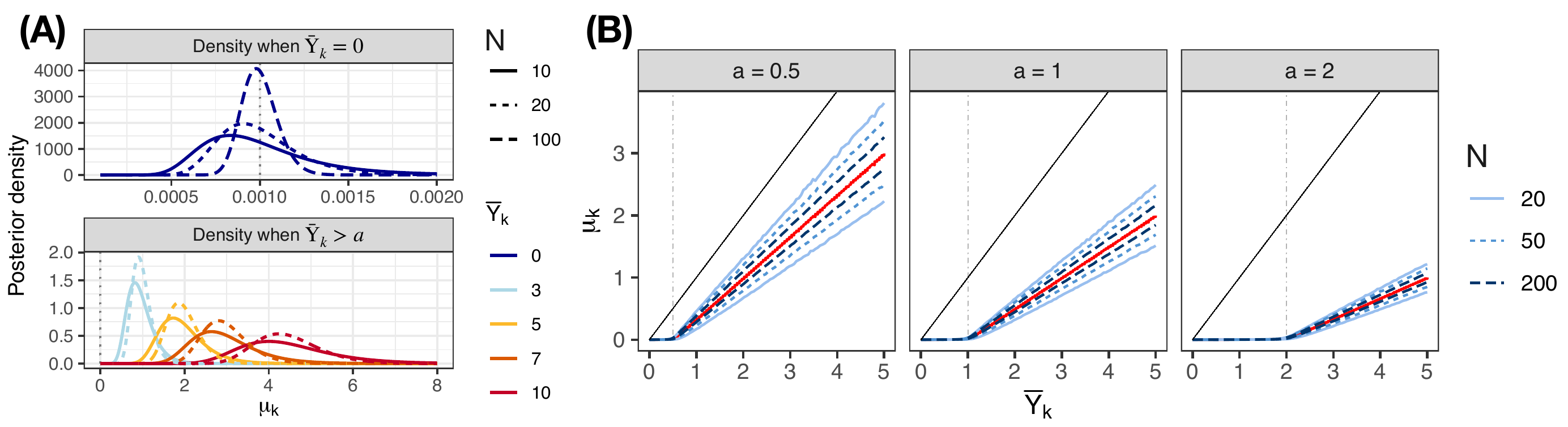}
    \caption{(A) Density of $\mu_k\mid Y$ \revision{for a range of values of $\overline{Y}_k$ and sample size $N$.} Here, $a = 1$ and $\varepsilon = 0.001$. (B) \revision{Distribution} of $\mu_k\mid Y$ as a function of $\overline{Y}_k$, for varying $N$ and $a$ (vertical line). For each distribution, we plot the 10th and 90th percentiles. \revision{The solid black line is $\mu_k = \overline{Y}_k$ and the red line is $E(\mu_k\mid\overline{Y}_k)$.} }
   \label{fig:kummer}
\end{figure}

As a function of $y_k$, the point ${\mu_k^*}$ at which concentration occurs follows an elbow-shaped curve, illustrated in \cref{fig:kummer}. This illuminating fact helps understand the sparsity-inducing effect: by a first-order Taylor approximation to the denominator, when $\varepsilon \ll |y_k - a|$,
\begin{equation}\label{eq:taylor-kummer}
    {\mu_k^*} \approx
    \begin{cases}
        \displaystyle\frac{y_k-a}{2} & \text{ if $y_k>a$} \\        \displaystyle\frac{\varepsilon a(a - y_k)}{(a - y_k)^2 + (a+y_k) \varepsilon}  & \text{ if $0\leq y_k<a$}.
    \end{cases}
\end{equation}
See Section~\ref{sec:additional_results} for the derivation. Hence, ${\mu_k^*}$ grows linearly with $y_k$ when $y_k > a$;  ${\mu_k^*}$ is close to~$\varepsilon$ when $0 < y_k < a$; and ${\mu_k^*}\approx \varepsilon a/(a + \varepsilon)$ when $y_k \approx 0$. This explains the elbow-shaped curves in \cref{fig:kummer}(B), which illustrates the distribution of $\mu_k \mid \overline{Y}_k$ for a range of $N$ and $a$ values. When $\overline{Y}_k$ is less than $a$, the inferred $\mu_k$ is negligible due to the compressive hyperprior. Since $\mu_k$ jointly controls all exposures $\theta_{kn}$, this effectively excludes signature $k$ from the decomposition. Note that in \cref{fig:kummer}(B), the distribution of $\mu_k\mid \bm{Y}$ contracts as $N$ grows, but $E(\mu_k\mid \bm{Y})$ is unaffected by $N$; see also Figure~\ref{fig:compressive_vs_fixed_elbow}. Hence, $a$ is a cutoff below which the relevance weights are shrunk to $\varepsilon$, regardless of $N$. Consequently, the following corollary provides a criterion for thresholding the relevance weights, which is helpful in practice.

\begin{corollary}\label{thm:selection}
Consider the model in \cref{eq:hierarchical-model}  with priors in \cref{eq:prior_R,eq:prior_theta,eq:hyperprior}. If $\overline{Y}_k = o(c_N N^{-1/2})$ for some $c_N\geq 0$ such that $c_N\to \infty$ and $c_N N^{-1/2}\to 0$, 
then
$$\pr(\mu_k > C\varepsilon \mid \bm{Y}) \xrightarrow[N \to \infty]{} 0 \qquad \textrm{for all} \ \ C>1.$$
\end{corollary}
Hence, when signature $k$ is not being used by the model, $\mu_k$ concentrates near small values. We refer to this as the \emph{compressive property} of the model.
This provides a natural criterion for selecting signatures for inclusion in the model, by using a threshold of $\mu_k > C\varepsilon$ to decide which signatures to keep and which to discard. Inspection of the proofs of \cref{thm:InvKumPost,thm:concentration} shows that they hold for any prior on signatures for which $\sum_{m} r_{m k} = 1$. Consequently, the above concentration results also hold for the relevance weights $\psi_k$ of the augmented model in \cref{eq:priors_IzziModel}, which employs informative priors on the signatures $\rho_k$. 

Using \cref{thm:concentration}, we establish the asymptotic distribution of $\theta_{kn}\mid Y$, both in the case of our compressive hyperprior and a fixed hyperprior, for comparison.
\begin{theorem}\label{pro:loadings_shrink}
Under the assumptions of \cref{thm:concentration}, it holds that
$$
(\theta_{kn} \mid \bm{Y}) \xrightarrow[N\to\infty]{\mathrm{d}} \mathrm{Ga}\big(a + Y_{nk},\, a/{\mu_k^*} + 1\big),
$$
with ${\mu_k^*}$ defined in \cref{thm:concentration}. 
When instead $\mu_k \sim \mathrm{InvGa}(a_0, b_0)$ for fixed $a_0, b_0>0$, we have 
$$
(\theta_{kn} \mid \bm{Y}) \xrightarrow[N\to\infty]{\mathrm{d}} \mathrm{Ga}\big(a + Y_{nk},\, a/y_k + 1\big).
$$
\end{theorem}
Note that $\theta_{kn}$ and $Y_{nk}$ do not depend on $N$, since they are specific to sample $n$ and signature~$k$. To see the difference, let $\theta_{kn}^{\textsc{c}}$ and $\theta_{kn}^{\textsc{f}}$ denote gamma random variables with these limiting distributions, \emph{compressive} (\textsc{c}) and \emph{fixed} (\textsc{f}), respectively. By \cref{eq:taylor-kummer},
\begin{equation}\label{eq:compressive_vs_fixed}
E(\theta_{kn}^{\textsc{c}}) \approx \frac{y_k-a}{y_k+a} (a + Y_{nk}), \qquad   E(\theta_{kn}^{\textsc{f}}) = \frac{y_k}{y_k+a} (a + Y_{nk}),
\end{equation}
when $y_k >a$ and $\varepsilon \ll |y_k-a|$.
\cref{eq:compressive_vs_fixed} justifies the claim that, for signatures having a significant contribution in the sense that $y_k = \lim_{N\to\infty} \overline{Y}_k$ is large relative to $a$, the exposures are not strongly biased due to using a compressive hyperprior rather than a fixed one.
In the fixed hyperprior case, the sparsity-inducing elbow shape of the curves goes away as $N$ grows since $\mu_k\mid \bm{Y}$ concentrates at $y_k$ (Theorem~\ref{thm:concentration_normality_fixed} and Figure~\ref{fig:compressive_vs_fixed_elbow}). This aggravates overestimation of the rank when the model is misspecified, as we discuss in Section~\ref{subsec:comparison_fixed_comp}.

\section{Simulations}\label{sec:simulation}
\subsection{Setup}
We conduct a simulation study to evaluate the performance of our method. Full details, including computational time and convergence diagnostics, are in Section~\ref{sec:Simulation_details}. We generate mutation counts from a negative binomial with overdispersion parameter $\tau$, using COSMIC signatures SBS1, SBS2, SBS5, and SBS13 ($K_{\mathrm{pre}}^0 = 4$) and an additional $K_{\mathrm{new}}^0$ signatures drawn from a symmetric Dirichlet distribution of dimension $M = 96$, and exposures simulated from independent gamma distributions. We consider two overdispersion settings: $\tau = 0$, in which case the negative binomial reduces to a Poisson \revision{(so the Poisson NMF model is correct)}, and $\tau = 0.15$, resulting in mild overdispersion. We consider $K_{\mathrm{new}}^0 = 2$ and $K_{\mathrm{new}}^0 = 6$, so that the total number $K^0 = K_{\mathrm{pre}}^0 + K_{\mathrm{new}}^0$ is either $6$ or $10$, and a range of sample sizes $N \in \{50,100,200\}$. For each combination of $\tau$, $M$, and $K^0$, we generate $20$ replicate sets of parameters and data matrices. 
We run: (i) CompNMF, as in \cref{eq:hierarchical-model,eq:prior_R,eq:prior_theta,eq:hyperprior}, with our default settings; (ii) CompNMF+COSMIC as in \cref{eq:IzziModel} with $K_{\mathrm{new}} = 15$ and $K_{\mathrm{pre}} = 67$ COSMIC v3.4 signatures not regarded as ``sequencing artifacts''; (iii)  PoissonCUSP \citep{Legramanti_2020}, whose structure and sampling algorithms are presented in Section~\ref{subsec:InfiniteFactors}; (iv) \textsc{signeR} \citep{Rosales_2016}; (v) SignatureAnalyzer \citep{kim2016somatic}; (vi) \textsc{SigProfilerExtractor} \citep{Islam_2022}; and (vii) BayesNMF \citep{Brouwer2017b}, which uses a Gaussian likelihood. Each is described in Section~\ref{subsec:competitors}.  

We highlight that, except for CompNMF+COSMIC, none of these can simultaneously use informative COSMIC priors and vague priors for \emph{de novo} analysis. While \textsc{signeR}, SignatureAnalyzer, and BayesNMF may potentially be adapted to perform such a task using the tuning approach of \citet{Grabski_2023} for unnormalized priors, their current implementation does not allow it. Additionally, because rank selection in PoissonCUSP relies on factors being exchangeable, using factor-specific priors would create an unnatural asymmetry that penalizes known signatures unevenly by index. Finally, SigProfilerExtractor is not Bayesian and provides no mechanism to incorporate prior information.

\subsection{Simulation results}
\begin{figure}%
    \centering
    \includegraphics[width=\linewidth]{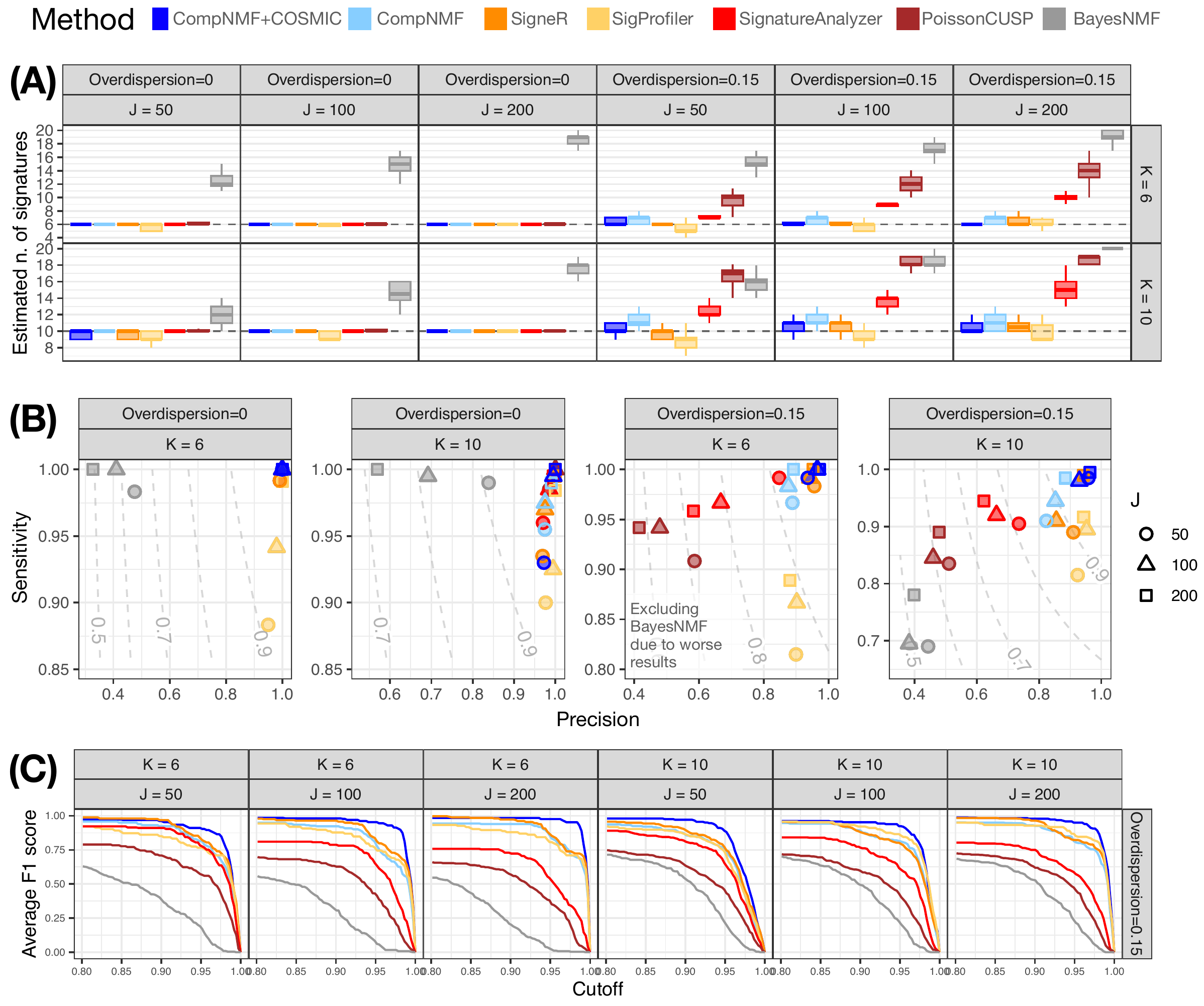}
    \caption{(A) Estimated number of signatures by each model, for varying $N$, overdispersion, and true number of signatures, across 20 replicated data set in each scenario. The horizontal grey line indicates the true number ($K^0 = 6$ or $K^0 = 10$). (B) Average precision and sensitivity across 20 replicate data sets in each scenario, with a $0.9$ cutoff for the cosine similarity. The dashed contour lines in the background indicate the $F_1$ score. (C) Average $F_1$ score as a function of the cosine similarity cutoff, across 20 replicates in each scenario, when the overdispersion is set to 0.15.
    }
    \label{fig:Simulation_results}
\end{figure}

\cref{fig:Simulation_results}(A) reports the estimated number of signatures $\widehat{K}$ for each method, for each combination of $\tau$, $N$, and $K^0$, across the 20 replicates; see Section~\ref{subsec:sim_setup} for how $\widehat{K}$ is chosen  in each case.
All methods but BayesNMF accurately estimate $K^0$ when $\tau = 0$ since the Poisson NMF model is correctly specified. Meanwhile, when $\tau = 0.15$ (mild overdispersion), we observe noticeable differences: CompNMF+COSMIC, which leverages informative priors, correctly recovers $K^0$ more often than all others both when $K^0 = 6$ and $K^0 = 10$.
CompNMF slightly overestimates $K^0$, likely due to the introduction of spurious signatures used by the model to accommodate the overdispersion. Interestingly, signeR works well even when the model is misspecified, as it tends to be conservative \citep{Rosales_2016}. 
SigProfiler also underestimates $K^0$. In contrast, SignatureAnalyzer and PoissonCUSP strongly overestimate $K^0$ under overdispersion. BayesNMF does poorly since it incorrectly uses a Gaussian likelihood.

Figure~\ref{fig:Simulation_results}(B) displays the precision and sensitivity, averaged over the 20 replicates, for each model in each setting. Here, we define \emph{precision} as the proportion of estimated signatures that have a cosine similarity $\geq 0.9$ with at least one of the ground truth signatures, and \emph{sensitivity} is the proportion of ground truth signatures for which there is an estimated signature with cosine similarity $\geq 0.9$ \citep[cutoff as in][]{Islam_2022}. 
The contour lines in  \cref{fig:Simulation_results}(B) represent the $F_1$ score, defined as $2 \times \mathrm{precision} \times \mathrm{sensitivity} / (\mathrm{precision} + \mathrm{sensitivity})$.

Points in the top-right corner of the plot in \cref{fig:Simulation_results}(B) indicate better performance. When the Poisson model is correct, all of the methods perform well.
In overdispersed settings, CompNMF+COSMIC consistently performs the best, while SignatureAnalyzer, PoissonCUSP, and BayesNMF struggle the most both in terms of precision and sensitivity. Note that here we consider values for the precision and sensitivity between 0.8 and 0.9 as low, but this interpretation is strictly dependent on the context of our cancer application.  

To see the range of performance as the cosine similarity cutoff varies, \cref{fig:Simulation_results}(C) shows the average $F_1$ score versus the cutoff value. As before, we see that CompNMF+COSMIC generally exhibits the best performance overall.
This demonstrates the benefits of our informative prior. Notably, since CompNMF+COSMIC includes all $K_{\mathrm{pre}} = 67$ COSMIC signatures---whereas only $K_{\mathrm{pre}}^0 = 4$ COSMIC signatures are used to generate the simulated data---this implies that $63$ of these signatures are correctly compressed out.%
The poor performance of SignatureAnalyzer and PoissonCUSP \revision{under misspecification} appears to be due to overfitting, since both yield the lowest root-mean-squared error for the count matrices (Section~\ref{sec:appendix_sim_results}). This is not surprising for PoissonCUSP due to its flexible nonparametric nature.

Section~\ref{sec:appendix_sim_results} reports additional results from the simulations, including estimation accuracy, computation time, and effective sample size for each method. In particular, we show that the computation time of CompNMF scales linearly with respect to each of $M$ and $N$ separately (Figure~\ref{fig:Computation_M_N}).
In Section~\ref{subsec:InfiniteFactors}, we provide further comparisons with Bayesian infinite factorization models, including the multiplicative gamma process of \citet{Bhattacharya_Dunson_2011} and the improved \textsc{cusp} proposed by \citet{Kowal_Canale2023}.
We find that CompNMF yields the best overall performance, especially under misspecification.
In Section~\ref{subsec:sparse_data}, we present a simulation under sparse indel mutation data.

\section{Application: 21 breast cancer dataset}\label{sec:application}

We apply our method to the 21 breast cancer dataset used by \citet{Nik_zainal_2012}, who first studied mutational signatures in cancer. We aim to assess the effect of our compressive hyperprior for small sample sizes and the practical advantage of informative priors. A further application on sparse pancreatic cancer indel mutations is presented in Section~\ref{sec:Indels}.

The dataset is based on whole-genome sequencing of $N = 21$ patients, and consists of mutation counts for each of the $M = 96$ channels for each patient.
The total mutation count is fairly homogeneous across patients, except for patient PD4120a, for whom a large number of mutations was detected ($70,690$) compared to the others ($18,871$ on average) due to higher sequencing coverage.
We evaluate the performance of
CompNMF and CompNMF+COSMIC 
compared to the Poisson NMF  alternatives as in \cref{sec:simulation}; see  Section~\ref{sec:additional_application_results} for details, including posterior effective sample sizes and sensitivity analyses to the CompNMF setting.

All of the methods fit the count matrix roughly equally well, with the exception of SigProfiler.
Specifically, the root-mean-squared error between the mutation count matrix~$\bm{X}$ and $\widehat{\boldsymbol{R}}\, \widehat{\boldsymbol{\Theta}}$ was 9.51 for CompNMF, 9.57 for CompNMF+COSMIC, 9.81 for signeR, 10.07 for PoissonCUSP, 10.08 for SignatureAnalyzer, and 37.08 for SigProfiler, based on the estimated signatures $\widehat{\bm{R}}$ and exposures $\widehat{\bm{\Theta}}$ for each method.

\begin{figure}[t]
    \centering
    \includegraphics[width =\linewidth]{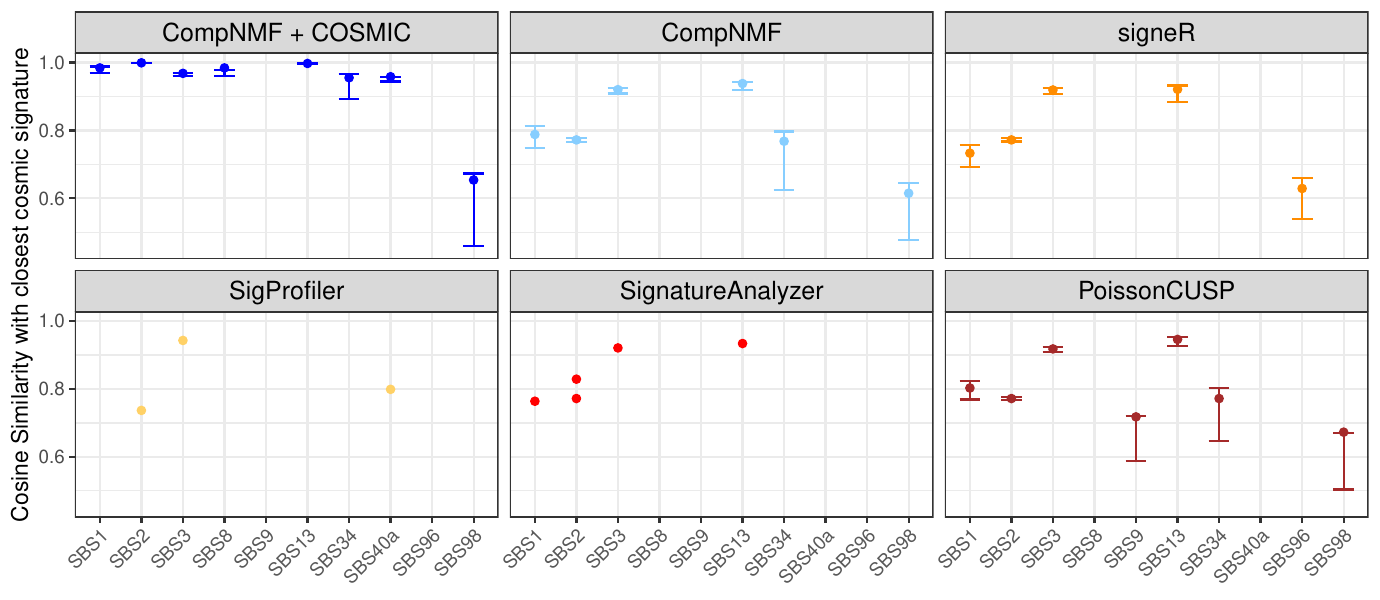}
    \caption{Cosine similarity between the signatures inferred by each method and the best matching signature in COSMIC. Vertical error bars indicate 90\% posterior credible intervals.}
    \label{fig:Uncertainty_methods}
\end{figure}
\begin{figure}[t]
    \centering
    \includegraphics[width =\linewidth]{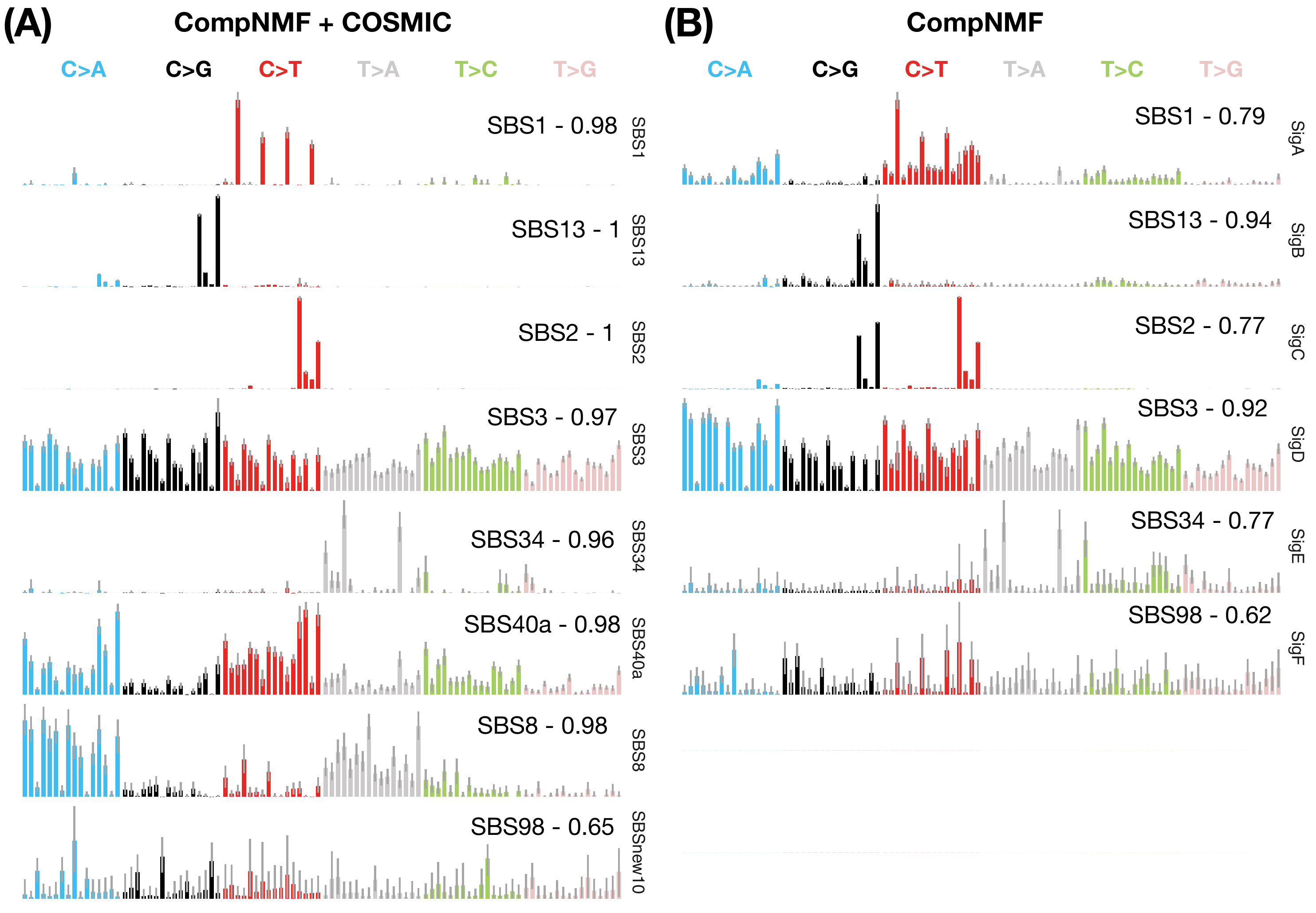}
    \caption{Signatures inferred by CompNMF+COSMIC (A) and CompNMF (B). The vertical grey lines indicate univariate 90\% posterior credible intervals for each mutational channel. 
    Model-based labels for each signature are shown to the right (vertical text), based on the informative prior for CompNMF+COSMIC or simply SigA, SigB, \dots~for CompNMF. The best matching COSMIC signature is shown in the top-right corner of each signature, along with the cosine similarity between it and the estimated signature.}
    \label{fig:Application_21br}
\end{figure}

However, major differences were observed in the estimated sets of signatures. For each method, \cref{fig:Uncertainty_methods} shows the cosine similarity between each estimated signature and the best matching one in COSMIC. Overall, CompNMF+COSMIC recovers the most signatures with the highest similarities, which is not surprising since COSMIC is used in the prior. Hence, using carefully tuned informative priors enables improved detection and estimation of signatures aligned with the prior information.
For instance, CompNMF+COSMIC detects SBS8 (associated with homologous recombination deficiency) and SBS40a \citep[ubiquitous across tumors; see][]{Alexandrov_2020}, which no other method recovers except SigProfiler, though with moderate cosine similarity.
Meanwhile, CompNMF recovers all but two of the signatures found by CompNMF+COSMIC, although with somewhat lower cosine similarities. PoissonCUSP yields similar results to CompNMF, but includes what appears to be a spurious match to SBS9 due to low cosine similarity.
Next, signeR is also similar to CompNMF, but does not recover SBS34, which does appear to be truly present since CompNMF+COSMIC detects it in many samples. SignatureAnalyzer recovers four of the same signatures as CompNMF;
however, it misses SBS34 and produces two signatures that are both matched to SBS2.
Lastly, SigProfiler recovers only three signatures.

Figures~\ref{fig:Application_21br}(A) and (B) show the estimated signature vectors $r_k$ and $\rho_k$ for our two methods (CompNMF+COSMIC and CompNMF); corresponding figures for all other methods are in Section~\ref{sec:additional_application_results}. 
The informative prior enables CompNMF+COSMIC to obtain clean signature estimates that are unambiguously matched to known COSMIC signatures, while still allowing for dataset-specific departures. 
In contrast, due to the small size of this dataset, all of the other methods produce ``merged'' signatures that are combinations of two or more COSMIC signatures.
For instance, CompNMF+COSMIC perfectly distinguishes SBS2 and SBS13 ---two APOBEC related signatures--- whereas these are merged into a single signature by all other methods; compare SBS2 and SBS13 in \cref{fig:Application_21br}(A) to CompNMF SigC in \cref{fig:Application_21br}(B). In Section~\ref{subsec:MergingSigs}, we discuss why SBS2 and SBS13 are often merged by NMF methods, and the theoretical conditions under which merged and unmerged signatures are indistinguishable. 

The vertical bars in \cref{fig:Uncertainty_methods,fig:Application_21br} indicate model-based uncertainty in the form of 90\% credible intervals, \revision{if provided}.
This uncertainty quantification is particularly useful for identifying signatures that may be spurious. For instance, for both CompNMF+COSMIC and CompNMF, the estimated signature matched to SBS98 has high uncertainty in the cosine similarity and in the signature vector itself, indicating that one should be skeptical about the validity of the estimated signature and whether it should be matched to SBS98.

\begin{figure}
    \centering
    \includegraphics[width = 0.95\linewidth]{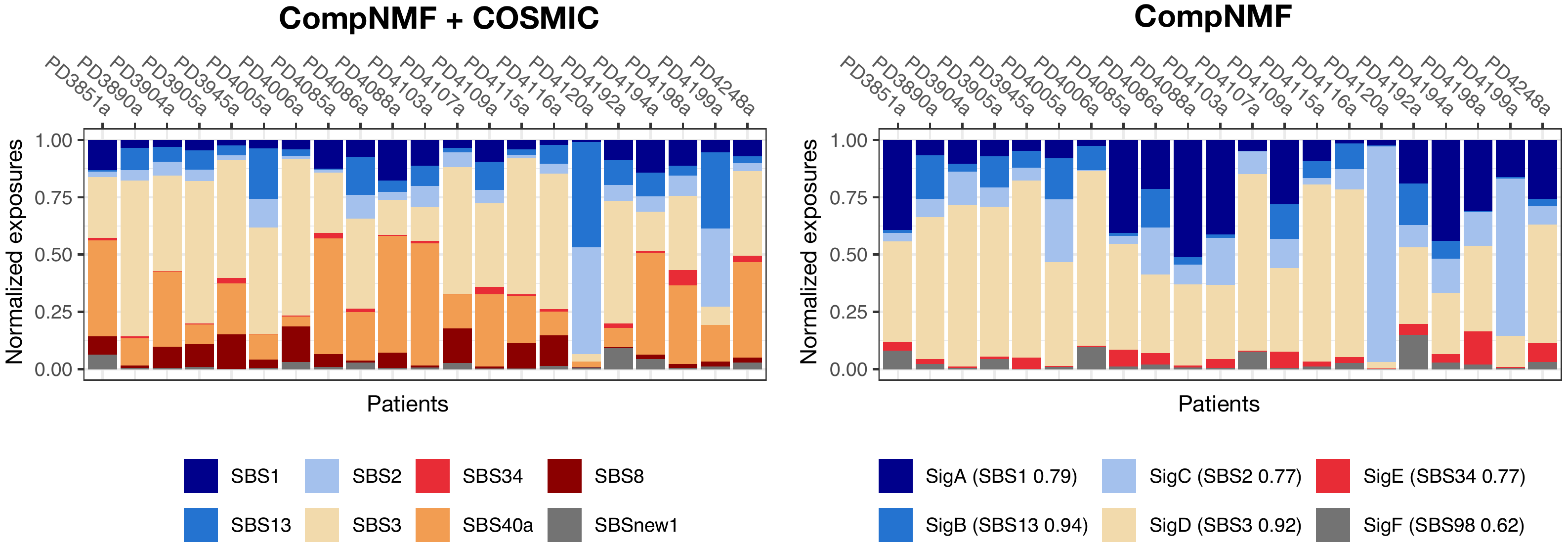}
    \caption{Estimate of exposure $\theta_{k n}$ for each patient $n$, normalized to sum to $1$ across signatures $k$, using posterior samples from CompNMF+COSMIC and CompNMF.}
\label{fig:Application_21br_weights}
\end{figure}

\cref{fig:Application_21br_weights} shows the normalized posterior estimates of the exposures $\theta_{k n}$ using our models.
The CompNMF+COSMIC exposures provide insight into the interpretation of the other methods' results. For instance, in \cref{fig:Application_21br_weights}, the exposures for SBS40a are split up and added to the exposures for SBS1 and SBS3 (\cref{fig:Application_21br}(A)) to make the CompNMF exposures for SigA and SigD, respectively (\cref{fig:Application_21br}(B)).
Similarly, SigC merges SBS2 and SBS13.  This is apparent in patient PD4120a, for whom CompNMF attributes almost all mutations to SigC, whereas CompNMF+COSMIC splits the exposure equally between SBS2 and SBS13.

\section{Discussion}\label{sec:discussion}

The compressive NMF model removes scaling ambiguities by normalizing the signatures via Dirichlet priors. When such priors are made informative, this helps disentangle correlated biological signals and reduce ambiguity in the factorization, thus aiding interpretability. However, informativeness \emph{per se} does not guarantee identifiability and uniqueness of the NMF solution, even under careful prior elicitation. 
Conditions for NMF identifiability have been developed based on requiring exposures to be \emph{separable} \citep{Donoho_2003} or \emph{sufficiently scattered} \citep{huang2014non}, but these criteria may be overly restrictive for mutational signatures analysis.  
Volume-regularized NMF has also been used for identifiability  \citep{lin2015identifiability,gillis2021nonnegative}, but such methods still lack a fully Bayesian counterpart.

There are several \revision{other} interesting follow-up directions \revision{as well}. Following \citet{Grabski_2023}, it would be interesting to jointly model multiple studies or multiple cancer types using a hierarchical model with study-specific parameters. Another useful extension would be to include sample-specific covariates, which could be helpful in improving targeted therapies \citep{Aguirre_2018}. Additionally, the applicability of the compressive hyperprior technique is potentially broader than Poisson NMF models, and might prove useful in other latent factorizations such as Gaussian ones, 
especially when having prior information on the factors.

Finally, while our method can handle mild misspecification in the form of small overdispersion, it is fundamentally based on the assumption that the counts are Poisson distributed -- like all of the leading methods considered in our empirical results.
Consequently, larger departures from the assumed Poisson NMF model can be expected to negatively impact the performance of all of these methods.
One approach is to handle overdispersion by modeling the data as negative binomial rather than Poisson \citep{Lyu_2020}. 
However, there are many other plausible sources of misspecification, and any parametric elaboration of the model will inevitably be misspecified in some way. Thus, an
area for future work is providing improved robustness to misspecification for mutational signatures analysis and NMF more generally.

\section*{Acknowledgments}
J.W.M. was supported in part by the Department of Defense grant W81XWH-18-1-0357 and by the National Cancer Institute of the NIH under award number R01CA240299. 
The content is solely the responsibility of the authors and does not necessarily represent the official views of the National Institutes of Health. 

\section*{Supplementary Materials}
\label{SM}
Web Appendices, Tables, and Figures referenced in \cref{sec:methodology,sec:theory,sec:simulation,sec:application}, and data and code to run the analyses are available with this paper at the Biometrics website on Oxford Academic.
\section*{Data Availability}
All code and data are publicly available at \url{github.com/alessandrozito/CompressiveNMF}.

\bibliographystyle{chicago}
\bibliography{references}


\newpage
\setcounter{page}{1}
\setcounter{section}{0}
\setcounter{table}{0}
\setcounter{figure}{0}
\setcounter{equation}{0}
\renewcommand{\theHsection}{SIsection.\arabic{section}}
\renewcommand{\theHtable}{SItable.\arabic{table}}
\renewcommand{\theHfigure}{SIfigure.\arabic{figure}}
\renewcommand{\thepage}{S\arabic{page}}
\renewcommand{\thesection}{S\arabic{section}}
\renewcommand{\theequation}{S\arabic{equation}}
\renewcommand{\thetable}{S\arabic{table}}
\renewcommand{\thefigure}{S\arabic{figure}}

\begin{center}
{\LARGE Supplementary Material for \\``Compressive Bayesian non-negative matrix factorization for mutational signatures analysis''}
\end{center}

\cref{sec:comparison_literature} presents an in-depth comparison of our method and existing Bayesian NMF models in the literature. \cref{sec:rationale} provides additional background on mutational signatures, including a first-principles derivation of the model.  \cref{sec:SamplerInform} gives a detailed description of the sampler for our NMF model with informative priors. \cref{sec:proof} contains the proofs of the theoretical results. \cref{sec:additional_results} provides additional results related to the inverse Kummer distribution, the marginal distribution of the latent counts, and the behavior of the model under a fixed hyperprior. \cref{sec:Simulation_details} contains additional details of the simulations in Section~\ref{sec:simulation}, including an adaptation of the Cumulative Shrinkage Process (CUSP) model of \citetSupp{Legramanti_2020} and \citetSupp{Kowal_Canale2023} and the multiplicative gamma process of \citetSupp{Bhattacharya_Dunson_2011} to the setting of Poisson NMF, and an additional simulation under sparser data. \cref{sec:sensitivity_analyses} presents sensitivity analyses for the results in the simulation, including a comparison between the compressive and the fixed hyperprior, and sensitivity to the choice of $K$, $\varepsilon$ $\alpha$, and $a$. Finally, \cref{sec:additional_application_results} provides additional results on the application, including sensitivity analysis for the 21 breast cancer dataset. \cref{sec:Indels} reports a further application to indels in pancreatic adenocarcinoma.

\counterwithin{figure}{section}
\counterwithin{table}{section}
\counterwithin{proposition}{section}
\counterwithin{theorem}{section}
\counterwithin{lemma}{section}
\counterwithin{corollary}{section}

\section{\revision{Differences with previous work}}\label{sec:comparison_literature}
\revision{In this section, we discuss the differences between our compressive NMF approach and other related methods in the literature. Specifically, \cref{subsec:tan_fevotte} highlights the differences with \citetSupp{Tan_Fevotte_2013};  \cref{subsec:NMF_ARD} discusses other automatic relevance determination (ARD) methods in the literature;  \cref{subsec:Pois_recommender} focuses on Bayesian Poisson factorization methods for recommender systems; \cref{subsec:NMF_BNP} presents an overview of Bayesian nonparametric approaches;  \cref{subsec:NMF_others} discusses other parametric Bayesian NMF models; \cref{subsec:global_local} discusses the relationship between the compressive hyperprior and continuous global-local shrinkage priors; and \cref{subsec:projectivity} discusses projectivity and Kolmogorov consistency.}

\subsection{\revision{Tan and F\'evotte's ARD-NMF algorithm}}\label{subsec:tan_fevotte}

\citetSupp{Tan_Fevotte_2013} introduce an ARD-based NMF algorithm that is used by SignatureAnalyzer \citepSupp{kim2016somatic,Alexandrov_2020}, one of the most prominent methods for mutational signatures analysis.
They present a general majorization-minimization algorithm for NMF under the penalized $\beta$-divergence loss, solving the following optimization problem:
$$
\min_{\bm{R}\geq \mathds{R}_+^{M\times K}, \boldsymbol{\Theta} \in \mathds{R}_+^{K\times N}, \lambda_k \in \mathds{R}_+} \frac{1}{\phi}\sum_{m = 1}^{M}\sum_{n = 1}^{N} d_\beta\big(X_{mn} \mid \bm{r}_k^\top \boldsymbol{\theta}_k\big) + \sum_{k = 1}^K\frac{1}{\lambda_k}\big(f(\bm{r}_k) + f(\boldsymbol{\theta}_k) + b_0\big) + c_0\log\lambda_k,
$$
where $d_\beta(x\mid y)$ is the $\beta$-divergence, $f$ is a penalty on $r_{mk}$ and $\theta_{kn}$, $\lambda_k$ are relevance weights needed for the selection, and $b$ and $c$ are constants. When $\phi = 1$ and $\beta = 1$, $d_1(x\mid y) = x \log(x/y) - x + y$ is the KL divergence, which is equivalent to the Poisson likelihood in Equation~\eqref{eq:Poisson}. Moreover, if $f$ is an $\ell_1$ penalty, then the penalty term corresponds to using exponential priors on $\theta_{k n}$ and $r_{m k}$, specifically,  $r_{mk}\mid \lambda_k \sim \mathrm{Exp}(1/\lambda_k)$ and $\theta_{kn} \mid \lambda_k \sim \mathrm{Exp}(1/\lambda_k)$, where $E(r_{mk}) = E(\theta_{kn}) = \lambda_k$, and $\lambda_k \sim \mathrm{InvGa}(a_0, b_0)$. This is an ARD structure in which $\lambda_k$ controls the mean of exposures and the signatures jointly.
In addition to developing an algorithm to solve the optimization above, they provide a data-driven approach for choosing $a_0$ and $b_0$ at every iteration. The rank $\widehat{K}$ is selected via a thresholding rule after optimization.
There are several important differences between this method and ours. 
First, \citetSupp{Tan_Fevotte_2013} use \emph{maximum a posteriori} (MAP) estimation, which only provides a point estimate of the parameters, without any uncertainty quantification. In contrast, we employ a Gibbs sampler to draw samples from the full posterior, providing uncertainty quantification for the signatures and the associated exposures. Second, the relevance weights $\lambda_k$ in the \citetSupp{Tan_Fevotte_2013} model simultaneously control $r_{mk}$ and $\theta_{kn}$, which are both unnormalized. This implies that the solution of \citetSupp{Tan_Fevotte_2013} suffers from scaling ambiguities. Meanwhile, in our model, $\mu_k$ only controls the exposures (Equation~\eqref{eq:prior_theta}) since the signatures are normalized to sum to one (Equation~\eqref{eq:prior_R}). This solves the scaling issue and allows us to easily incorporate prior knowledge on the COSMIC signatures in our framework, which are themselves normalized. It is not clear whether the \citetSupp{Tan_Fevotte_2013} approach can be extended to use such prior information, especially since their exponential priors for the signatures depend on $\lambda_k$ alone. 
The theoretical derivations that permit our use of this prior information (specifically, Theorems~\ref{thm:concentration} and \ref{thm:normalityCLT}, and Corollary~\ref{thm:selection}) are unique to our framework; for instance, their $\lambda_k$ does not have a closed-form distribution in the same way as our $\mu_k$ (given latent counts), to our knowledge. 
For these reasons, we argue that the compressive hyperprior in \ref{eq:hyperprior} has a clearer interpretation than the prior over $\lambda_k$ in \citetSupp{Tan_Fevotte_2013}, who need to tune $a_0$ and $b_0$ at every iteration of the optimization to perform their selection. Finally, the \citetSupp{Tan_Fevotte_2013} estimates degrade rapidly when the model is not exactly correct (see the SignatureAnalyzer results in  Figures~\ref{fig:Simulation_results}, \ref{fig:LambdaCount_015} and \ref{fig:SigTheta_015}), whereas our framework provides robustness to small departures from the model.

\subsection{\revision{Other ARD-based NMF methods}}\label{subsec:NMF_ARD}

\citetSupp{Brouwer2017b} consider a class of Bayesian NMF models with ARD priors that we compare with in simulations; see the BayesNMF results in  Figure~\ref{fig:Simulation_results}, \ref{fig:LambdaCount_015}, and \ref{fig:SigTheta_015}. While their relevance weights have gamma priors, their model is based on a Gaussian likelihood, which is not well-suited for our application since mutation count data is nonnegative integer-valued and may be sparse.
Consequently, BayesNMF does not perform well for mutational signatures analysis.
Moreover, \citetSupp{Brouwer2017b} do not provide an explicit rule to prune the unneeded signatures. For this reason, in our simulations, we exclude the signatures that appear relatively flat in the posterior, since empirically we observe that any extra signatures tend to be flat. 

\revision{\citetSupp{morup_hansen_kai_2009} present an NMF algorithm that normalizes the sum of squares of the signature entries, that is, $\sum_{m = 1}^M r_{mk}^2 =1$, but does not impose a non-negativity constraint $r_{m k} \geq 0$. Hence, this model is not suitable for mutational signature analysis, since the true mutation probabilities cannot be negative.}

\revision{Relatedly, \citetSupp{Rahiche_2022} present a Bayesian orthogonal NMF model for continuous data, using a Gaussian likelihood. Specifically, the prior distribution on the signature matrix is uniform on the space of orthogonal matrices, so that $R^\top R = I$ almost surely, and the model dimension is regulated via gamma ARD priors over the exposures. 
This framework is not suitable for our application, since there is no biological reason why mutational signatures should be orthogonal. Moreover, the prior on $R$ may take on negative values, which is dealt with by setting them to zero in their estimating algorithm.}

\subsection{Bayesian Poisson factorization for recommender systems}\label{subsec:Pois_recommender}
\citetSupp{Gopalan_2015} propose a model for recommendation systems that employs the same Poisson NMF likelihood as our model (Equation~\eqref{eq:Poisson}).  However, they use a different hierarchical prior structure in which $r_{mk}\mid \xi_m \sim \mathrm{Ga}(a, \xi_m)$ with $\xi_m \sim \mathrm{Ga}(a', a'/ b')$, and $\theta_{kn}\mid  \eta_n \sim \mathrm{Ga}(c, \eta_n)$  with $ \eta_n \sim \mathrm{Ga}(c', c'/ d')$,
where $m = 1, \ldots, M$ are \emph{users} and $n = 1, \ldots, N$ are \emph{items} in a recommendation system. Here, $\xi_m$ and $\eta_n$ are global weights specific to user $m$ and item $n$, respectively, and there is no weight controlling the global contribution of each $k$.  See also \citetSupp{NIPS2014_97d01458} for a closely related model. While they mention that their hierarchical gamma priors are helpful in capturing sparse factors when the shape parameters are set to small values, their model is not specifically designed for rank selection. Rather, their contribution is centered around obtaining fast inferences for the model via variational Bayes methods, and they do not explore the distributional properties of $r_{mk}$ and $\theta_{kn}$.

\citetSupp{Gopalan_2014} introduce a  nonparametric version of the model presented by \citetSupp{Gopalan_2015}.  This version is designed to infer the rank $\widehat{K}$ from the data by setting $K = \infty$ in Equation~\eqref{eq:Poisson} and employing a Bayesian nonparametric prior. Specifically, they adopt a gamma process prior: $r_{mk} = s_m \cdot v_{ik}\prod_{\ell = 1}^{k-1} (1 - v_{\ell k})$ where $v_{ik} \sim \mathrm{Beta}(1, \alpha)$, $s_m \sim \mathrm{Ga}(\alpha, c)$, and $\theta_{kn} \sim \mathrm{Ga}(a, b)$ for $k = 1, \ldots, \infty$. Notice that this structure could be equivalently imposed on the exposures $\theta_{kn}$, leaving $r_{mk}$ unconstrained. Again, \citetSupp{Gopalan_2014} rely on variational inference for posterior computation, using a finite truncation to the stick-breaking construction. Neither of these models \citepSupp{Gopalan_2014,Gopalan_2015} has an ARD structure, and they do not consider any normalization across mutational channels. 
Consequently, they exhibit the same scaling ambiguities as the method of \citetSupp{Tan_Fevotte_2013} and do not facilitate the use of informative priors based on COSMIC signatures. 

\subsection{Bayesian nonparametric factorization approaches}\label{subsec:NMF_BNP}

\revision{Along the same lines as \citetSupp{Gopalan_2014}, several methods employ a finite truncation of the gamma process to develop Bayesian nonparametric priors and select the number of factors in an NMF model. For example, \citetSupp{Hoffman_2010} consider $X_{mn} \sim \mathrm{Exp}(\sum_{k = 1}^K \vartheta_k r_{mk}\theta_{kn})$ with $r_{mk} \sim \mathrm{Ga}(a,a)$, $\theta_{kn} \sim \mathrm{Ga}(b,b)$, and  $\vartheta_k\sim \mathrm{Ga}(\alpha/K, \alpha c)$. In this case, $K$ is a truncation level for the gamma process prior, which is obtained when $K\to\infty$  \citepSupp{kingman-poisson-processes}. They mention briefly that only a few $\vartheta_k$ are large when $K$ is large, but no theoretical justification is provided. Moreover, $r_{mk}$ and $\theta_{kn}$ are all independent and identically distributed \textit{a priori}, so in particular, they do not normalize the signatures. Similar finite approximations of the gamma process for count data are considered by \citetSupp{Zhou_2018} in the context of negative binomial factorizations, and by \citetSupp{pmlr-v38-zhou15a} in an infinite community detection problem where communities are overlapping. Additionally, \citetSupp{Ayed_Caron_2021} extend the model of \citetSupp{pmlr-v38-zhou15a} by using priors based on completely random measures to perform similar community detection inference in count networks. Refer to \citetSupp{Zhou_Carin_2015} for a general account.}

\revision{In a separate thread, several Bayesian nonparametric priors based on multiplicative processes have been developed for factor analysis \citepSupp{Bhattacharya_Dunson_2011, DURANTE2017198, Legramanti_2020, Schiavon_2022, FruFruSnatta_2023}. These infer the appropriate rank of the factorization by increasingly penalizing the number of latent dimensions through the prior. A member of this family is the CUSP prior \citepSupp{Legramanti_2020}, which we compare with in our simulations; see \cref{sec:cusp} for details.}

\revision{It is worth stressing that the approaches discussed above are all finite approximations of an infinite stochastic process. Instead, our framework is a purely parametric one where $K$ is a large but fixed upper bound that does not influence the prior on $\theta_{kn}$ or $\mu_k$. A closer analogy to our compressive approach would be the mechanisms underlying the emptying out of extra components in overfitted mixture models \citepSupp{Rousseau_Mengersen_2011}. Furthermore, the introduction of informative priors based on COSMIC would result in an unnatural asymmetry in the nonparametric models described above, given the required exchangeability of the mechanism governing the rank selection. }

\revision{In a separate instance, a recent contribution from  \citetSupp{Townes_Engelhardt_2023} describes a Poisson NMF model with Gaussian process priors over the exposures to account for spatial covariates in a spatial transcriptomics application. An interesting direction would be to combine this spatial model with our ARD-based approach, especially since ARD has long been used with Gaussian processes.}

\subsection{Other Bayesian NMF models}\label{subsec:NMF_others}

\revision{The original formulation of Poisson NMF was described in \citetSupp{Cemgil_2009}, and later used in \citetSupp{Rosales_2016} and \citetSupp{SigneR_2023} for mutational signature analysis in single-study settings. These papers treat $K$ as a tuning hyperparameter and run a separate NMF decomposition for each of a range of choices of $K$, choosing the final decomposition that minimizes some information criteria. Related to this, \citetSupp{Grabski_2023} consider a multi-study version of \citetSupp{Rosales_2016} where signatures are excluded from the model through Bernoulli random variables. Similar to our framework, they also incorporate COSMIC priors in their framework. See the recent contribution of \citetSupp{hansen2025bayesianprobitmultistudynonnegative} for a further extension using probit-hyperprior structures.}

\revision{More recently, \citetSupp{lu2022_robust} and \citetSupp{lu2022_flexible} study the general NMF (both Gaussian or exponential) problem under various types of penalization and choices of hierarchical priors, respectively. Specifically, \citetSupp{lu2022_robust} presents a general class of priors that enforce a desired $L_p$ penalization for any $p\geq 1$, as protection against overfitting. Instead, \citeSupp{lu2022_flexible} considers a rectified-normal prior over (unnormalized) signatures and exposures that ensure sparsity in the final solution. Both contributions do not consider count data.}

\subsection{\revision{Global-local shrinkage priors}}\label{subsec:global_local}
\revision{Global-local shrinkage priors such as the horseshoe \citepSupp{Polson_scott_2011} bear a resemblance to our compressive hierarchical model, but there are some fundamental differences. 
In such models, there is a local parameter for each effect of interest and global parameters that govern a set of effects as a group.
Sparsity is induced by giving each global and local parameter a heavy-tailed prior with considerable mass near zero, effectively approximating spike-and-slab mixtures in a continuous way \citepSupp{Polson_scott_2011, Bhadra_2019}. 
In this way, sparsity can occur at either the individual (local) or group (global) level. See \citetSupp{Datta_Dunson_2016} and \citetSupp{Hamura_2022} for examples of horseshoe-type priors applied to Poisson likelihoods.}

Similarly, our compressive hierarchical model features both global and local parameters, in the form of relevance weights $\mu_k$ and exposures $\theta_{k n}$, respectively.
However, we do not employ spike-and-slab priors or continuous approximations thereof.
Instead, sparsity is induced via our compressive hyperprior on $\mu_k$, which shrinks the relevance weights (global parameters) and, in turn, the exposures (local parameters) for unneeded factors down to $\varepsilon$.
Finally, unlike global-local approaches, our model does not require further hierarchical structure among the $\mu_k$'s, since the values for the hyperparameters in Equation~\eqref{eq:hyperprior} are sufficient to drive the selection.

\subsection{Projectivity and Kolmogorov consistency}\label{subsec:projectivity}

In the compressive hyperprior used by our CompNMF model, the dependence on the sample size $N$ in Equation \eqref{eq:hyperprior} is a departure from the traditional Bayesian paradigm since it breaks the projectivity property of Kolmogorov consistent sequences \citepSupp[e.g.,][]{Lee2013}. In other words, marginalizing the $N$th sample does not leads to the same specification that would be obtained under $N-1$ samples. This is a limitation when extrapolating to a population is of substantial interest, but less so in problems where out-of-sample prediction is not the focus; see \citetSupp{Betancourt_2020, Zito_2026} for examples. Since mutational signatures analysis falls in this second category, lack of projectivity is not a major concern.

\section{Rationale for the model likelihood}\label{sec:rationale}

\subsection{Types of base pair substitutions}\label{subsec:substitutions}

In DNA, there are four bases: cytosine (C), thymine (T), adenine (A), and guanine (G). Considering both strands of the double helix, cytosine always pairs with guanine, and thymine always pairs with adenine. Thus, if we distinguish one of the two strands of a given DNA molecule, there are four possible base pairs at each point: C-G, G-C, T-A, and A-T.  

When considering base pair substitutions at a given point, the convention is to distinguish the strand containing the pyrimidine (C or T) before the substitution has been made. Recall that cytosine (C) and thymine (T) are \emph{pyrimidines}, whereas adenine (A) and guanine (G) are \emph{purines}. With this convention, there are six possible types of substitutions at any given point:

\begin{table}[H]
\centering
\begin{tabular}{ l c c c}
  & before & after & abbreviation\\
1 & C-G & A-T & C\textgreater A\\
2 & C-G & G-C & C\textgreater G\\
3 & C-G & T-A & C\textgreater T\\
4 & T-A & A-T & T\textgreater A\\
5 & T-A & C-G & T\textgreater C\\
6 & T-A & G-C & T\textgreater G\\
\end{tabular}
\end{table}

Sometimes, these are abbreviated denoting only the pre-substitution pyrimidine and what it changes to, as seen above. 

These six classes can be further divided by considering the trinucleotide context, that is, the bases directly adjacent to the base undergoing substitution. The convention is to label the context in terms of the bases (C, T, A, or G) on the 5' and 3' sides on the strand containing the pre-substitution pyrimidine. For instance, in a substitution C\textgreater A, the C may be flanked by a T on the 5' side and a G on the 3' side: 

\begin{table}[H]
\centering
\begin{tabular}{ c c c }
  before & after & abbreviation\\
 TCG & TAG & T[C\textgreater A]G\\
 ~5'~~~3' & ~5'~~~3' \\
\end{tabular}
\end{table}

There are $4\times 4 = 16$ different contexts for each of the original six substitution types. Therefore, there are $16 \times 6 = 96$ single-base substitution types when the trinucleotide context is taken into account. At each position in the genome, one of the two strands contains a pyrimidine C or T, flanked by bases on the 5' and 3' sides, say, X and Y, respectively: so the trinucleotide context is either XCY or XTY. Thus, each position in the genome can be in one of  $2\times 4\times 4 = 32$ possible states. Since there are three possible single-base substitutions at every position, we arrive at a total of $32 \times 3 = 96$ \emph{mutational channels}. 

\subsection{Continuous-time Markov process for substitutions}

Focusing on one position $\ell$ in the genome, let us assume mutations at $\ell$ occur as a time-homogeneous continuous-time Markov process, holding the neighboring bases fixed. More precisely, when the current state is $a$, it remains $a$ for an $\mathrm{Exponential}(|\Lambda_{a a}|)$ amount of time and then transitions to $b\neq a$ with probability $\Lambda_{a b}/|\Lambda_{a a}|$, where $\Lambda$ is a $32\times 32$ matrix such that (i) $\Lambda_{a b}\geq 0$ for $a\neq b$, and (ii) $\sum_b \Lambda_{a b} = 0$. This is equivalent to saying that transitions from $a$ to $b$ occur with rate $\Lambda_{a b}$; thus, $\Lambda$ is called the transition rate matrix.
See \citetSupp{lawler2018introduction} for background.

Let $S^t_\ell$ denote the state at locus $\ell$ at time $t$, and let $S^0_\ell$ be the state at $\ell$ for the normal (germline) genome of the individual under consideration. Let $P^t_{a b} = \pr(S^t_\ell = b \mid S^0_\ell = a)$ be the probability that the state is $b$ at time $t$ given that the state is $a$ at time 0. From the theory of continuous-time Markov processes, we have that
$$ P^t = \exp(t \Lambda) = \sum_{k = 0}^\infty \frac{(t \Lambda)^k}{k!} $$
where $\exp(\cdot)$ denotes the matrix exponential. Since the mutation rates $\Lambda_{a b}$ are very small, it is reasonable to use a first-order Taylor approximation, $P^t \approx I + t \Lambda$.

\subsection{Substitution counts are approximately Poisson distributed}

Let $a_m$ and $b_m$ denote the starting and ending states, respectively, for each of the substitution types $m = 1,\ldots,96$. Let $\lambda_m = \Lambda_{a_m b_m}$, and define $X^t_m = \#\{\ell : S^0_\ell = a_m, S^t_\ell = b_m\}$, that is, $X^t_m$ is the number of positions in the genome that undergo substitution $m$, starting at state $a_m$ at time $0$ and ending at state $b_m$ at time $t$.

Now, consider all of the positions $\ell$ that are in state $a$ at time 0, and to simplify the math, let us assume that (a) no two of these positions are adjacent, and (b) that substitions occur independently across positions.  Of the 32 states, only four of them can be reached from $a$: the state can remain at $a$, or one of three substitutions can occur. Suppose these three substitutions are $m = 1,2,3$, so that the starting states are $a_1 = a_2 = a_3 = a$ and the ending states are $b_1,b_2,b_3$, respectively.  Let $s^0 = (s_\ell^0 : \ell=1,\ldots,L)$ be a fixed vector of starting states for all positions $\ell$, and let $n = \#\{\ell : s^0_\ell = a\}$ be the number of positions starting in state $a$ at time $0$.  Let $\lambda_0 = \lambda_1+\lambda_2+\lambda_3$ be the sum of the rates for substitution types $1,2,3$. 
By the definition of $P^t_{a b}$ and the assumption of independence across positions, letting $X_0^t = X_1^t + X_2^t + X_3^t$, the vector $(X_1^t, X_2^t, X_3^t, n-X_0^t)$ follows a multinomial distribution. Specifically, for non-negative integers $x_1,x_2,x_3$ such that $x_0 := x_1+x_2+x_3 \leq n$, we have
\begin{align}\label{equation:multinomial}
\pr(X^t_{1:3} = x_{1:3} \mid S^0 = s^0) 
&= \frac{n!}{(n-x_0)! x_1! x_2! x_3!}
(P^t_{a a})^{n-x_0} \prod_{m=1}^3 (P^t_{a_m b_m})^{x_m}\notag \\
&\approx \frac{n!}{(n-x_0)! x_1! x_2! x_3!} (1-t\lambda_0)^{n-x_0} (t\lambda_1)^{x_1} (t\lambda_2)^{x_2} (t\lambda_3)^{x_3}
\end{align}
by the first-order Taylor approximation, $P^t \approx I + t \Lambda$.
Since the genome is large and mutation rates are small, it is natural to assume that $n$ is large and $t\lambda_0 = O(1/n)$.  Thus, letting $c = n t\lambda_0$ we have
$(1-t\lambda_0)^n = (1-c/n)^n \approx e^{-c} = \exp(-n t \lambda_0)$ 
and $(1-t\lambda_0)^{-x_0} = (1-c/n)^{-x_0} \approx 1$ when $x_0 \ll n$, which is the case with high probability. Plugging these approximations into \cref{equation:multinomial} yields
\begin{align*}%
&\approx \frac{n!}{(n-x_0)! x_1! x_2! x_3!} \exp(-n t \lambda_0) (t\lambda_1)^{x_1} (t\lambda_2)^{x_2} (t\lambda_3)^{x_3}\\
&= \frac{n!\, n^{-x_0}}{(n-x_0)!} \prod_{m = 1}^3 \exp(-n t\lambda_m) \frac{(n t\lambda_m)^{x_m}}{x_m!}
\end{align*}
since $x_0 = x_1 + x_2 + x_3$ and $\lambda_0 = \lambda_1 + \lambda_2 + \lambda_3$ by definition.
By Stirling's approximation, 
$$ \frac{n!\, n^{-x_0}}{(n-x_0)!} \sim 
\frac{\{2 \pi n\}^{1/2}\, (n/e)^n  n^{-x_0}}{\{2 \pi (n - x_0)\}^{1/2} \,((n - x_0)/e)^{n - x_0}} = \left(\frac{n}{n - x_0}\right)^{1/2} \frac{e^{- x_0} n^n}{(n - x_0)^n} \frac{(n - x_0)^{x_0}}{n^{x_0}} \longrightarrow 1$$
as $n\to\infty$ with $x_0$ fixed, since $(1 - x_0/n)^n \to e^{-x_0}$.  Hence, we have 
\begin{align*}%
\pr(X^t_{1:3} = x_{1:3} \mid S^0 = s^0) 
\approx \prod_{m = 1}^3 \mathrm{Poisson}(x_m \mid n t \lambda_m)
\end{align*}
when $n$ is large, $x_0\ll n$, and $t\lambda_0 = O(1/n)$.

For each of the 32 distinct possible starting states $a$, the same approximation applies to the set of positions starting in state $a$. Modeling these 32 sets of positions independently, we have 
\begin{align*}%
\pr(X^t_{1:96} = x_{1:96} \mid S^0 = s^0) 
\approx \prod_{m = 1}^{96} \mathrm{Poisson}(x_m \mid n_m t \lambda_m)
\end{align*}
where $n_m = \#\{\ell : s^0_\ell = a_m\}$. In other words, the counts of the $96$ substitution types are approximately distributed as independent Poisson random variables with rates $n_m t \lambda_m$.

The preceding derivation ignores the fact that a substitution at one position changes the context of the two adjacent positions. However, since it is rare for single-base substitutions to occur at two adjacent positions, the effect of ignoring this should be negligible.

\subsection{Multiple mutational processes}

Suppose $X_{m n}$ is the number of mutations of substitution type $m$ for subject $n$, for $m = 1,\ldots,M$ and $n = 1,\ldots,N$, where $M = 96$.  The derivation above justifies modeling these mutation counts as 
$$ X_{m n} \sim \mathrm{Poisson} (n_{m n} \lambda_{m n} t_n) $$
independently, 
where $t_n$ is the age or exposure time of subject $n$, 
$\lambda_{m n}$ is the mutation rate for substitution type $m$ in subject $n$, and $n_{m n}$ is the number of positions that are in state $a_m$ in the normal genome of subject $n$, out of all positions that were measured. The positions measured may be a subset of the genome due to whole-exome/targeted sequencing or low sequencing depth, for example. 

From birth, each subject is exposed to many mutational processes, such as environmental exposures, replication errors, defective DNA repair mechanisms, and so on.  Each mutational process causes each substitution type to occur at a given rate, and the profile of rates across the 96 substitution types can be expected to vary depending on the mutational process.
Since rates are additive in a continuous-time Markov process, it is natural to model the subject-specific mutation rates $\lambda_{m n}$ as linear combinations of these mutational process rate profiles, with non-negative weights depending on the exposure of the subject to each process. 
Further, assuming the opportunity counts $o_{m n}$ are constant (or nearly constant) across all subjects $n$, one can absorb $o_{m n}$ into $\lambda_{m n}$, which changes the interpretation of $\lambda_{m n}$ by reparametrizing it.
This leads to using a representation of the form
$$n_{m n}\lambda_{m n}t_n = \sum_{k = 1}^K r_{m k}\theta_{k n}$$
where the weight $\theta_{k n} \geq 0$ is the exposure of subject $n$ to process $k$, and $(r_{1 k},\ldots,r_{m k})$ is the mutation rate profile for mutational process $k$, which is referred to as its mutational signature.  Thus, we arrive at the Poisson non-negative matrix factorization model in Equation~\eqref{eq:Poisson},
$$ X_{m n} \sim \mathrm{Poisson} \Big(\sum_{k = 1}^K r_{m k}\theta_{k n}\Big). $$

A statistical issue with this representation is that there is a non-identifiability between the $r_{m k}$'s and $\theta_{k n}$'s, since arbitrary multiplicative constants $c_k$ can be moved between them. We deal with this by normalizing the mutational signatures to sum to $1$, that is, by enforcing the constraint $\sum_{m = 1}^M r_{m k} = 1$ for all $k$. 

\section{Computational details of the model}\label{sec:SamplerInform}
\subsection{Gibbs sampler for NMF with informative priors}\label{subsec:SamplerInform}
In this section, we present the general Gibbs sampler for the model in Equation~\eqref{eq:IzziModel} with priors as in Equation~\eqref{eq:priors_IzziModel}. Each step follows from simple semi-conjugate prior updates, so we omit the derivations. 

First, we tune the value of $\beta_k$ depending on the level of sparsity of each COSMIC signature. Specifically, for a given $s_k$, we draw 1000 samples $\boldsymbol{\rho}_k \sim \mathrm{Dir}(\beta_k s_{1 k}, \ldots, \beta_k s_{M k})$ for a range of plausible $\beta_k$ values (from 10 to 5000, evenly spaced on a log scale), and we select the value of $\beta_k$ for which the median cosine similarity between $\bm{s}_k$ and the sampled $\boldsymbol{\rho}_k$ vectors is closest to 0.975. This ensures that all signatures have approximately equal variance under the prior. Our tuning approach differs from \citetSupp{Grabski_2023}, who use informative COSMIC priors as well. In particular, \citetSupp{Grabski_2023} specify a model where the columns of $\bm{R}$ are not normalized, but follow independent gamma priors with hyperparameters elicited to represent COSMIC signatures. Their strategy involves estimating a model with $K = 1$ on synthetic data generated from the signature itself, with a constant exposure of 100, and then using the posterior parameters of the gamma to construct the priors of the larger model. Our approach is much simpler, because it involves the selection of a single hyperparameter $\beta_k$ that depends on the relative flatness of the signature.
As an alternative strategy, one can also leverage the recent approach of \citeSupp{xue2024}, which elicits the hyperparameters of a Dirichlet distribution based on a target location and a notion of distance, such as cosine similarity. 

Inference in the CompNMF+cosmic model is performed by iterating the following steps.
\begin{enumerate}[leftmargin=*, align=left]
    \item For $m = 1, \ldots, M$ and $n = 1,\ldots, N$, update the latent mutation counts by drawing
    $$(\bm{Y}_{m n} \mid -) \sim \mathrm{Multinomial}\big(X_{m n}, (\widetilde{q}_{m n 1}, \ldots, \widetilde{q}_{m n K_{\mathrm{pre}}} , q_{m n 1}, \ldots, q_{m n K_{\mathrm{new}}})\big)$$ where $\widetilde{q}_{m n k} =  \rho_{m k} \omega_{k n}/Q_{m n}$ and $q_{m n k} = r_{m k} \theta_{k n}/Q_{m n}$, with $Q_{m n} = \sum_{k = 1}^{K_{\mathrm{pre}}} \rho_{m k} \omega_{k n} + \sum_{k = 1}^{K_{\mathrm{new}}} r_{m k} \theta_{k n}$.
    \item For $k = 1,\ldots, K_{\mathrm{pre}}$, update the COSMIC signatures by drawing
    $$(\boldsymbol{\rho}_k \mid - )\sim \mathrm{Dir}\bigg(\beta_k s_{1 k} + \sum_{n=1}^N Y_{1 n k}, \ldots, \beta_k s_{M k} + \sum_{n=1}^N Y_{M n k}\bigg).$$
    \item For $k = K_{\mathrm{new}} + 1,\ldots, K_{\mathrm{pre}} + K_{\mathrm{new}}$, update the \emph{de novo} signatures by drawing
    $$(\bm{r}_k \mid - )\sim \mathrm{Dir}\bigg(\alpha + \sum_{n=1}^N Y_{1 n k},\,\ldots, \,\alpha + \sum_{n=1}^N Y_{M n k}\bigg).$$
    \item For $k = 1,\ldots, K_{\mathrm{pre}}$ and $n = 1, \ldots, N$, update the exposures associated to the COSMIC signatures by drawing
    $$(\omega_{k n}\mid-) \sim\mathrm{Ga}\bigg(b + \sum_{m=1}^M Y_{m n k},\; \frac{b}{\psi_{k}} + 1\bigg).$$
    \item For $k = K_{\mathrm{new}} + 1,\ldots, K_{\mathrm{pre}} + K_{\mathrm{new}}$ and $n = 1,\ldots, N$, update the exposures associated to the \emph{de novo} signatures by drawing
    $$(\theta_{k n}\mid-) \sim\mathrm{Ga}\bigg(a + \sum_{m=1}^M Y_{m n k},\; \frac{a}{\mu_{k}} + 1\bigg).$$
    \item For $k =  1,\ldots, K_{\mathrm{pre}}$, update the relevance weights associated to the COSMIC signatures by drawing 
    $$(\psi_k \mid -)\sim\mathrm{InvGa}\bigg(2bN + 1,\; \varepsilon bN + b\sum_{n=1}^{N} \omega_{k n}\bigg).$$
    \item For $k = K_{\mathrm{new}} + 1,\ldots, K_{\mathrm{pre}} + K_{\mathrm{new}}$, update the relevance weights associated to the \emph{de novo} signatures  by drawing 
    $$(\mu_k \mid -)\sim\mathrm{InvGa}\bigg(2aN + 1,\; \varepsilon aN + a\sum_{n=1}^{N} \theta_{k n}\bigg).$$
\end{enumerate}

\subsection{Prior-posterior alignment for the informative prior}\label{subsec:PriorPosterioMismatch}
When using the sampler described above, we occasionally observe a mismatch between the informative prior and the posterior samples for some signatures. 
To explain, consider a model signature with prior $\boldsymbol{\rho}_k \sim \mathrm{Dir}(\beta_{k}s_{1k}, \ldots, \beta_{k} s_{Mk})$ centered at known COSMIC signature $\bm{s}_k$.
It can happen that (a) the posterior samples for $\boldsymbol{\rho}_k$ concentrate near another COSMIC signature $\bm{s}_{k'}$ rather than $\bm{s}_k$, or (b) $\boldsymbol{\rho}_k$ is inactivated by the model ($\psi_k \approx \varepsilon$) and another signature $\bm{\rho}_{k'}$ or $\bm{r}_{k'}$ concentrates near $\bm{s}_k$.
These issues hinder interpretation of the results in terms of the COSMIC signatures, and may hurt estimation performance since the informative prior does not have the desired effect on posterior samples of signatures that are mismatched in this way.

While tuning each $\beta_k$ as above does help, such mismatches can sometimes still occur for relatively flat COSMIC signatures, such as SBS3, SBS5, or SBS40a,b.
To properly align the posterior samples to the informative prior, 
we apply a prior-posterior matching step at 2/3 of the burn-in phase, where the signatures that have not been compressed out of the model are re-matched to the COSMIC signatures via the Hungarian algorithm \citepSupp{Hungarian_cit}. 
The Hungarian algorithm is  designed to solve the minimum cost bipartite matching problem in $O(\hat{K}^3)$ time, where we set the cost function to be the cosine distance between any pair of non-compressed signatures and $\hat{K}$ is the estimated number of signatures. Note that this matching step does not invalidate the sampling algorithm since it is only performed in the burn-in phase; hence, it can be viewed as an initialization step designed to ensure proper alignment of the posterior to the informative prior.

\begin{figure}[t]
    \centering
    \includegraphics[width=\linewidth]{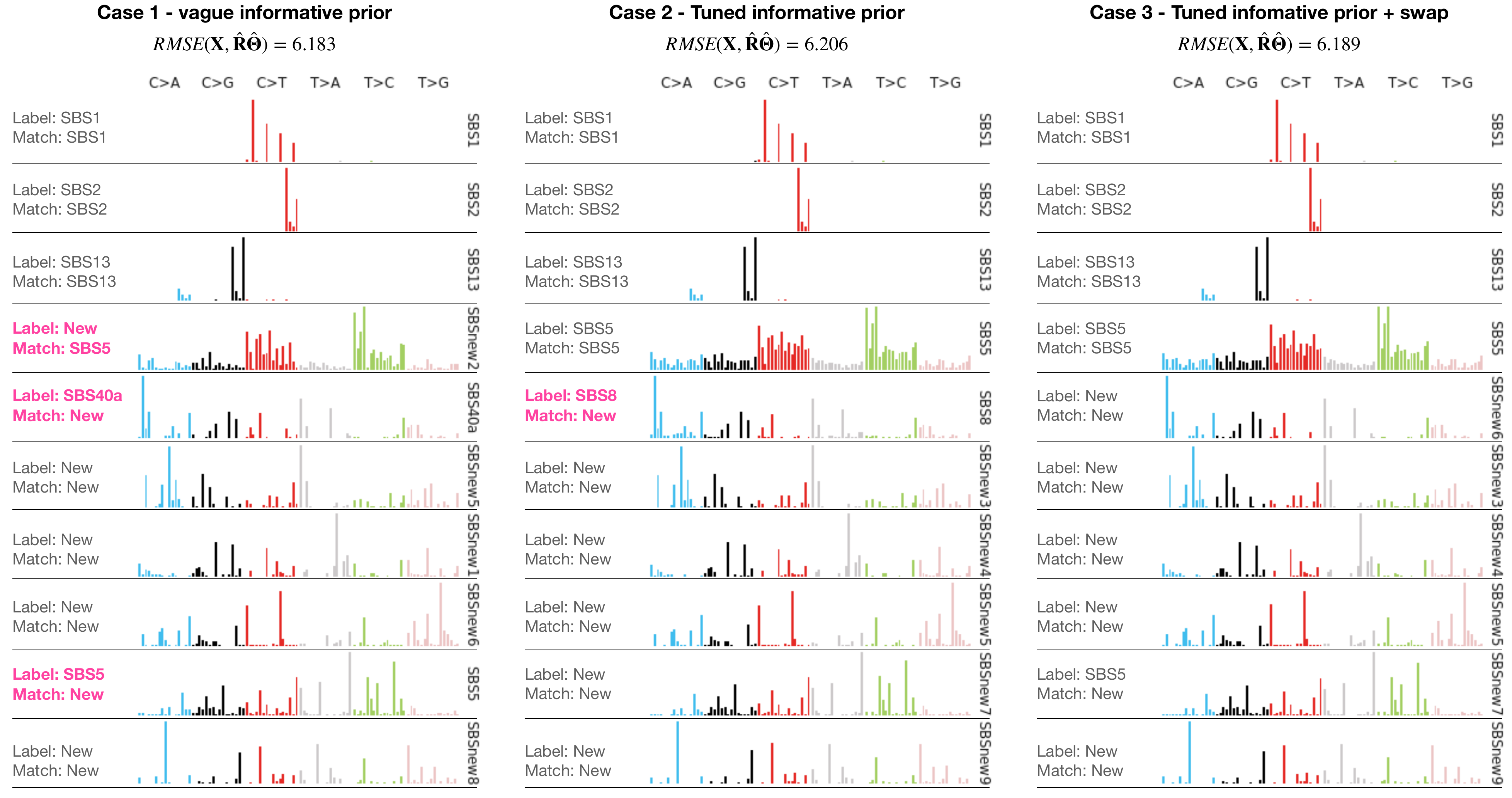}
    \caption{Posterior estimate of $\bm{R}$ from CompNMF+cosmic on one simulated dataset with overdispersion $\tau = 0.15$, $K_{\textrm{new}} = 6$, $N = 100$, and $K^{\textrm{pre}} =4$ that include SBS1, SBS2, SBS5 and SBS13. Left: value informative prior where $\beta_k = 100$ in every signature; Center: solution under carefully elicited $\beta_k$; Right: solution when $\beta_k$ is tuned and we perform a prior-swap during burn-in. Vertical text on the right indicates the label of the signature, while the text in pink on the left highlights a mismatch between the prior and the posterior mean.}
    \label{fig:Mismatch}
\end{figure}

\cref{fig:Mismatch} provides an example of how the procedure works on simulated data. We consider one simulated dataset under the misspecified case of Section~\ref{sec:simulation}, with $N = 100$,  $K_{\textrm{new}} = 6$, and where signatures SBS1, SBS2, SBS5, and SBS13 are used to simulate the data. We fit the model in Section~\ref{subsec:infoPriors} by letting $K_{\textrm{new}} = 10$ and $K_{\textrm{pre}} = 8$, using signatures SBS1, 2, 13, 3, 5, 8, 34, and 40a in the prior. We consider three cases: (1) a model where each $\beta_k$ is set to 100 for each signature; (2) a model where each $\beta_k$ is tuned with the approach described above, but no matching step is performed; and (3) the same as case (2), but with the prior-posterior matching step at 2/3 of the burn-in phase. For each model, we run the Gibbs sampler for 3,000 iterations, using the last 1,000 samples for inference.

The posterior estimate of $\bm{R}$ under each of the three models is displayed in \cref{fig:Mismatch}, for the signatures that are active. 
For notational convenience, here we write $\bm{R}$ to denote the matrix containing all of the $\bm{\rho}_k$'s and $\bm{r}_k$'s; likewise, $\boldsymbol{\Theta}$ denotes the matrix containing all of the $\bm{\omega}_k$'s and $\bm{\theta}_k$'s.
In this example, all three models have approximately the same reconstruction error, since the RMSEs between $\bm{X}$ and $\widehat{\bm{R}}\widehat{\boldsymbol{\Theta}}$ are almost identical. Moreover, the same signature profiles are retrieved in the three cases. The major difference lies in the prior index for each signature. In particular, model (1) recovers SBS5 but starting from a flat signature in the prior (SBSnew2). Meanwhile, it takes the signatures whose priors are centered at SBS5 and SBS40a and morphs them into two of the novel simulated signatures, respectively. 
In model (2), this behavior occurs less frequently but remains present for SBS8, which is morphed into a new signature.  
In model (3), these issues are no longer observed, due to the prior-posterior matching step. Finally, we notice that these prior-posterior swaps tend to happen for relatively flat signatures such as SBS5 and SBS40s, and not for sparser signatures such as SBS1, 2, and 13.

\subsection{Diagnostic check for label switching}\label{subsec:label_switching}

We describe and demonstrate a technique for detecting label switching in the posterior samples from the model. The likelihood of the NMF model is invariant to permutation of the factors, since the product $\bm{R}\boldsymbol{\Theta}$ can be equivalently written as $\widetilde{\bm{R}}\widetilde{\boldsymbol{\Theta}}$, where $\widetilde{\bm{R}} = \bm{R}\bm{P}^\top$ and 
$\widetilde{\boldsymbol{\Theta}} = \bm{P}\boldsymbol{\Theta}$ for any $K\times K$ permutation matrix $\bm{P}$; see \citetSupp{gillis2021nonnegative}, Ch.\ 4. This implies that label switching may occur while running the Gibbs sampler, that is, signatures and their corresponding exposures may swap indices in the middle of the chain without any effect on the likelihood. Label switching is a recurring issue in Bayesian factor analysis and mixture models \citepSupp{Jasra_2005}, and can be dealt with using \emph{post hoc} alignment steps \citepSupp[e.g.,][]{Stephens2002, Poworoznek2025} or priors that impose specific constraints \citepSupp{FruttaSnatta_2024} to form meaningful posterior estimates.

We use the following procedure to detect potential label-switching issues in CompNMF.
First, we estimate the signature matrix $\widehat{\bm{R}}^*$ using the last $Q = 0.01 T$ samples of a chain of $T$ post burn-in iterations; then, we calculate the cosine similarity (column-by-column) between $\widehat{\bm{R}}^*$ and each posterior sample $\bm{R}^{(1)}, \ldots, \bm{R}^{(T - Q)}$. If there is no label switching, then we expect 
$\mathrm{cosine}(\widehat{\bm{r}}^*_k, \bm{r}_{k}^{(t)}) > \mathrm{cosine}(\widehat{\bm{r}}^*_k, \bm{r}_{k'}^{(t)})$ for all $k'\neq k$, for all $t$ after burn-in; otherwise, label switching may have occurred.

\begin{figure}[!h]
    \centering
    \includegraphics[width=\linewidth]{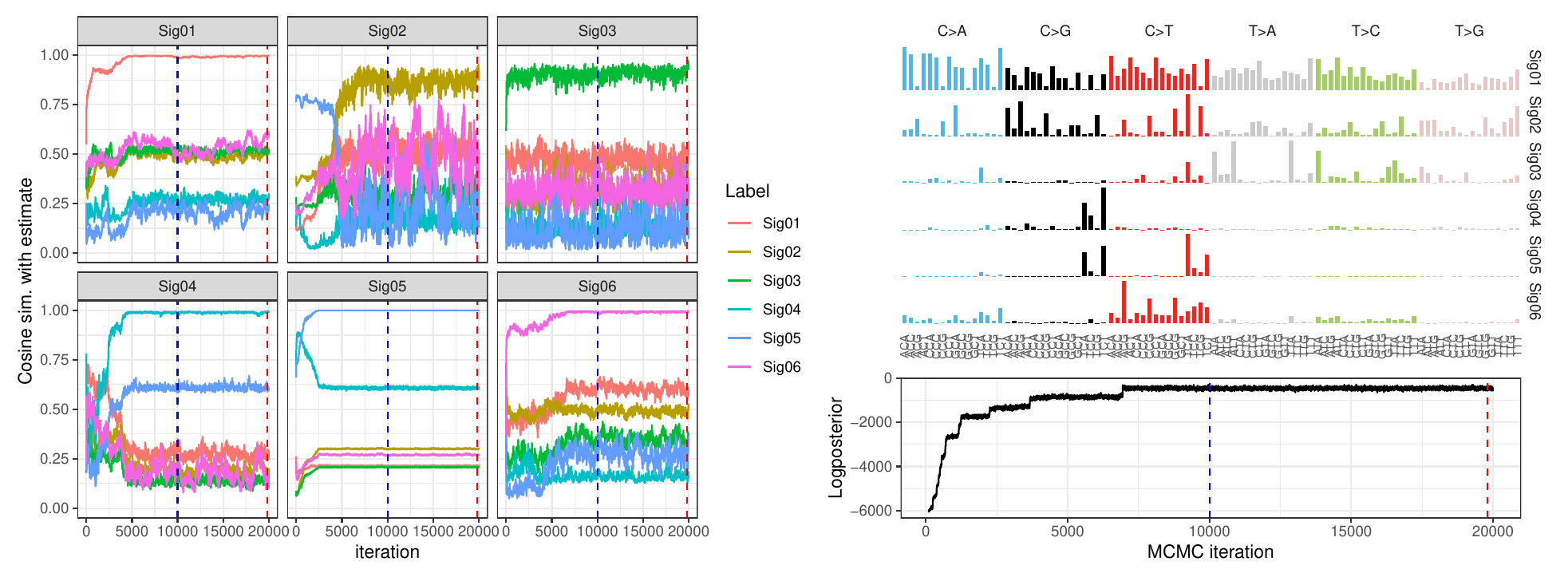}
    \caption{Label switching detection in the 21 breast cancer dataset under CompNMF. Left: cosine similarity between the columns of the signature matrix estimated on the last 1\% of the chain and the preceding MCMC samples. Panel names indicate the column index of $\widehat{\bm{R}}^*$, and line colors indicate the labels of columns of each MCMC sample. Right: top panel is the matrix $\widehat{\bm{R}}^*$, bottom panel is a traceplot of the log posterior. }
    \label{fig:LabelSwitching21breast}
\end{figure}

\cref{fig:LabelSwitching21breast} displays the results of applying this procedure to the CompNMF output for the 21 breast cancer dataset, for a chain of 20,000 iterations. The dashed blue line on the left panel indicates the burn-in phase, while the dashed red line marks the part of the chain used to estimate $\widehat{\bm{R}}^*$, shown in the top-right panel. We also display the log posterior to assess convergence. For every signature $k$, the cosine similarity between the estimate $\widehat{\bm{r}}^*_k$ and the corresponding posterior samples $\bm{r}_k^{(t)}$ 
consistently exceeds the similarity with samples $\bm{r}_{k'}^{(t)}$ for any other $k' \neq k$.
This indicates that the chain maintains the index of each signature throughout sampling, with no evidence of label switching. We observe similar results across all datasets simulated in Section~\ref{sec:simulation}. 

\section{Proofs}\label{sec:proof}

\noindent\textbf{Proof of Theorem~\ref{thm:InvKumPost}.}
The proof proceeds by directly marginalizing $\bm{r}_k$ and $\boldsymbol{\theta}_k$ from the joint distribution. For notational simplicity, we first handle the general case of $\mu_k\sim\mathrm{InvGa}(a_0, b_0)$, and then plug in the values of $a_0$ and $b_0$ for the compressive hyperprior. Under the model in Equation~\eqref{eq:hierarchical-model} with priors as in Equations~\eqref{eq:prior_R}, \eqref{eq:prior_theta}, and \eqref{eq:hyperprior},
the joint density of $\bm{Y} = (Y_{m n k})$, $\boldsymbol{\Theta} = (\theta_{k n})$, $\bm{R} = (r_{m k})$, and $\boldsymbol{\mu} = (\mu_k)$ is
\begin{equation}\label{eq:joint-latent}
\begin{aligned}
\pi(\bm{Y}, \boldsymbol{\Theta}, \bm{R}, \boldsymbol{\mu}) \propto 
     &\bigg\{\prod_{m, n, k} e^{-r_{m k}\theta_{k n}}  \frac{(r_{m k}\theta_{k n})^{Y_{m n k}}}{Y_{m n k}!}\bigg\} \times  &\textrm{(Latent Poisson counts)} \\
 & \bigg\{\prod_{m, k} r_{m k}^{\alpha - 1}  \bigg\}\times  &\textrm{(Dirichlet prior on $\bm{r}_k$)} \\
 & \bigg\{\prod_{n, k} \Big(\frac{a}{\mu_k}\Big)^a \theta_{k n}^{a -1} e^{- a\theta_{k n}/\mu_k} \bigg\}\times  &\textrm{(Gamma prior on $\boldsymbol{\theta}_k$)} \\
 & \bigg\{\prod_{k} \mu_{k}^{-a_0 - 1} e^{-b_0/\mu_k} \bigg\}, &\textrm{(InvGamma prior on $\mu_k$)}
\end{aligned}
\end{equation}
dropping constants of proportionality.
Since $\sum_{m} r_{m k} = 1$ for all $k$, we have $\prod_{m} e^{-r_{m k}\theta_{k n}} = e^{-\theta_{k n}}$.  Thus, by \cref{eq:joint-latent}, we have
\begin{align}
    \pi(\bm{Y}, \boldsymbol{\Theta}, \bm{R}, \boldsymbol{\mu}) = f(\bm{Y}, \bm{R})\, 
    \bigg\{\prod_{n,k} \theta_{k n}^{\sum_{m} Y_{m n k}+a-1} 
 e^{-\theta_{k n} - a\theta_{k n}/\mu_k} 
    \bigg\}
    \bigg\{\prod_{k} \mu_{k}^{-N a - a_0 - 1} e^{-b_0/\mu_k}\bigg\}
\end{align}
where $f(\bm{Y}, \bm{R})$ is a function that does not depend on $\boldsymbol{\Theta}$ or $\boldsymbol{\mu}$. Hence,
\begin{align*}
    \pi(\boldsymbol{\mu} \mid \bm{Y}) &\underset{\boldsymbol{\mu}}{\propto} \pi(\bm{Y},\boldsymbol{\mu}) = \int \int \pi(\bm{Y},\boldsymbol{\Theta},\bm{R},\boldsymbol{\mu}) \, d\boldsymbol{\Theta}\, d \bm{R} \\
    &= \bigg\{\int f(\bm{Y},\bm{R})\,d \bm{R}\bigg\} \bigg\{\prod_{k} \mu_{k}^{-N a - a_0 - 1} e^{-b_0/\mu_k}\bigg\} \bigg\{ 
    \prod_{n,k}\int \theta_{k n}^{\sum_{m} Y_{m n k}+a-1} 
 e^{-\theta_{k n} - a\theta_{k n}/\mu_k} 
     d\theta_{k n} \bigg\} \\
    &\underset{\boldsymbol{\mu}}{\propto}  \bigg\{\prod_{k} \mu_{k}^{-N a - a_0 - 1} e^{-b_0/\mu_k}\bigg\} \bigg\{ 
    \prod_{n,k} \frac{\Gamma(a +\sum_{m} Y_{m n k})}{(1 + a/\mu_k)^{a + \sum_{m} Y_{m n k}}} \bigg\}.
\end{align*}
Since this factors over $k$ into products that depend on $\mu_1,\ldots,\mu_K$, respectively, it follows that
\begin{align}
    \pi(\mu_k\mid \bm{Y}) &\propto \mu_{k}^{-N a - a_0 - 1} e^{-b_0/\mu_k} \Big(1 + \frac{a}{\mu_k}\Big)^{-N a - \sum_{m,n} Y_{m n k}} \nonumber \\
    &\propto \mu_{k}^{-(a_0 - \sum_{m,n}Y_{m n k}) - 1}  \Big(\frac{\mu_k}{a} + 1\Big)^{-N a - \sum_{m,n} Y_{m n k}} e^{-b_0/\mu_k} \nonumber\\
    &\propto \mathrm{InvKummer}\Big(\mu_k\;\Big\vert\; a_0 + N a,\, b_0, \, N a + \sum_{m,n} Y_{m n k},\, a\Big) \label{eq:invKumm_post_generic}
\end{align}
by Equation~\eqref{def:InvKummer}. The proof is completed by letting $a_0 = a N + 1$ and $b_0 = \varepsilon a N$. 

\vspace{1em}
\noindent\textbf{Proof of Theorem~\ref{pro:PostTheta}.}
Recall that $(\theta_{kn}\mid \bm{Y}, \mu_k) \sim \mathrm{Ga}(a + Y_{nk}, a/\mu_k + 1)$. Therefore, $E(\theta_{kn}\mid \bm{Y}, \mu_k) = (a + Y_{nk}) \mu_k/(\mu_k + a)$. By the law of iterated expectation,
$$
E(\theta_{kn}\mid \bm{Y}) = E\big(E(\theta_{kn}\mid \bm{Y},\mu_k)\,\big\vert\,\bm{Y}\big) = (a + Y_{nk})E\bigg(\frac{\mu_k}{\mu_k + a}\,\bigg\vert\,\bm{Y}\bigg).
$$
This last expectation is available analytically in terms of Kummer hypergeometric functions, as follows.
As in the proof of Theorem~\ref{thm:InvKumPost}, we first consider $\mu_k\sim\mathrm{InvGa}(a_0,b_0)$ for generic $a_0$ and $b_0$.
Define $C = a^{N\overline{Y}_k -a_0}\Gamma(a_0 + aN)U(a_0 + aN, a_0 + 1 - N \overline{Y}_k, b_0/a)$, noting that by Equation~\eqref{eq:InvKumm-definition}, $C$ can be interpreted as the normalizing constant of the density of $\mu_k\mid \bm{Y}$, that is, the constant of proportionality in \cref{eq:invKumm_post_generic}.  Then
\begin{align*}
E(\theta_{kn}\mid \bm{Y}) &= (a + Y_{nk})E\bigg(\frac{\mu_k/a}{\mu_k/a + 1}\,\bigg\vert\,\bm{Y}\bigg) \\
&= \frac{a + Y_{nk}}{aC} \int_{0}^{\infty} \mu_k^{-(a_0 - N\overline{Y}_k - 1)-1}\bigg(\frac{\mu_k}{a} + 1\bigg)^{-aN - N\overline{Y}_k - 1}e^{-b_0/\mu_k} \mathrm{d}\mu_k \\
&= \frac{a + Y_{nk}}{aC} a^{N\overline{Y}_k + 1 - a_0}\Gamma(a_0 + aN)U(a_0 + aN, a_0 - N\overline{Y}_k, b_0/a) \\
&= (a + Y_{nk})\frac{U(a_0 + aN, a_0 - N\overline{Y}_k, b_0/a)}{U(a_0 + aN, a_0 + 1 - N \overline{Y}_k, b_0/a)}.
\end{align*}
The proof is completed by letting $a_0 = a N + 1$ and $b_0 = \varepsilon a N$. 

The following preliminary results are used in the proof of Theorems \ref{thm:concentration} and \ref{thm:normalityCLT}.

\begin{lemma}\label{thm:Kummer_normality}
\revision{Let $\varepsilon >0$, $a > 0$, and $y_N\geq 0$ such that $y_N \to y$ for some $y\in[0, \infty)$. Let $T_N$ be the random variable on $(0,\infty)$ with probability density function
\begin{align}\label{eq:f_n}
f_N(t) \propto t^{2aN} (1+t)^{-Ny_N - aN}e^{-\varepsilon N t}
\end{align}
for $t > 0$. Then
$$
N^{1/2}(T_N - t_N^*)\ \xrightarrow[N \to \infty]{\mathrm{d}} \ \mathrm{N}\left(0,\; \frac{1 + t^*}{2 a / (t^*)^2 + \varepsilon}\right),
$$
where $t_N^* = \big[\{(y_N - a + \varepsilon_N)^2 + 8a\varepsilon_N\}^{1/2} - (y_N - a + \varepsilon_N)\big]/(2\varepsilon_N)$, $\varepsilon_N = \varepsilon - 1/N$, and $t^* = \lim_{N\to\infty} t_N^*$. }
\end{lemma}
\begin{proof}
The proof follows by direct application of the results in \citetSupp{Miller_2021}. We write $f_N(t) \propto \exp(-Ng_N(t))e^{-t}$, where $g_N(t) = -2 a\log(t) + (y_N + a)\log(1+t) + (\varepsilon - 1/N) t$, to put the density of $T_N$ in the form considered by  \citetSupp{Miller_2021}, with $\Theta = (0,\infty)$ and $\pi(t) = e^{-t}$.
In \cref{lem:miller_checklist}, we show that the functions $g_N(t)$ and the point $t^*$ satisfy the assumptions of Theorem 5 in \citetSupp{Miller_2021}.
Thus, the density of $N^{1/2}(T_N - t^*_N)$ converges in total variation to $\mathrm{N}(0, 1/g''(t^*))$, where $g''(t^*) = (2 a / (t^*)^2 + \varepsilon)/(1 + t^*)$ by \cref{lem:miller_checklist}, part 5. Convergence in total variation implies convergence in distribution, by the portmanteau lemma. 
\end{proof}

For $x\in\mathds{R}$ and $\delta \geq 0$, define $B_{\delta}(x) = \{t\in \mathds{R} : |t - x| < \delta\}$.

\begin{lemma}\label{lem:miller_checklist}
Let $\varepsilon >0$, $a > 0$, and $y_N\geq 0$ such that $y_N \to y$ for some $y\in[0, \infty)$, and define
\begin{align}
g_N(t) &= -2 a\log(t) + (y_N + a)\log(1+t) + (\varepsilon - 1/N) t, \label{eq:g_n_new}\\
g(t) &= -2 a\log(t) + (y + a)\log(1+t) + \varepsilon t, \label{eq:g_new}
\end{align}
for $t > 0$. Define $t^* = \big[\{(y - a + \varepsilon)^2 + 8a\varepsilon\}^{1/2} - (y - a + \varepsilon)\big]/(2\varepsilon)$ and let $\delta \in (0, t^*)$. Then
\begin{enumerate}[label=\textup{(\arabic*)}, leftmargin=*, align=left]
    \item $g_N$ has continuous third derivatives and $(g_N''')$ is uniformly bounded on $B_{\delta}(t^*)$,
    \item $g_N \to g$ pointwise,
    \item $g(t) > g(t^*)$ for all $t \in (0,\infty) \setminus \{t^*\}$,
    \item $\mathrm{liminf}_{N\to\infty} \inf_{t \in (0,\infty)\setminus K} g_N(t) > g(t^*)$ where $K = \{t\in\mathds{R} : |t - t^*| \leq \delta/2\}$,    
    \item $g''(t^*) = (2 a / (t^*)^2 + \varepsilon)/(1 + t^*) > 0$. 
\end{enumerate}
\end{lemma}

\begin{proof}
(1) \revision{The derivatives of $g_N$ are
\begin{align}
    g_N'(t) &= -\frac{2a}{t} + \frac{y_N + a}{1 + t} + \varepsilon - \frac{1}{N}, \label{eq:gn_deriv1}\\
    g_N''(t) &= \frac{2a}{t^2} - \frac{y_N + a}{(1 + t)^2}, \label{eq:gn_deriv2}\\
    g_N'''(t) &= -\frac{4a}{t^3} + \frac{2(y_N + a)}{(1 + t)^3}. \label{eq:gn_deriv3}
\end{align}
Thus, $g_N'''(t)$ is continuous at all $t\in (0, \infty)$ and is uniformly bounded on $B_\delta(t^*)$ since
$$
|g_N'''(t)| \leq \left|-\frac{4a}{t^3}\right| + \left|\frac{2(y_N + a)}{(1 + t)^3}\right| \leq  \frac{4 a}{t^3} + \frac{2 (y_N + a)}{t^3} \leq \frac{2\sup_N y_N + 6a}{(t^* - \delta)^3} < \infty
$$
for all $N\geq 1$ and all $t\in B_{\delta}(t^*)$, since $y_N \to y$ and $\delta < t^*$ by assumption.} 

(2) \revision{This follows directly from the assumption that $y_N \to y$ as $N \to \infty$.}

(3) \revision{Similarly to \cref{eq:gn_deriv1}, $$g'(t) = -2a/t + (y + a)/(1+t) + \varepsilon.$$ Setting $g'(t) = 0$ and solving via the quadratic formula, we find that any critical point must occur at 
$\widehat{t} = (-b \pm \{b^2 + 8a\varepsilon\}^{1/2})/(2\varepsilon)$, where $b = y-a+\varepsilon$. Since $|-b| = (b^2)^{1/2} < \{b^2 + 8a\varepsilon\}^{1/2}$, the only solution on $(0,\infty)$ is $\widehat{t} = t^*$, where $t^*$ is defined in the statement of the lemma. Thus, $g'(t)$ has a unique zero on $(0,\infty)$.  Since $g'(t) \to -\infty$ as $t\to 0$, and $g'(t) \to \varepsilon$ as $t\to\infty$, the intermediate value theorem implies that $g'(t) < 0$ for all $t\in(0,t^*)$ and $g'(t) > 0$ for all $t\in(t^*,\infty)$. Therefore, by the fundamental theorem of calculus, $g(t)$ is strictly monotone decreasing on $(0,t^*]$ and strictly monotone increasing on $[t^*,\infty)$. This implies that $g(t)$ is uniquely minimized at $t^*$, as claimed.}

(4) \revision{Let $\varepsilon_N = \varepsilon - 1/N$, and suppose $N$ is large enough that $\varepsilon_N > 0$.  By applying the argument in (3) to $g_N$ instead of $g$, we have that $g_N$ is strictly monotone decreasing on $(0,t_N^*)$ and strictly monotone increasing on $(t_N^*,\infty)$, where $t_N^* = \big(\{(y_N - a + \varepsilon_N)^2 + 8a\varepsilon_N\}^{1/2} - (y_N - a + \varepsilon_N)\big)/(2\varepsilon_N)$. 
Note that $t_N^* \to t^*$ as $N \to \infty$, since $y_N \to y$ and $\varepsilon_N \to \varepsilon$. Let $k_1 = t^* - \delta/2$ and $k_2 = t^* + \delta/2$, so that $K = [k_1,k_2]$.
Choose $N_0$ such that for all $N \geq N_0$, we have (i) $\varepsilon_N > 0$, (ii) $t_N^*\in K$, (iii) $g_N(k_1) \geq (g(k_1) + g(t^*))/2$, and (iv) $g_N(k_2) \geq (g(k_2) + g(t^*))/2$, using (3) and the fact that $g_N \to g$ pointwise.
Let $c = (\min\{g(k_1),g(k_2)\} + g(t^*))/2$, noting that $c > g(t^*)$.  Then for all $N \geq N_0$, we have (a) if $0 < t < k_1$ then $g_N(t) \geq g_N(k_1) \geq c$ since $g_N$ is monotone decreasing on $(0,t_N^*)$, and (b) if $t > k_2$ then $g_N(t) \geq g_N(k_2) \geq c$ since $g_N$ is monotone increasing on $(t_N^*,\infty)$.  Therefore, $$\liminf_{N\to\infty} \inf_{t\in(0,\infty)\setminus K} g_N(t) \geq c > g(t^*).$$}

(5) From the proof of (3), we have $g'(t^*) = 0$, and thus, $(y + a)/(1 + t^*) = 2 a / t^* - \varepsilon$. Therefore, 
\begin{align*}
g''(t^*) = \frac{2a}{(t^*)^2} - \frac{y + a}{(1 + t^*)^2} = \frac{2a}{(t^*)^2} - \frac{2 a/t^* - \varepsilon}{1 + t^*} = \frac{2 a/(t^*)^2 + \varepsilon}{1 + t^*} > 0.
\end{align*}

\end{proof}

\begin{theorem}\label{thm:Kummer_concentration}
In the setting of $\cref{thm:Kummer_normality}$,
for any sequence $d_1, d_2, \ldots \in [0, \infty)$ such that $d_N \to\infty$ and $|y_N - y| = o(d_N/N^{1/2})$, we have
$$
\pr(|T_N - t^*| \geq d_N/N^{1/2}) \xrightarrow[N \to \infty]{} 0
$$
where $t^*$ is defined as in \cref{thm:Kummer_normality}.
\end{theorem}

\begin{proof}
    \revision{By \cref{lem:convergence_tstar}, $t_N^* - t^* = C_N (y_N - y) + q_N$ where $C_N$ converges to a constant and $q_N = O(1/N)$. Thus,
    $$ \frac{N^{1/2}}{d_N}|t_N^* - t^*| \leq |C_N|\frac{N^{1/2}}{d_N}|y_N - y| + \frac{N^{1/2}}{d_N}|q_N| \longrightarrow 0 $$
    as $N\to\infty$, since $|y_N - y| = o(d_N/N^{1/2})$ and $d_N\to\infty$ by assumption. This implies that $d_N - N^{1/2}|t_N^* - t^*| = d_N(1 - (N^{1/2}/d_N)|t_N^* - t^*|) \to \infty$, since $d_N\to\infty$.} 
    By the triangle inequality, 
    $|T_N - t^*| \leq |T_N - t_N^*|+ |t_N^* - t^*|$. Thus, 
    \begin{equation*}
    \begin{split}
    \pr(|T_N - t^*| \geq d_N/N^{1/2}) &\leq  \pr(|T_N - t_N^*|+ |t_N^* - t^*| \geq d_N/N^{1/2}) \\
    &=  \pr(|N^{1/2}(T_N - t_N^*)| \geq d_N - N^{1/2}|t_N^* - t^*|) \xrightarrow[N\to\infty]{} 0,
    \end{split}
    \end{equation*}
    because $N^{1/2}(T_N - t_N^*)$ converges in distribution to a normal distribution. 
\end{proof}
\begin{lemma}\label{lem:convergence_tstar}
In the setting of \cref{thm:Kummer_normality}, we have
\begin{align}\label{eq:convergence_tstar}
    t_N^* - t^* = C_N(y_N - y) + q_N
\end{align}
where $C_N \to -t^*/\{(y-a+\varepsilon)^2 + 8 a\varepsilon\}^{1/2}$ and $q_N = O(1/N)$.
\end{lemma}
\begin{proof}
\revision{
Define $x_N = y_N - a + \varepsilon$ and $x = y - a + \varepsilon$. Then $x_N \to x$ and $x_N - x = y_N - y$. In terms of $x_N$ and $x$, we have
\begin{equation*}
\begin{split}
t_N^* &= \frac{\{(x_N - 1/N)^2 + 8a(\varepsilon - 1/N)\}^{1/2} - x_N + 1/N}{2(\varepsilon - 1/N)}, \\
t^* &= \frac{\{x^2 + 8a\varepsilon\}^{1/2} - x}{2\varepsilon}.
\end{split}
\end{equation*}
For any $z_1, z_2\geq 0$, we have $(z_1- z_2)(z_1+ z_2) = z_1^2 - z_2^2$ and $(z_1^{1/2} - z_2^{1/2})(z_1^{1/2} + z_2^{1/2}) = z_1 - z_2$. Letting $k_N = \{(x_N - 1/N)^2 + 8a(\varepsilon - 1/N)\}^{1/2} + \{x^2 + 8a\varepsilon\}^{1/2}$, we manipulate $t_N^* - t^*$ as follows:
\begin{align*}
t_N^* - t^* &= \frac{1}{2(\varepsilon-1/N)}\left(\{(x_N - 1/N)^2 + 8a(\varepsilon - 1/N)\}^{1/2} - x_N + 1/N - 2(\varepsilon -1/N) t^* \right) 
\\
&= \frac{1}{2(\varepsilon-1/N)}\left(\{(x_N - 1/N)^2 + 8a(\varepsilon - 1/N)\}^{1/2} - \{x^2 + 8a\varepsilon\}^{1/2} - (x_N - x) + \frac{2t^* + 1}{N}\right) 
\\
&= \frac{1}{2(\varepsilon-1/N)}\left( \frac{(x_N - 1/N)^2 - x^2 - 8a/N}{k_N} - (x_N - x) + \frac{2t^* + 1}{N} \right) 
\\
&= \frac{1}{2(\varepsilon-1/N)}\left(\frac{x_N^2 - x^2 -2x_N/N + 1/N^2 - 8a/N}{k_N} - (x_N - x) + \frac{2t^* + 1}{N} \right) \\
&= \frac{1}{2(\varepsilon-1/N)}\left( \frac{(x_N - x)(x_N + x) -2x_N/N + 1/N^2 - 8a/N}{k_N} - (x_N - x) + \frac{2t^* + 1}{N} \right) \\
&= \frac{1}{2(\varepsilon-1/N)}\left(\Big( \frac{x_N + x}{k_N} - 1\Big)(x_N - x) - \frac{2x_N + 8a}{k_NN} + \frac{1}{k_NN^2} +\frac{2t^* + 1}{N} \right) \\
&= \frac{1}{2(\varepsilon-1/N)}\Big( \frac{x_N + x}{k_N} - 1\Big)(y_N - y) + q_N \\
&= C_N (y_N - y) + q_N
\end{align*}
where 
$$ C_N = \frac{1}{2(\varepsilon-1/N)}\Big( \frac{x_N + x}{k_N} - 1\Big) $$
and
\begin{equation}\label{eq:additional_drift}
q_N =  \frac{1}{2(\varepsilon-1/N)}\left(- \frac{2x_N + 8a}{k_NN} + \frac{1}{k_NN^2} +\frac{2t^* + 1}{N}\right).
\end{equation}
}
Since $x_N\to x$, it follows that $k_N \to 2\{x^2 + 8a\varepsilon\}^{1/2} > 0$. Therefore, $C_N \to -t^* / \{x^2 + 8 a \varepsilon\}^{1/2}$ and $q_N = O(1/N)$.
\end{proof}

\vspace{1em}
\begin{theorem}\label{thm:concentration_normality_mu}
Let $\varepsilon >0$, $a > 0$, and $y_N\geq 0$ such that $y_N\to y$ for some $y\in[0,\infty)$.
If $\mu_N\sim \mathrm{InvKummer}(2aN + 1, \varepsilon a N, Ny_N + aN, a)$ then we have the following results. 
\begin{enumerate}[label=\textup{(\arabic*)}, leftmargin=*, align=left]
    \item For any sequence $c_1, c_2, \ldots \in [0, \infty)$ such that $c_N \to\infty$,
    $c_N/N^{1/2}\to 0$, and $|y_N - y| = o(c_N/N^{1/2})$,
    we have $$\pr(|\mu_N - \mu_*| \geq c_N/N^{1/2}) \xrightarrow[N \to \infty]{} 0$$
    where $\mu_* = 2 a \varepsilon / ( \{(y - a+ \varepsilon)^2 +  8a\varepsilon\}^{1/2}- (y - a+ \varepsilon))$.
    \item There exist constants $D_1,D_2,\ldots \in \mathds{R}$ and $v_1, v_2, \ldots \in \mathds{R}$ such that
    $$
    N^{1/2}(\mu_N - \mu_*) - \Delta_N \xrightarrow[N \to \infty]{\mathrm{d}} \mathrm{N}\left(0,\; \frac{\mu_*^3(\mu_* + a)}{2a\mu_*^2 + a^2\varepsilon}\right),
    $$
    where $\Delta_N = D_NN^{1/2}(y_N - y) + v_N$, $\lim_{N \to \infty} D_N = \mu_*^2/(2a\varepsilon + \mu_*(y - a + \varepsilon))$, and $v_N \to 0$.
\end{enumerate} 
\end{theorem}
\begin{proof} \revision{Let $\mu_N\sim \mathrm{InvKummer}(2aN + 1, \varepsilon a N, Ny_N + aN, a)$ and define $T_N = a/\mu_N$.  Then $T_N$ has the density $f_N(t)$ studied in  \cref{thm:Kummer_normality} and \cref{thm:Kummer_concentration}; see \cref{eq:f_n}.}

(1)
\revision{
Pick $N$ large enough so that $\mu_* - c_N/N^{1/2} > 0$. Then, $|\mu_N - \mu_*| \leq c_N/N^{1/2}$ if and only if $\mu_* - c_N/N^{1/2} \leq \mu_N \leq \mu_* + c_N/N^{1/2}$, or equivalently, $a/(\mu_* + c_N/N^{1/2}) \leq a/\mu_N \leq a/(\mu_* - c_N/N^{1/2})$. Define $d_1, d_2, \ldots$ such that
$$
d_N = \left(\frac{a/\mu_*}{\mu_* + c_N/N^{1/2}}\right)c_N,
$$
and observe that $d_N\to\infty$ and $d_N/N^{1/2} \to 0$. Moreover, $|y_N - y| = o(d_N/N^{1/2})$ since 
\begin{align}\label{eq:yn_little_o}
\frac{|y_N - y|}{d_N/N^{1/2}} = \frac{\mu_* + c_N/N^{1/2}}{a/\mu_*}\,\frac{|y_N - y|}{c_N/N^{1/2}} \xrightarrow[N\to\infty]{} 0
\end{align}
since $c_N/N^{1/2}\to 0$ and $|y_N - y|=o(c_N/N^{1/2})$ by assumption.
Then for all $N$ sufficiently large,
\begin{equation}\label{eq:ineq_sequences_chain}
\frac{a}{\mu_* + c_N/N^{1/2}} \stackrel{(\mathrm{a})}{=} \frac{a}{\mu_*} - \frac{d_N}{N^{1/2}} \stackrel{(\mathrm{b})}{=} t^* - \frac{d_N}{N^{1/2}} \leq t^* + \frac{d_N}{N^{1/2}} \stackrel{(\mathrm{c})}{=} \frac{a}{\mu_*} + \frac{d_N}{N^{1/2}}  \stackrel{(\mathrm{d})}{\leq} \frac{a}{\mu_* - c_N/N^{1/2}}
\end{equation}
where $t^*$ is defined as in \cref{thm:Kummer_normality}.
Step (a) holds since $c_N = \mu_* d_N / (a/\mu_* - d_N/N^{1/2})$.  Steps (b) and (c) hold since $t^* = a/\mu_*$. To justify step (d), first note that for any $x\in(0,1/2)$, we have $1 + x \leq (1-x)/(1 - 2 x)$ because $(1 + x) (1 - 2 x) = 1 - x - 2 x^2 \leq 1 - x$.  Choosing $x = \mu_* d_N / (aN^{1/2})$, we have $x \in (0,1/2)$ for all $N$ sufficiently large because $d_N/N^{1/2} \to 0$, and thus,
\begin{align*}
\frac{a}{\mu_*} + \frac{d_N}{N^{1/2}} =  \frac{a}{\mu_*}\Big(1 + \frac{\mu_* d_N}{a N^{1/2}}\Big)
\leq  \frac{a}{\mu_*}\Big( \frac{1 - \mu_* d_N / (aN^{1/2})}{1 - 2\mu_* d_N / (aN^{1/2})}\Big) 
= \frac{a}{\mu_* - c_N/N^{1/2}}.
\end{align*}
\cref{thm:Kummer_concentration} applies since $y_N\to y$, $d_N\to\infty$, and $|y_N - y| = o(d_N/N^{1/2})$ by \cref{eq:yn_little_o}.
Together, \cref{thm:Kummer_concentration} and \cref{eq:ineq_sequences_chain} imply that
\begin{equation*}
1 = \lim_{N\to\infty}  \pr(|T_N - t^*|\leq d_N/N^{1/2}) \leq \lim_{N\to\infty} \pr(|\mu_N - \mu_*| \leq c_N/N^{1/2})
\end{equation*}
concluding the proof of (1).}

(2)
By \cref{thm:Kummer_normality},
$$
N^{1/2}(T_N - t_N^*) \xrightarrow[N \to \infty]{\mathrm{d}} \ \mathrm{N}\left(0, \;\frac{1 + t^*}{2 a / (t^*)^2 + \varepsilon}\right)
$$
and $t_N^* \to t^*$.
Hence, by the delta method \citepSupp[Theorem 3.8]{Van_der_Vaart_2000},
\begin{align}\label{eq:delta_method}
N^{1/2}(h(T_N) - h(t_N^*)) \xrightarrow[N \to \infty]{\mathrm{d}} \ \mathrm{N}\left(0, \;h'(t^*)^2\frac{1 + t^*}{2 a / (t^*)^2 + \varepsilon}\right)
\end{align}
where $h(t) = a/t$ for $t > 0$.
Since $\mu_N = a/T_N = h(T_N)$ and $\mu_* = a/t^* = h(t^*)$,
\begin{align}\label{eq:delta_addsubtract}
N^{1/2}(\mu_N - \mu_*) = N^{1/2}(h(T_N) - h(t^*)) = N^{1/2}(h(T_N) - h(t_N^*)) + N^{1/2}(h(t_N^*) - h(t^*)).
\end{align}
Therefore, combining \cref{eq:delta_addsubtract,eq:delta_method} along with $h'(t) = -a/t^2$ and writing the variance in terms of $\mu_*$ yields
\begin{align}\label{eq:delta_conclusion}
N^{1/2}(\mu_N - \mu_*) - \Delta_N  \xrightarrow[N \to \infty]{\mathrm{d}} \ \mathrm{N}\left(0, \;\frac{\mu_*^3(\mu_* + a)}{2a\mu_*^2 + a^2\varepsilon}\right)
\end{align}
where $\Delta_N = N^{1/2}(h(t_N^*) - h(t^*)) =N^{1/2}(a/t_N^* - a/t^*) = -aN^{1/2}(t_N^* - t^*)/(t_N^*t^*)$. We can further characterize $\Delta_N$ using the proof of \cref{lem:convergence_tstar}. Specifically, 
$$
t^*_N - t^* = \frac{1}{2(\varepsilon-1/N)}\Big( \frac{y_N + y - 2 a + 2\varepsilon}{k_N} - 1\Big)(y_N - y) + q_N, 
$$ 
where $q_N$ is defined in \cref{eq:additional_drift} and satisfies $q_N = O(1/N)$. Hence, 
$$\Delta_N = N^{1/2}(y_N - y)\underbrace{\frac{-a}{2t_N^*t^*(\varepsilon - 1/N)}\left(\frac{y_N + y - 2 a + 2\varepsilon}{k_N}-1\right)}_{\displaystyle D_N} + \underbrace{\frac{-aN^{1/2}\,q_N}{t_N^*t^*}}_{\displaystyle v_N}.$$
Note that $v_N\to 0$ since $q_N = O(1/N)$ and $t_N^* \to t^*$.  Finally, since $a/t_N^* \to a/t^* = \mu_*$ and $k_N \to 2\{(y - a + \varepsilon)^2 + 8 a \varepsilon\}^{1/2}$,
\begin{align*}
\lim_{N\to\infty} D_N &= - \frac{\mu_*^2}{2a\varepsilon}\Big(\frac{y - a + \varepsilon}{\{(y - a + \varepsilon)^2 + 8a\varepsilon\}^{1/2}} - 1\Big) \\
&= \frac{\mu_*}{\{(y - a + \varepsilon)^2 + 8a\varepsilon\}^{1/2}} \\
&= \frac{\mu_*^2}{2a\varepsilon + \mu_*(y - a + \varepsilon)}.
\end{align*}

\end{proof}

\noindent\textbf{Proof of Theorems~\ref{thm:concentration} and \ref{thm:normalityCLT}.}
\cref{thm:concentration_normality_mu} is simply a restatement of Theorem~\ref{thm:concentration} and of Theorem~\ref{thm:normalityCLT} in more familiar notation, where $y_N = \overline{Y}_k$ and $\mu_N$ is distributed according to $\mu_k\mid \bm{Y}$. 

\vspace{1em}

\noindent\textbf{Proof of Corollary~\ref{thm:selection}.}
Suppose $\overline{Y}_k = o(c_N/N^{1/2})$ where $c_N\to\infty$ and $c_N/N^{1/2}\to 0$.
Then $y = 0$ in Theorem~\ref{thm:concentration}, and therefore the concentration point is $\mu_* = 2a\varepsilon/(\{(\varepsilon - a)^2 + 8 a\varepsilon\}^{1/2} - (\varepsilon-a))$. It holds that $\mu_* < \varepsilon$, since 
$(\varepsilon - a)^2 + 8 a \varepsilon > (\varepsilon - a)^2 + 4 a \varepsilon = (\varepsilon + a)^2$, and hence,
$\{(\varepsilon - a)^2 + 8 a \varepsilon\}^{1/2} - (\varepsilon - a) > 2 a$.
Let $d = C\varepsilon - \mu_*$, noting that $d>0$ since $C>1$ and $\mu_* < \varepsilon$. Then by Theorem~\ref{thm:concentration}, for all $N$ sufficiently large that $c_N/N^{1/2} \leq d$,
$$
\pr(\mu_k > C\varepsilon\mid \bm{Y}) 
\leq  \pr(|\mu_k - \mu_*| > d\mid \bm{Y})
\revision{\, \leq \pr(|\mu_k - \mu_*| \geq c_N/N^{1/2} \mid \bm{Y})} \longrightarrow 0
$$
as $N\to\infty$. 

\noindent\textbf{Proof of Theorem~\ref{pro:loadings_shrink}.}
From Section~\ref{sec:inference}, we have $(\theta_{kn}\mid \bm{Y}, \mu_k) \sim \mathrm{Ga}(a + Y_{nk}, a/\mu_k + 1)$ where $Y_{n k} = \sum_{m=1}^M Y_{m n k}$. We use the Laplace transform of $(\theta_{kn}\mid \bm{Y})$ to characterize its limit as $N\to\infty$. For all $t>0$,
\begin{align*}
E(e^{-t\theta_{kn}} \mid \bm{Y}) &= E\big(E(e^{-t\theta_{kn}} \mid \bm{Y}, \mu_k)\,\big\vert\, \bm{Y}\big) \\
&= E\Big(\Big(\frac{a/\mu_k + 1}{t + a/\mu_k +1}\Big)^{a + Y_{nk}}
\,\Big\vert\, \bm{Y}\Big) \\
&\xrightarrow[N\to\infty]{} \Big(\frac{a/{\mu_k^*} + 1}{t + a/{\mu_k^*} +1}\Big)^{a + Y_{nk}}.
\end{align*}
The first equality holds by the law of iterated expectations. The second equality follows from the Laplace transform of a gamma distribution. The limit is a consequence of the portmanteau lemma (since $x \mapsto (a/x + 1) / (t + a/x + 1)$ is a bounded continuous function for $x > 0$) and the fact that $(\mu_k \mid \bm{Y})\stackrel{\mathrm{d}}{\to} {\mu_k^*}$ by Theorem~\ref{thm:concentration}.
Notice that the limit is the Laplace transform of $\mathrm{Ga}(a + Y_{nk},\, a/{\mu_k^*} + 1)$.
Thus, by the continuity theorem for Laplace transforms \citepSupp[Chapter 13, Theorem 2]{feller_book}, 
$(\theta_{k n}\mid \bm{Y}) \stackrel{d}{\to} \mathrm{Ga}(a + Y_{n k},\, a/{\mu_k^*} + 1)$, as desired.
The second part of the statement concerning the fixed hyperprior can be proved in the same manner, but using the fact that $(\mu_k \mid \bm{Y})\stackrel{\mathrm{d}}{\to}  y_k$ by \cref{thm:concentration_normality_fixed}.  

\section{Additional theoretical results}\label{sec:additional_results}

\subsection{Further properties of the Inverse Kummer}
\label{sec:further-properties}
We study the relationship between the hyperparameter $\gamma \in \mathds{R}$ of the inverse Kummer distribution (Definition~\ref{def:InvKummer}), and the first moment of the distribution.  We begin by showing that when $\gamma= 0$, the density of the inverse Kummer simplifies to the density of an inverse gamma.
\begin{proposition}\label{pro:InvGamma}
When $\gamma = 0$, then for any $\delta > 0$ we have $\mathrm{InvKummer}(\alpha, \beta, 0, \delta) = \mathrm{InvGa}(\alpha, \beta)$.
\end{proposition}
\begin{proof}
Recall that the density of the inverse Kummer is given by
$$
\pi(\mu) = \frac{\mu^{-(\lambda - \gamma) - 1}(1 + \mu/\delta)^{-\gamma}e^{-\beta/\mu}}{\delta^{\gamma - \lambda}\,\Gamma(\lambda) \, U(\lambda,\lambda + 1 - \gamma, \beta/\delta)}.
$$
When $\gamma = 0$, then $(1 + \mu/\delta)^{-\gamma} = 1$. Moreover, 
$$
U(\lambda, \lambda + 1, \beta/\delta) = \frac{1}{\Gamma(\lambda)}\int_0^\infty t^{\lambda -1}e^{-\beta t /\delta}\mathrm{d}t = (\delta/\beta)^{\lambda}, 
$$
which leads to $\pi(\mu) = \beta^{\lambda}/\Gamma(\lambda) \mu^{-\lambda - 1} e^{-\beta/\mu}.$
This is the density of $\mathrm{InvGa}(\lambda, \beta)$. 

\end{proof}
The following proposition implies that the mean of an inverse Kummer with $\gamma > 0$ is larger than the mean of the corresponding inverse gamma distribution, when $\lambda > 2$.

\begin{proposition}\label{pro:gamma_mean}
    Let $\mu \sim \mathrm{InvKummer}(\lambda, \beta, \gamma, \delta)$. If $\lambda>2$, then $E(\mu)$ is monotone increasing as a function of $\gamma$, for $\gamma\in(0,\infty)$.
\end{proposition}
\begin{proof}
We recall that from Equation (6.7) in \citetSupp{Armero_Bayarri_1997}, for $q < \lambda$, the moments of an inverse-Kummer are
\begin{align}\label{eq:invKum-moments}
E(\mu^q) = \delta^q \frac{\Gamma(\lambda-q)}{\Gamma(\lambda)}\frac{U(\lambda-q,\, \lambda-q + 1-\gamma,\, \beta/\delta)}{U(\lambda,\, \lambda + 1-\gamma,\, \beta/\delta)}.
\end{align}.  
Fix $\epsilon\in(0,1)$. Let $\mu_\gamma \sim \mathrm{InvKummer}(\lambda, \beta, \gamma, \delta)$ and $\mu_{\gamma + \epsilon} \sim \mathrm{InvKummer}(\lambda, \beta, \gamma + \epsilon, \delta)$, and define $f(t) = t^{\lambda-1}(1+t)^{-\gamma}e^{-\beta t/\delta}$. Our aim is to show that $E(\mu_{\gamma + \epsilon}) \geq E(\mu_{\gamma})$. By Equation~\eqref{eq:invKum-moments} and the definition of confluent hypergeometric function, this is true if and only if 
\begin{equation}\label{eq:ineq_moment_gamma}
\frac{\int_{0}^{\infty}\frac{1}{t(1+t)^\epsilon}f(t)\mathrm{d}{t}}{\int_{0}^{\infty}\frac{1}{(1+t)^\epsilon}f(t)\mathrm{d}{t}}\geq \frac{\int_{0}^{\infty}\frac{1}{t}f(t)\mathrm{d}{t}}{\int_{0}^{\infty}f(t)\mathrm{d}{t}}.
\end{equation}
Let $X$ be the continuous random variable on $(0,\infty)$ with probability density function $p(x) = f(x)/\int_{0}^{\infty}f(t)\mathrm{d}{t}$.
Then, after multiplying and dividing the left-hand side by $\int_{0}^{\infty}f(t)\mathrm{d}{t}$, \cref{eq:ineq_moment_gamma} can be written in terms of expectations as 
\begin{equation}
    \frac{E\Big(\frac{1}{X(1+X)^\epsilon}\Big)}{E\Big(\frac{1}{(1+X)^\epsilon}\Big)} \geq E\Big(\frac{1}{X}\Big).
\end{equation}
or equivalently, 
\begin{equation}\label{eq:cov}
\mathrm{Cov}\Big(\frac{1}{X}, \frac{1}{(1+X)^\epsilon}\Big) \geq 0.
\end{equation}
Define $g(x) = -1/x$ and $h(x) = -1/(1+x)^\epsilon$ for $x\in(0,\infty)$. Observe that, by transformation of random variables, $1/X \sim \mathrm{InvKummer}(\lambda,\beta,\gamma,\delta)$.  Thus, by Equation~\eqref{eq:invKum-moments}, since $\lambda > 2$ by assumption,
\begin{align*}
E|g(X)|^2 &= E\Big(\frac{1}{X^2}\Big) < \infty, \\
E|h(X)|^2 &= E\bigg(\frac{1}{(1+X)^{2\epsilon}}\bigg) \leq E\Big(\frac{1}{X^{2\epsilon}}\Big) < \infty.
\end{align*}
Therefore, since $g(x)$ and $h(x)$ are monotone increasing and have finite second moments, we have $\mathrm{Cov}(g(X),h(X)) \geq 0$
by \citetSupp{schmidt2003covariance}, which is equivalent to \cref{eq:cov}. This completes the proof. 
\end{proof} 

\subsection{Characterizing the concentration point of inverse Kummer}\label{sec:concentration-point}

We characterize the point at which the inverse Kummer concentrates, ${\mu_k^*}$, in Theorem~\ref{thm:concentration}.  Define 
\begin{equation}\label{eq:mu-function}
    {\mu_k^*}(\varepsilon,y, a) = \frac{2 a\varepsilon}{\{(y-a+\varepsilon)^2 + 8 a\varepsilon\}^{1/2} - (y - a + \varepsilon)}
\end{equation}
for $\varepsilon>0$, $y\geq0$, and $a>0$.

\begin{proposition}
    For all $\varepsilon>0$ and $a > 0$, ${\mu_k^*}(\varepsilon,y, a)$ is monotone increasing as a function of $y$.
\end{proposition}
\begin{proof}
    Fix $\varepsilon>0$ and define 
    \begin{equation*}
        g(y) = \{(y-a+\varepsilon)^2 + 8 a\varepsilon\}^{1/2},
    \end{equation*}
    so that ${\mu_k^*}(\varepsilon,y,a) = 2a\varepsilon/(g(y) - (y-a+\varepsilon))$.
    Then 
    \begin{equation*}
        \frac{\partial {\mu_k^*}}{\partial y} = \frac{-2a\varepsilon \big(g'(y) - 1\big)}{\big(g(y) - (y-a+\varepsilon)\big)^2}.
    \end{equation*}
    Differentiating $g$, we find that
    \begin{equation*}
        g'(y) = \frac{y - a + \varepsilon}{\{(y-a+\varepsilon)^2 + 8a\varepsilon\}^{1/2}} < 1.
    \end{equation*}
 
    Therefore, $\partial {\mu_k^*}/\partial y> 0$,

    showing that ${\mu_k^*}$ is monotone increasing as a function of $y$.
\end{proof}

Next, we derive Equation~\eqref{eq:taylor-kummer} using a first-order Taylor approximation. Fix $y \geq 0$ and $a > 0$, and define 
\begin{equation}
    h(\varepsilon) = \{(y-a+\varepsilon)^2 + 8 a\varepsilon\}^{1/2}
\end{equation}
for $\varepsilon > 0$.  Differentiating and simplifying, we find that
\begin{equation}
    h'(\varepsilon) = \frac{y + \varepsilon + 3 a}{h(\varepsilon)}.
\end{equation}
Thus, $h(0) = |y-a|$ and $h'(0) = (y + 3 a)/|y-a|$.
Hence, a first-order Taylor approximation to $h$ at $\varepsilon = 0$ yields
\begin{equation}
    h(\varepsilon) \approx h(0) + h'(0) \varepsilon = |y - a| + \frac{y + 3a}{|y - a|} \varepsilon
\end{equation}
when $\varepsilon$ is small relative to $|y - a|$.  Plugging this into the definition of ${\mu_k^*}$ in \cref{eq:mu-function}, we obtain
\begin{equation}\label{eq:mu-approx}
    {\mu_k^*}(\varepsilon,y,a) \approx \frac{2 a\varepsilon}{|y - a| + \frac{y + 3 a}{|y - a|} \varepsilon - (y-a+\varepsilon)}.
\end{equation}
When $y > a$, we have $|y - a| = y - a$, so in this case \cref{eq:mu-approx} becomes
$$ {\mu_k^*}(\varepsilon,y,a)  \approx \frac{y - a}{2}. $$
Meanwhile, when $0 \leq y < 1$, we have $|y - a| = a - y$, so in this case
$$ {\mu_k^*}(\varepsilon,y, a)  \approx \frac{\varepsilon a(a - y)}{(a - y)^2 + (a+y) \varepsilon} $$
by collecting and rearranging terms.
Therefore, the first-order Taylor approximation is
\begin{equation}
    {\mu_k^*}(\varepsilon,y,a)  \approx
    \begin{cases}
        \displaystyle\frac{y-a}{2} & \text{ if $y>a$} \\        \displaystyle\frac{\varepsilon a(a - y)}{(a - y)^2 + (a+y) \varepsilon}  & \text{ if $0\leq y<a$}
    \end{cases}
\end{equation}
as claimed in Equation~\eqref{eq:taylor-kummer}. Plugging the above formula into the mean of the gamma distributions of Theorem~\ref{pro:loadings_shrink} yields the approximation in Equation~\eqref{eq:compressive_vs_fixed}. 
    
\subsection{Relationship between the relevance weights and latent counts}
We further characterize the relationship between the latent counts $Y_{m n k}$ and relevance weights $\mu_k$. In particular, we derive the distribution of $Y_{m n k} \mid \mu_k$, integrating out $\bm{r}_k$ and $\boldsymbol{\theta}_k$ from Equation~\eqref{eq:hierarchical-model}, using Equations~\eqref{eq:prior_R} and \eqref{eq:prior_theta}. We show that, appealingly, this distribution has a closed-form expression in terms of hypergeometric functions. First, the distribution of $Y_{m n k}\mid\mu_k,\bm{r}_k$, integrating out $\theta_{k n}$ in Equation~\eqref{eq:hierarchical-model}, is easily seen to be 
$Y_{m n k}\mid\mu_k,\bm{r}_k \sim \mathrm{NegBin}\big(a, a/(a+ r_{m k}\mu_{k})\big)$, where the negative binomial is parametrized such that the mean and variance are $E(Y_{m n k}\mid \mu_k, \bm{r}_k) = \mu_k r_{m k}$ and $\mathrm{var}(Y_{m n k}\mid \mu_k, \bm{r}_k) = \mu_k r_{m k} (a + \mu_k r_{m k})$. When both $\bm{r}_k$ and $\boldsymbol{\theta}_k$ are integrated out, we obtain the following result, where 
$\prescript{}{2}{F}_1(a, b; c, z) = \sum_{j=0}^\infty \frac{(a)_j (b)_j}{(c)_j} \frac{z^j}{j!}$
denotes the Gauss-hypergeometric function and $(a)_j = \Gamma(a+j)/\Gamma(a)$ is the ascending factorial \citepSupp{abramowitz_stegun_1972}. 

\begin{proposition}\label{lem:marginal_Y}
The probability mass function of $Y_{m n k}\mid \mu_k$ under the model in Equation~\eqref{eq:hierarchical-model} with priors as in Equations~\eqref{eq:prior_R}, \eqref{eq:prior_theta}, and \eqref{eq:hyperprior} is
\begin{align*}
\pr(Y_{m n k} &= y\mid \mu_k)= \left(\frac{\mu_k}{a}\right)^y\frac{(a)_y(\alpha)_y}{y! (\alpha M)_y}\prescript{}{2}{F}_1\Big(y+a,\,y+\alpha,\, y+\alpha M,\, -\frac{\mu_k}{a}\Big),
\end{align*}
for $y\in\{0, 1, 2, \ldots\}$. Furthermore, the mean of this distribution is $E(Y_{m n k}\mid \mu_k) = \mu_k/M$.
\end{proposition}

\begin{proof}
Recall that $Y_{m n k}\mid \mu_k,\bm{r}_k,\boldsymbol{\theta}_k \sim \mathrm{Poisson}(r_{m k}\theta_{k n})$ and $\theta_{k n}\mid\mu_k\sim \mathrm{Ga}(a, a/\mu_k)$. Also, $r_{m k} \sim \mathrm{Beta}(\alpha,\, \alpha M - \alpha)$ by marginalization property of the Dirichlet distribution. Hence, 
\begin{align*}
&\pi(Y_{m n k} = y \mid \mu_k) = \int_{0}^{1}\bigg\{\int_{0}^{\infty} \pr(Y_{m n k} = y \mid \mu_k, \bm{r}_k, \boldsymbol{\theta}_k) p(\theta_{k n}\mid \mu_k) \mathrm{d}\theta_{k n}\bigg\}  p(r_{m k}) \mathrm{d} r_{m k} \\
&= \int_{0}^{1}\bigg\{\int_{0}^{\infty} (r_{m k}\theta_{k n})^{y}\frac{e^{-r_{m k}\theta_{k n}}}{y!} \frac{(a/\mu_k)^a}{\Gamma(a)} \theta_{k n}^{a-1} e^{-(a/\mu_k)\theta_{k n}} d\theta_{k n}\bigg\} \frac{\Gamma(\alpha M)}{\Gamma(\alpha)\Gamma(\alpha M - \alpha)} r_{m k}^{\alpha-1}(1-r_{m k})^{\alpha M - \alpha - 1} \mathrm{d} r_{m k} \\
&=  \frac{(a/\mu_k)^a}{y!\,\Gamma(a)} \frac{\Gamma(\alpha M)}{\Gamma(\alpha)\Gamma(\alpha M - \alpha)} \int_{0}^{1}\bigg\{\int_{0}^{\infty} \theta_{k n}^{y+a-1} e^{-(r_{m k} + a/\mu_k)\theta_{k n}} d\theta_{k n}\bigg\}  r_{m k}^{y+\alpha-1}(1-r_{m k})^{\alpha M - \alpha - 1} \mathrm{d} r_{m k} \\
&=  \frac{(a/\mu_k)^a}{y!\,\Gamma(a)} \frac{\Gamma(\alpha M)}{\Gamma(\alpha)\Gamma(\alpha M - \alpha)}  \int_{0}^{1}\frac{\Gamma(y+a)}{(r_{m k} + a/\mu_k)^{y + a}} r_{m k}^{y+\alpha-1}(1-r_{m k})^{\alpha M - \alpha - 1} \mathrm{d} r_{m k} \\
&=  \frac{(a)_{y} (\mu_k/a)^y}{y!} \frac{\Gamma(\alpha M)}{\Gamma(\alpha)\Gamma(\alpha M - \alpha)}  \int_{0}^{1} (t\mu_k/a + 1)^{-(y + a)} t^{y+\alpha-1}(1-t)^{(y+\alpha M) - (y+\alpha) - 1} \mathrm{d} t \\
&= \Big(\frac{\mu_k}{a}\Big)^y\frac{(a)_y(\alpha)_y}{y! (\alpha M)_y}\prescript{}{2}{F}_1\big(y+a,\, y+\alpha,\, y+\alpha M,\, -\mu_k/a\big),
\end{align*}
where $(x)_j = \Gamma(x + j)/\Gamma(x)$ is the ascending factorial, and 
$$
\prescript{}{2}{F}_1(a, b, c, z) = \frac{\Gamma(c)}{\Gamma(b)\Gamma(c-b)} \int_0^1 t^{b-1} (1-t)^{c-b-1} (1 - tz)^{-a} \mathrm{d}t, 
$$
with $c>b>0$ and $a, z \in \mathds{R}$ is an alternative representation of the Gauss-hypergeometric function; see \citetSupp{abramowitz_stegun_1972}. 
\end{proof}
This result further explains the role of each $\mu_k$ in the mutational process: it directly controls the contribution of signature $k$ in determining the number of mutations $X_{m n}$ in channel $m$ for patient $n$, for given values of hyperparameters $a$ and $\alpha$.

\subsection{Distribution of the exposures}
We now characterize the density of the exposures $\theta_{kn}$ given the tensor of latent counts $\bm{Y}$. Interestingly, the density has an analytical form in terms of Kummer hypergeometric functions and belongs to the class of gamma-Kummer continuous mixtures. Note that the expected value of this distribution is presented in Theorem~\ref{pro:PostTheta} in the main manuscript.
\begin{theorem}\label{thm:post_theta_dist}
In the setting of Theorem~\ref{thm:InvKumPost}, let $Y_{nk} = \sum_{m = 1}^M Y_{m n k}$ and $\overline{Y}_k = \frac{1}{N}\sum_{n = 1}^{N} Y_{nk}$. Then, the density of the exposures $\theta_{kn}$ is 
$$\pi(\theta_{kn} \mid \bm{Y}) = \theta_{kn}^{a + Y_{nk} - 1} e^{-\theta_{kn}}\frac{U\{2 a N + 1, N(a - \overline{Y}_k) + a + 2 + Y_{nk}, \varepsilon N + \theta_{kn}\}}{\Gamma(a + Y_{nk}) U\{2aN + 1, N(a - \overline{Y}_k) + 2, \varepsilon N\}}.$$
\end{theorem}
\begin{proof}
From Section~\ref{sec:inference}, recall that $(\theta_{kn} \mid \bm{Y}, \mu_k) \sim \mathrm{Ga}(a + Y_{n k}, a/\mu_k + 1)$. 
For notational simplicity, we first consider the general case where the hyperprior is $\mu_k\sim\mathrm{InvGa}(a_0,b_0)$, and then we plug in the values of $a_0$ and $b_0$ for the compressive hyperprior.
In the proof of Theorem~\ref{thm:InvKumPost}, we showed that $(\mu_k \mid \bm{Y}) \sim \mathrm{InvKummer}\big(a_0 + N a,\, b_0,\, N a + N\overline{Y}_k,\, a\big)$. As in the proof of Theorem~\ref{pro:PostTheta}, let $C = a^{N\overline{Y}_k -a_0}\Gamma(a_0 + aN)U(a_0 + aN, a_0 + 1 - N \overline{Y}_k, b_0/a)$  denote the normalizing constant of the density of $\mu_k \mid \bm{Y}$, that is, the constant of proportionality in \cref{eq:invKumm_post_generic}. Then, using the inverse Kummer density in Equation~\eqref{eq:InvKumm-definition} to integrate,
\begin{align*}
&\pi(\theta_{kn} \mid \bm{Y}) = \int_{0}^{\infty} \pi(\theta_{kn} \mid \bm{Y}, \mu_k) \pi(\mu_k\mid \bm{Y}) \mathrm{d}\mu_k\\
&= \frac{a^{a + Y_{nk}}}{C\Gamma(a + Y_{nk})} \theta_{kn}^{a + Y_{nk} -1} e^{-\theta_{kn}}\int_{0}^{\infty} \mu_{k}^{-(a_0 - N\overline{Y}_k + a + Y_{nk}) - 1}\bigg(1 + \frac{\mu_k}{a}\bigg)^{-aN - N\overline{Y}_k + a + Y_{nk}} e^{-(b_0 + a \theta_{kn})/\mu_k}\mathrm{d}\mu_k \\
&=  \frac{a^{a + Y_{nk}}}{C\Gamma(a + Y_{nk})} \theta_{kn}^{a + Y_{nk} -1} e^{-\theta_{kn}} a^{N\overline{Y}_k -a -Y_{nk} -a_0}\Gamma(a_0 + aN)U(a_0 + aN, a_0 + 1 - N \overline{Y}_k + a + Y_{nk}, b_0/a + \theta_{k n})\\
&= \theta_{kn}^{a + Y_{nk} - 1} e^{-\theta_{kn}}\frac{U(a_0 + aN, a_0 + 1 - N \overline{Y}_k + a + Y_{nk}, b_0/a + \theta_{k n})}{\Gamma(a + Y_{nk}) U(a_0 + aN, a_0 + 1 - N \overline{Y}_k, b_0/a)} 
\end{align*}
Substituting $a_0 = aN+1$ and $b_0 = \varepsilon aN$ completes the proof. 
\end{proof}

\subsection{\revision{Theoretical results for the fixed  hyperprior}}\label{subsec:fixed_case_prior}
\revision{In this section, we derive concentration and asymptotic normality results when using the fixed hyperprior, that is, $\mu_k \sim \mathrm{InvGa}(a_0, b_0)$ for fixed choices of $a_0$ and $b_0$. In our model, the resulting conditional distribution is $\mu_k\mid \bm{Y} \sim \mathrm{InvKummer}(a_0 + N a,\, b_0, \, N a + N\overline{Y}_k,\, a)$. 
As before, to ease readability we write $y_N$ and $\mu_N$ in place of $\overline{Y}_k$ and $\mu_k$, respectively. We discuss the interpretation of this result in comparison with the compressive case in \cref{subsec:comparison_fixed_comp}.}
\begin{theorem}\label{thm:concentration_normality_fixed}
Let $a>0$, $a_0 > 0$, $b_0>0$, and $y_N> 0$ such that $y_N \to y$ for some $y>0$. If $\mu_N\sim \mathrm{InvKummer}(a_0 +N a,\, b_0, \, N a + Ny_N,\, a)$, then:
\begin{enumerate}[label=\textup{(\arabic*)}, leftmargin=*, align=left]
    \item for any sequence $c_1, c_2, \ldots \in [0, \infty)$ such that $c_N \to\infty$ with $c_N/N^{1/2} \to 0$ and $|y_N - y| = o(c_N/N^{1/2})$, we have $$\pr(|\mu_N - y| \geq c_N/N^{1/2}) \xrightarrow[N \to \infty]{} 0,$$
    \item defining $\Delta_N = N^{1/2}(y_N - y)$, we have
    $$
    N^{1/2}(\mu_N - y) - \Delta_N \xrightarrow[N \to \infty]{\mathrm{d}} \mathrm{N}\left(0,\;\frac{y(y+a)}{a}\right).
    $$
\end{enumerate}  
\end{theorem}
\begin{proof}
\revision{Letting $\mu_N\sim \mathrm{InvKummer}(a_0 +N a,\, b_0, \, N a + Ny_N,\, a)$ and $T_N = a/\mu_N$, the density of $T_N$ is $f_N(t) \propto t^{a_0 + aN -1}(1+t)^{-Ny_N-aN}e^{-b_0 t}$. The rest of the proof follows the same steps used in the proofs of \cref{thm:Kummer_concentration} and \cref{thm:concentration_normality_mu}.}
(1)
\revision{Similarly to \cref{thm:Kummer_normality}, we have $f_N(t) \propto \exp\left(-Ng_N(t)\right)\pi(t)$, where $\pi(t) = t^{a_0 - 1}e^{-b_0t}$, and 
\begin{align*}
g_N(t) = -a\log(t) + (y_N + a)\log(1+t), \\  
g(t) = -a\log(t) + (y + a)\log(1+t).
\end{align*}
Call $t^*_N = a/y_N$ and $t^* = a/y$. It is easy to verify that $g'_N(t_N^*) = 0$ and $g'(t^*) = 0$. Moreover, the conditions of Theorem 5 
in \citetSupp{Miller_2021} are met for $g_N(t)$ and $g(t)$, by the same arguments used in \cref{thm:Kummer_normality}. In particular, since $g''(t)= a/t^2 - (y+a)/(1+t)^2$, we have $g''(t^*) = g''(a/y) = a(a+y)/y^3 >0$ and so
\begin{equation}\label{eq:kummer_normality_fixed}
N^{1/2}(T_N - t^*_N) = N^{1/2}(T_N - a/y_N)\ \xrightarrow[N \to \infty]{\mathrm{d}} \ \mathrm{N}\left(0, \frac{a(y + a)}{y^3}\right).
\end{equation}
Then, by the same reasoning as \cref{thm:Kummer_concentration}, it holds that $\pr(|T_N - t_N^*| \geq d_N/N^{1/2}) \to 0$ for any $d_N \to \infty$ and $ d_N/N^{1/2} \to 0$. The proof is completed by mirroring the steps detailed in point (1) of \cref{thm:concentration_normality_mu}, defining $$
d_N = \left(\frac{a/y}{y + c_N/N^{1/2}}\right) c_N.
$$}
(2) 
As in \cref{thm:concentration_normality_mu}, part (2), we can apply the transformation $h(t) = a/t$ to \cref{eq:kummer_normality_fixed}. The delta method \citepSupp{Van_der_Vaart_2000} then yields 
$$
N^{1/2}(h(T_N) - h(a/y_N)) \xrightarrow[N \to \infty]{\mathrm{d}} \ \mathrm{N}\left(0, h'(a/y)^2\frac{a(y + a)}{y^3}\right).
$$
Hence, we write
$$
N^{1/2}(\mu_N - y) = N^{1/2}(h(T_N) - h(a/y)) = N^{1/2}(h(T_N) - h(a/y_N)) + N^{1/2}(h(a/y_N) - h(a/y)).
$$
Simplifying the variance and recalling that $h'(a/y)^2 = y^4/a^2$, we obtain
$$
N^{1/2}(\mu_N - y) - \Delta_N \xrightarrow[N \to \infty]{\mathrm{d}} \ \mathrm{N}\left(0, \frac{y(y + a)}{a}\right),
$$
with $\Delta_N = N^{1/2}(y_N - y)$. This concludes the proof. 
\end{proof}

\subsection{Characterization of signature merging in theory and in application}\label{subsec:MergingSigs}
In this section, we illustrate the general conditions under which the Poisson NMF model cannot distinguish between two distinct signatures and their ``merged'' counterpart. We also note that this is not problematic from an interpretive point of view, provided that sufficient biological knowledge is available. 
We begin with the following result.
\begin{proposition}\label{pro:merged_signatures}
Let $X_{m n} \sim \mathrm{Poisson}\big(\sum_{k = 1}^K r_{mk}\theta_{kn}\big)$ for $m = 1, \ldots, M$ and $n= 1, \ldots, N$ be an NMF model. Assume $\sum_{m = 1}^M r_{mk} = 1$  and $\theta_{k n}>0$ for all $k = 1, \ldots, K$ and all $n$, and $\bm{r}_{k}\neq \bm{r}_{k'}$ for all pairs $k,k'$ such that  $k\neq k'$.  Then for any pair $k\neq k'$, the following conditions are equivalent:
\begin{enumerate}[leftmargin=*, align=left]
    \item There exist a signature $\widetilde{\bm{r}}$ and exposure vector  $\widetilde{\boldsymbol{\theta}}$ such that $X_{m n} \sim \mathrm{Poisson}\big(\widetilde{r}_{m}\widetilde{\theta}_{n} + \sum_{\ell \in S} r_{m\ell}\theta_{\ell n}\big),$ with $S = \{1, \ldots, K\}/\{k, k'\}$;
    \item $\boldsymbol{\theta}_k$ and $\boldsymbol{\theta}_{k'}$ are linearly dependent, that is, there exists $c >0$ such that $\theta_{k'n} =c \, \theta_{kn}$ for all $n$.
\end{enumerate}
\end{proposition}

\begin{proof}
Let $\boldsymbol{\Lambda} = \sum_{k=1}^K \bm{r}_k \boldsymbol{\theta}_k^\top$, that is, $\boldsymbol{\Lambda}$ is the matrix in which entry $(m,n)$ is $\lambda_{m n} = \sum_{k=1}^K r_{m k} \theta_{k n}$, for all $m$, $n$.
For simplicity, we first consider  the case when $K = 2$, so that  
$\boldsymbol{\Lambda} = \bm{r}_1\boldsymbol{\theta}_1^\top + \bm{r}_2\boldsymbol{\theta}_2^\top$.\\
\emph{(2) $\Rightarrow$ (1):} If condition (2) holds, then 
\begin{align*}
\boldsymbol{\Lambda} &= \bm{r}_1\boldsymbol{\theta}_1^\top + \bm{r}_2\boldsymbol{\theta}_2^\top  = \bm{r}_1\boldsymbol{\theta}_1^\top + c\,\bm{r}_2\boldsymbol{\theta}_1^\top =  \frac{\bm{r}_1 + c\, \bm{r}_2}{1 + c} (1+c)\boldsymbol{\theta}_1^\top = \widetilde{\bm{r}}\widetilde{\boldsymbol{\theta}}^\top,
\end{align*}
where $\widetilde{\bm{r}} = (\bm{r}_1 + c\, \bm{r}_2)/(1 + c)$ is a weighted average of $\bm{r}_1$ and $\bm{r}_2$, and $\widetilde{\boldsymbol{\theta}}= (1+c)\boldsymbol{\theta}_1$ is its exposure. This implies that $X_{m n} \sim \mathrm{Poisson}(\sum_{k =1}^2 r_{m k}\theta_{k n}) = \mathrm{Poisson}(\widetilde{r}_{m}\widetilde{\theta}_{n})$ for all $m$, $n$, proving (1). \\
\emph{(1) $\Rightarrow$ (2):} Suppose condition (1) holds.  Since $\sum_{m = 1}^M r_{mk} = 1$, $\bm{r}_1$ and $\bm{r}_2$ must be linearly independent. To see this, suppose $\bm{r}_2 = c\, \bm{r}_1$ for some $c > 0$. The sum-to-one constraint implies $c = 1$, because $1 = \sum_{m= 1}^M r_{m2} = c \sum_{m= 1}^M r_{m1} = c$. But then $\bm{r}_2 = \bm{r}_1$, contrary to assumption.
Thus, the rank of the factorization is $\mathrm{rank}(\boldsymbol{\Lambda}) = \dim(\mathrm{span}\{\boldsymbol{\theta}_1, \boldsymbol{\theta}_2\})$. By condition (1), there exists $\widetilde{\bm{r}}$ such that $\sum_{m = 1}^M \widetilde{r}_{m} = 1$ and $\boldsymbol{\Lambda} = \widetilde{\bm{r}}\widetilde{\boldsymbol{\theta}}^\top$, which means that $\mathrm{rank}(\boldsymbol{\Lambda}) = 1$. Then, it must be that $\boldsymbol{\theta}_1$ and $\boldsymbol{\theta}_2$ are linearly dependent, which is condition (2).

Now, we consider the case of $K > 2$. Define $\boldsymbol{\Lambda}_{k,k'} = \boldsymbol{\Lambda} - \sum_{\ell \not\in\{ k, k'\}}\bm{r}_\ell\boldsymbol{\theta}_\ell^\top 
= \bm{r}_k\boldsymbol{\theta}_k^\top + 
\bm{r}_{k'}\boldsymbol{\theta}_{k'}^\top$. Condition (1) is equivalent 
to $\boldsymbol{\Lambda}_{k,k'} = \widetilde{\bm{r}}\widetilde{\boldsymbol{\theta}}^\top$ 
since the terms indexed by $\ell\not\in\{k, k'\}$ are identical on both sides 
of the equation for $\boldsymbol{\Lambda}$ and cancel. The assumption 
$\bm{r}_k \neq \bm{r}_{k'}$ ensures, by the argument above, that 
the pair $\bm{r}_k, \bm{r}_{k'}$ is linearly independent. Applying 
the $K = 2$ result to $\boldsymbol{\Lambda}_{k,k'}$ gives the equivalence 
between conditions (1) and (2) for the pair $(k, k')$. This completes the proof. 

\end{proof}

\begin{figure}
    \centering
    \includegraphics[width=\linewidth]{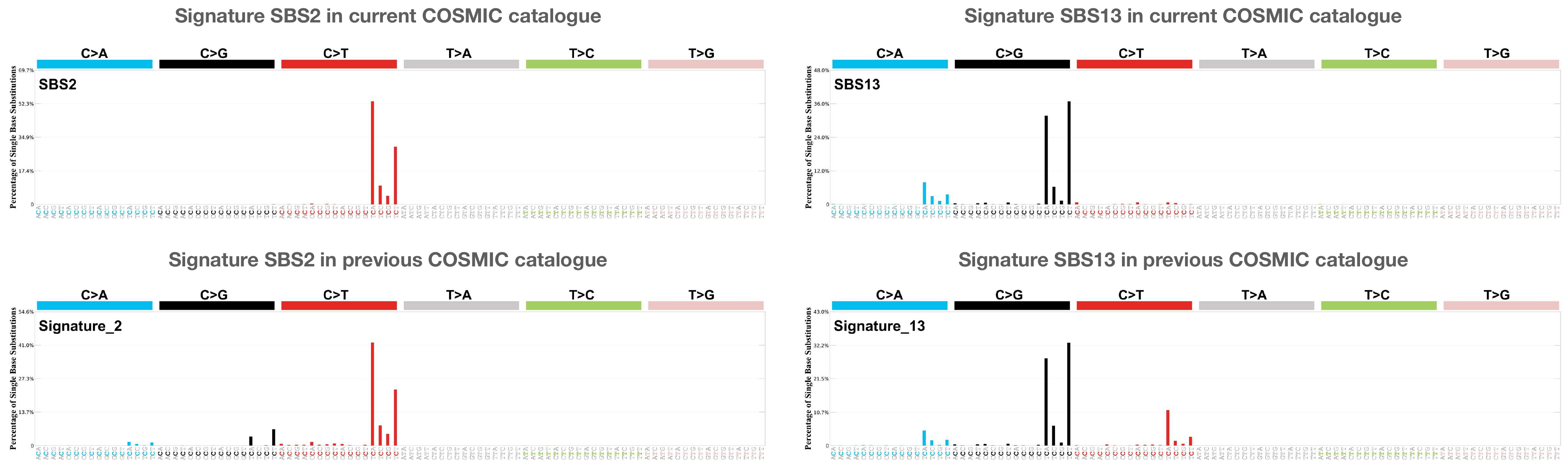}
    \caption{Examples for SBS2 and SBS13 in two versions of the COSMIC catalog, with top row pertaining to COSMIC v3.5. Images are downloaded from \url{https://cancer.sanger.ac.uk/signatures/sbs/sbs2/} and \url{https://cancer.sanger.ac.uk/signatures/sbs/sbs13/}. }
    \label{fig:Signature_SBS2_13}
\end{figure}

The preceding result gives necessary and sufficient conditions under which a pair of signatures can be merged by the model. However, in practice, exact linear dependence is a rather stringent condition. The more likely scenario is that exposures are highly correlated, which often occurs when multiple signatures are associated with the same biological phenomenon. For instance, SBS2 and SBS13 are associated with APOBEC, and SBS1 and SBS5 are associated with aging \citepSupp{Alexandrov_2020}. As a result, NMF models have a tendency to merge them more than others, due to their correlated exposures  \citepSupp{Wu2022}. Nevertheless, it is worth noting that COSMIC signatures are designed and updated by considering NMF results from multiple datasets, knowledge of biological processes, and experimental validation. As such, previous versions of the COSMIC catalogue contained merged---or partially merged---versions of signatures that were later distinguished as separate. For example, \cref{fig:Signature_SBS2_13} shows how signatures SBS2 and SBS13 have been updated over different editions of the COSMIC database: the top row displays the current version, while the bottom row shows an older version \citepSupp{Nik_zainal_2012}. Here, it is easy to see that the two bottom signatures are (approximately) linear combinations of the two top ones.
This occurs because both signatures are associated with APOBEC activity and therefore have correlated exposures. 
Similar examples are SBS40 \citepSupp{Jin2024} as well as SBS7, which is now divided into SBS7a, b, c, and d, which are all associated with UV light exposure.

\section{Additional simulation details and results}\label{sec:Simulation_details}

We present additional details and results for the simulation study in Section~\ref{sec:simulation}, as well as results from additional simulations. 
\cref{subsec:sim_setup} provides details on the simulation procedure and setup;
\cref{subsec:competitors} describes the methods that we compare with; \cref{sec:appendix_sim_results} presents results on estimation accuracy, as well as details on computation time and effective sample size of the sampling algorithms; 
\cref{subsec:InfiniteFactors} presents an additional simulation comparing with Bayesian infinite factorization models, including specification of the PoissonCUSP model and sampler in \cref{sec:cusp}; and \cref{subsec:sparse_data} presents an additional simulation under sparse data representing indel counts.

\subsection{Simulation setup}\label{subsec:sim_setup}
This section discusses the setup of the simulation in Section~\ref{sec:simulation}.

We generate mutation counts
as $X_{m n} \sim \mathrm{NegBin}\{1/\tau,\, 1/(1 + \tau\lambda_{m n}^0)\}$, where $\mathrm{NegBin}$ denotes a negative binomial distribution and $\lambda^0_{m n} = \sum_{k = 1}^{K_{\mathrm{pre}}^0} \rho^0_{m k}\omega^0_{k n} + \sum_{k = 1}^{K_{\mathrm{new}}^0} r^{0}_{m k}\theta_{k n}^0$.
We parametrize the negative binomial such that the mean and variance of $X_{m n}$ are $\lambda_{m n}^0$ and $\lambda_{m n}^0(1 + \tau\lambda_{m n}^0)$, respectively, where $\tau>0$ is a parameter controlling overdispersion. 
We set $K_{\mathrm{pre}}^0 = 4$, and for $k = 1,\ldots,4$, we define $\rho_{k}^0 = (\rho_{1 k}^0,\ldots,\rho_{M k}^0)$ to be  COSMIC signatures SBS1, SBS2, SBS5, and SBS13, respectively. SBS1 is a sparse signature present in every cancer type, arising from the spontaneous deamination of 5-methylcytosine. SBS5 is a rather flat signature that has been shown to appear in every cancer type. SBS2 and SBS13 are commonly occurring signatures associated with APOBEC activity.
Meanwhile, we randomly generate $r_{k}^0 = (r_{1 k}^0,\ldots,r_{M k}^0)$ as $r_{k}^0 \sim \mathrm{Dir}(0.25,\ldots,0.25)$, independently for $k = 1,\ldots,K_{\mathrm{new}}^0$.
We generate exposures using a multiplicative global-local structure by setting $\omega^0_{k n} = w_k \xi_{k n}$ where $w_k \sim \mathrm{Ga}(100, 1)$ and $\xi_{k n} \sim \mathrm{Ga}(0.5, 0.5)$ independently, and $\theta^0_{k n}$ in the same way as $\omega^0_{k n}$. This implies that $\omega^0_{k n}\mid w_k \sim \mathrm{Ga}(0.5, 0.5/w_k)$, mirroring our hierarchical structure.

We consider two overdispersion settings: $\tau = 0$, in which case the negative binomial reduces to a Poisson (so the Poisson NMF model is correct), and $\tau = 0.15$, resulting in mild misspecification. For the number of non-COSMIC signatures $K_{\mathrm{new}}^0$, we consider $K_{\mathrm{new}}^0 = 2$ and $K_{\mathrm{new}}^0 = 6$, so that the total number $K^0 = K_{\mathrm{pre}}^0 + K_{\mathrm{new}}^0$ is either $6$ or $10$. We consider a range of sample sizes $N \in \{50,100,200\}$.
For each combination of $\tau$, $M$, and $K^0$, we generate $20$ replicate sets of parameters and data matrices.
On each simulated data matrix, we run:

\begin{enumerate}[itemsep=0pt, parsep=0pt, label=(\roman*), leftmargin=*, align=left]
    \item\label{item:CompNMF} CompNMF: our compressive NMF model in Equations~\eqref{eq:prior_R} to \eqref{eq:hierarchical-model}, with our default settings,
    \item\label{item:CompNMF+cosmic} CompNMF+cosmic: our model in Equation~\eqref{eq:IzziModel} with $K_{\mathrm{new}} = 15$ \emph{de novo} signatures and the $K_{\mathrm{pre}} = 67$ COSMIC v3.4 signatures not regarded as ``sequencing artifacts'',%
    \item\label{item:PoissonCUSP} PoissonCUSP: the CUSP \revision{infinite spike-and-slab} model of \citetSupp{Legramanti_2020} using their Algorithm 2 to adaptively tune $K$,
    \item\label{item:signeR} signeR: the \textsc{signeR} model \citepSupp{Rosales_2016} with default parameters
    and with $K$ ranging from $2$ to $20$,
    \item\label{item:SignatureAnalyzer} SignatureAnalyzer: as implemented in the \texttt{sig\_auto\_extract} function of the \texttt{sigminer} package \citepSupp{sigminer_2020}, with selection method set to \texttt{L1KL} and $K = 20$, 
    \item\label{item:SigProfiler} SigProfiler: \textsc{SigProfilerExtractor}  v1.1.23 \citepSupp{Islam_2022} with 5 replicates for each $K \in\{ 2,\ldots,20\}$, using the  \texttt{sigprofiler\_extract} wrapper in \texttt{sigminer}. 
    \item\label{item:BayesNMF} BayesNMF: the Bayesian NMF model based on Gaussian likelihood and exponential automatic relevance determination priors from \citetSupp{Brouwer2017b}, with $K = 20$.%
\end{enumerate}

For the CompNMF models, we use our recommended default settings of $\varepsilon = 0.001$, $a = 1$, and $\alpha = 0.5$, along with $K = 20$ for CompNMF and our $\beta_k$ tuning procedure for CompNMF+cosmic.
We provide a description of each of these previous methods in Section~\ref{subsec:competitors}. Moreover, in \cref{subsec:InfiniteFactors}, we also present an extended comparison of our method against other infinite factorization methods, including the extension of CUSP based on \citetSupp{Kowal_Canale2023} and the multiplicative gamma process of \citetSupp{Bhattacharya_Dunson_2011}. For methods \ref{item:CompNMF}, \ref{item:CompNMF+cosmic}, and \ref{item:PoissonCUSP}, we run the sampler for $5000$ iterations, discarding the first $4000$ as burn-in.  In method \ref{item:PoissonCUSP}, adaptation of $K$ was started after 500 iterations.

Each method uses a different strategy to estimate the number of signatures $\widehat{K}$: for \ref{item:CompNMF} and \ref{item:CompNMF+cosmic}, $\widehat{K}$ is the number of signatures for which the posterior estimate of $\mu_k$ is greater than $C\varepsilon = 5\varepsilon = 0.005$; for \ref{item:PoissonCUSP}, $\widehat{K}$ is the number of signatures for which the posterior probability of being assigned to the spike is less than $0.05$; and for \ref{item:signeR}, \ref{item:SignatureAnalyzer}, and \ref{item:SigProfiler}, $\widehat{K}$ is the suggested solution returned by each package. For \ref{item:BayesNMF}, we include signatures for which the posterior mean cosine similarity between the signature and the uniform distribution is less than 0.95.

\subsection{Description of competing methods}\label{subsec:competitors}

We now describe the NMF mutational signatures methods that we use for comparisons in our simulation, most of which appear in the recent review by \citetSupp{Islam_2022}. 
Currently, the most prominent method is \textsc{SigProfiler} \citepSupp{Alexandrov_2020} and its successor, \textsc{SigProfilerExtractor} \citepSupp{Islam_2022}. \textsc{SigProfilerExtractor} uses a bootstrap-like procedure to resample the mutation count matrix from a Poisson model, and fits an NMF model to each resampled data matrix by minimizing the Poisson Kullback--Leibler divergence. The number of signatures is selected by applying the algorithm for a range of $K$ values and using a neural network to choose $K$.%

A leading Bayesian method is \textsc{signeR} \citepSupp{Rosales_2016, SigneR_2023} which is based on the Poisson NMF model in Equation~\eqref{eq:Poisson}, but employs independent hierarchical gamma priors over both the signatures and the exposures. Selection of $K$  is performed using the Bayesian information criteria (BIC) after running a separate model for each $K$. Hence, both \textsc{SigProfilerExtractor} and  \textsc{signeR} are particularly slow when the range of possible $K$ values is moderate to large.

A faster alternative is offered by \textsc{SignatureAnalyzer} \citepSupp{kim2016somatic}, which fits the Poisson NMF model in Equation~\eqref{eq:Poisson} using a  \emph{maximum a posteriori} estimation algorithm \citepSupp{Tan_Fevotte_2013}.
Similarly to our approach, \textsc{SignatureAnalyzer} uses ARD with inverse-gamma hyperpriors on the relevance weights to determine the number of signatures.
While the method is efficient and flexible in terms of the choice of the objective function (Kullback--Leibler or squared error) and prior (exponential or half-normal), it does not provide any uncertainty quantification.

A natural Bayesian approach to selecting the number of latent factors is to use spike-and-slab priors, where the relevance weight of each of the $K$ signatures has some probability of being sampled from a spike close to zero. This is the approach taken in the elegant nonparametric factorization model proposed by \citetSupp{Legramanti_2020}, which infers the number of factors using a cumulative shrinkage spike-and-slab process prior (CUSP). 
This model assumes an infinite number of factors \emph{a priori}, but the probability that $\mu_k$ comes from the spike---effectively removing that factor from the model---increases with $k$; Posterior inference is performed via an adaptive Metropolis algorithm;  
we defer to \cref{sec:cusp} for details. A further comparison of our model against other infinite factorization methods is presented in Section~\ref{subsec:InfiniteFactors}. Finally, a fully Bayesian approach to ARD using a Gaussian likelihood, referred to as BayesNMF, is presented by \citetSupp{Brouwer2017b}.

\subsection{Additional simulation results}
\label{sec:appendix_sim_results}

\subsubsection{Estimation accuracy and reconstruction error}\label{subsec:rmse_simulations}
We compare the models by assessing the root mean squared error (RMSE) for (i) the observed counts $\bm{X}$, (ii) the true mean matrix $\boldsymbol{\Lambda}^0 = (\lambda^0_{mn})$, (iii) the true matrix of signatures $\bm{R}^0$, and (iv) the true exposures matrix $\boldsymbol{\Theta}^0$. 

\begin{figure}
    \centering
    \includegraphics[width = \linewidth]{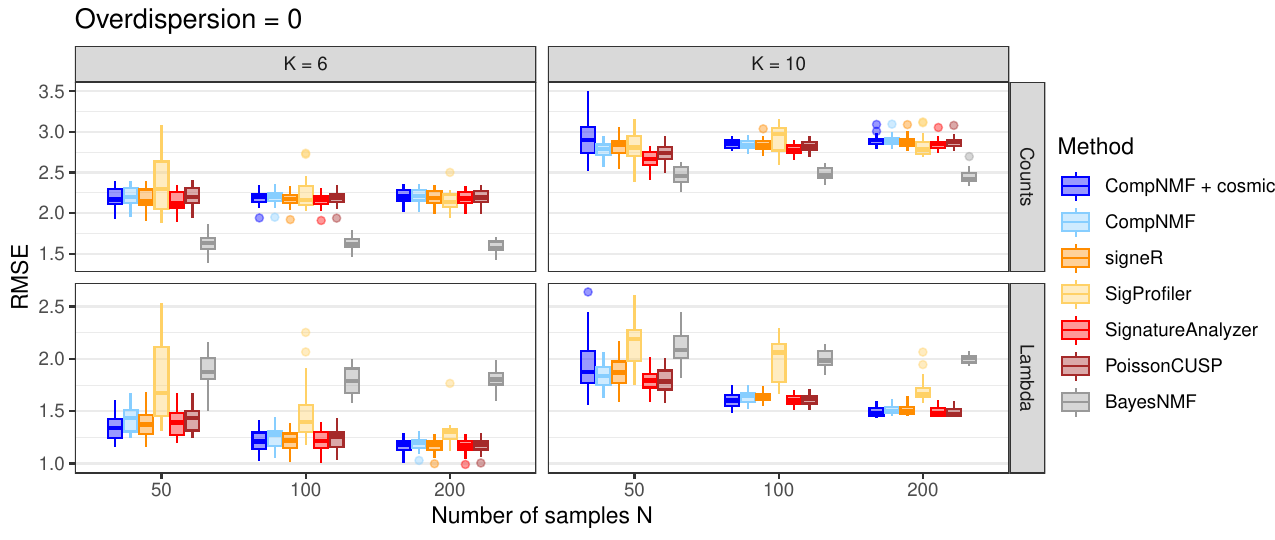}
    \caption{RMSE between (top) $\widehat{\boldsymbol{\Lambda}}$ and the count matrix $\bm{X}$, and (bottom) $\widehat{\boldsymbol{\Lambda}}$ and the true mean matrix $\boldsymbol{\Lambda}^0$, for each method across 20 replicates, when data are generated with overdispersion $\tau = 0$, i.e., the Poisson model.}
    \label{fig:LambdaCount_0}
\end{figure}

\begin{figure}
    \centering
    \includegraphics[width = \linewidth]{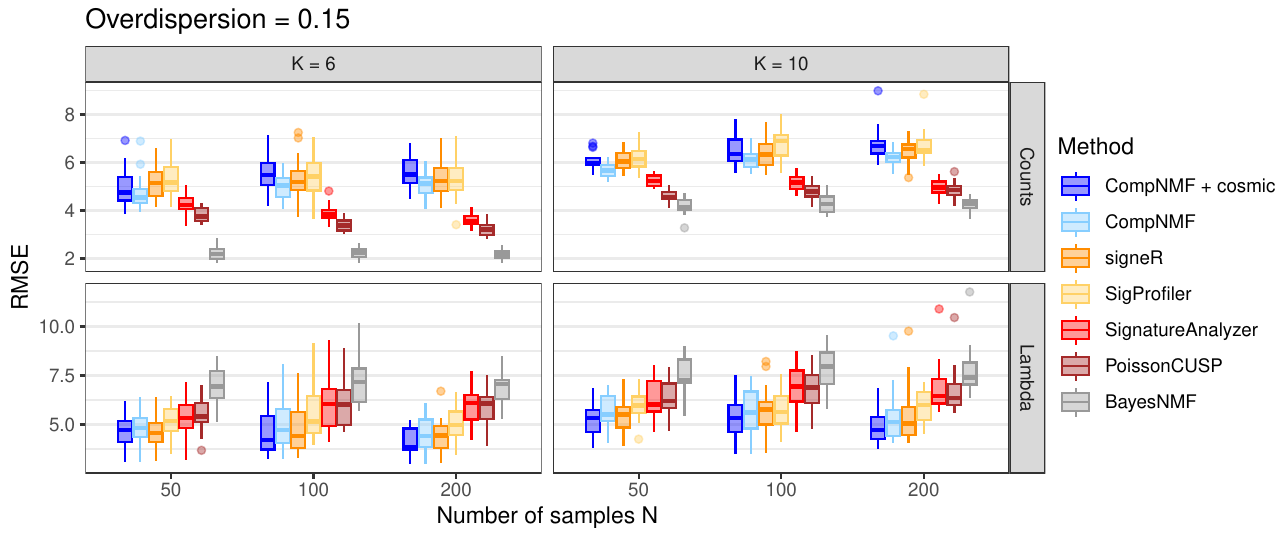}
     \caption{RMSE between (top) $\widehat{\boldsymbol{\Lambda}}$ and the count matrix $\bm{X}$, and (bottom) $\widehat{\boldsymbol{\Lambda}}$ and the true mean matrix $\boldsymbol{\Lambda}^0$, for each method across 20 replicates, when data are generated with overdispersion $\tau = 0.15$, i.e., negative binomial. }
    \label{fig:LambdaCount_015}
\end{figure}

First, the  RMSE for $\bm{X}$ and $\boldsymbol{\Lambda}^0$ is calculated as $\textsc{rmse}(\bm{X}, \widehat{\boldsymbol{\Lambda}}) = \big(\sum_{mn} (X_{mn} - \widehat{\lambda}_{mn})^2/MN\big)^{1/2}$ and $\textsc{rmse}(\boldsymbol{\Lambda}^0, \widehat{\boldsymbol{\Lambda}}) = \big(\sum_{mn} (\lambda^0_{mn} - \widehat{\lambda}_{mn})^2/MN\big)^{1/2}$, respectively, where $\widehat{\lambda}_{mn} = \sum_{k = 1}^{K^*} \widehat{r}_{mk}\widehat{\theta}_{kn}$ and $\widehat{K}$ is the estimated number of signatures for each method. \cref{fig:LambdaCount_0,fig:LambdaCount_015} display the comparison across all models and replicate data sets for overdispersion $\tau = 0$ and $\tau = 0.15$, respectively.  In the correctly specified case ($\tau 
= 0$), no method performs overwhelmingly better than all the others. This is expected since all models correctly estimate the true number of signatures in this case; see Figure~\ref{fig:Simulation_results}(A). One exception is SigProfiler, which performs noticeably worse than the others, particularly in terms of the RMSE for $\boldsymbol{\Lambda}^0$. The performance of SigProfiler does improve somewhat as the sample size increases, suggesting that the algorithm might struggle in small dimensions or on relatively low mutation counts.

The situation changes, however, in the overdispersed (negative binomial) case with $\tau = 0.15$, displayed in \cref{fig:LambdaCount_015}. Here, SignatureAnalyzer and PoissonCUSP obtain the lowest $\textsc{rmse}(\bm{X}, \widehat{\boldsymbol{\Lambda}})$ across all values of $N$, however, their corresponding $\textsc{rmse}(\boldsymbol{\Lambda}^0, \widehat{\boldsymbol{\Lambda}})$ shows the opposite trend. This is a clear indication of overfitting, which is reinforced by the fact that both models overestimate the number of signatures; see Figure~\ref{fig:Simulation_results}(A). 

\begin{figure}
    \centering
    \includegraphics[width = \linewidth]{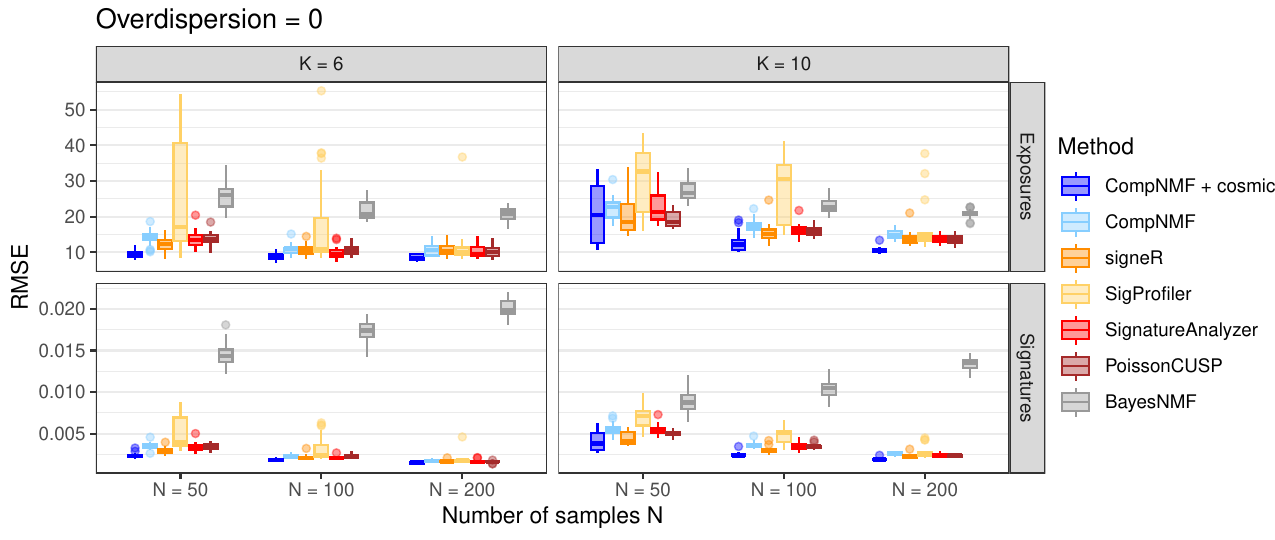}
    \caption{RMSE between (top) the estimated exposures $\widehat{\boldsymbol{\Theta}}$ and true exposures $\boldsymbol{\Theta}^0$, and (bottom) the estimated signatures $\widehat{\bm{R}}$ and true signatures $\bm{R}^0$, over 20 replicate datasets, when data are generated with overdispersion $\tau = 0$, i.e., the Poisson model. }
    \label{fig:SigTheta_0}
\end{figure}

\begin{figure}
    \centering
    \includegraphics[width = \linewidth]{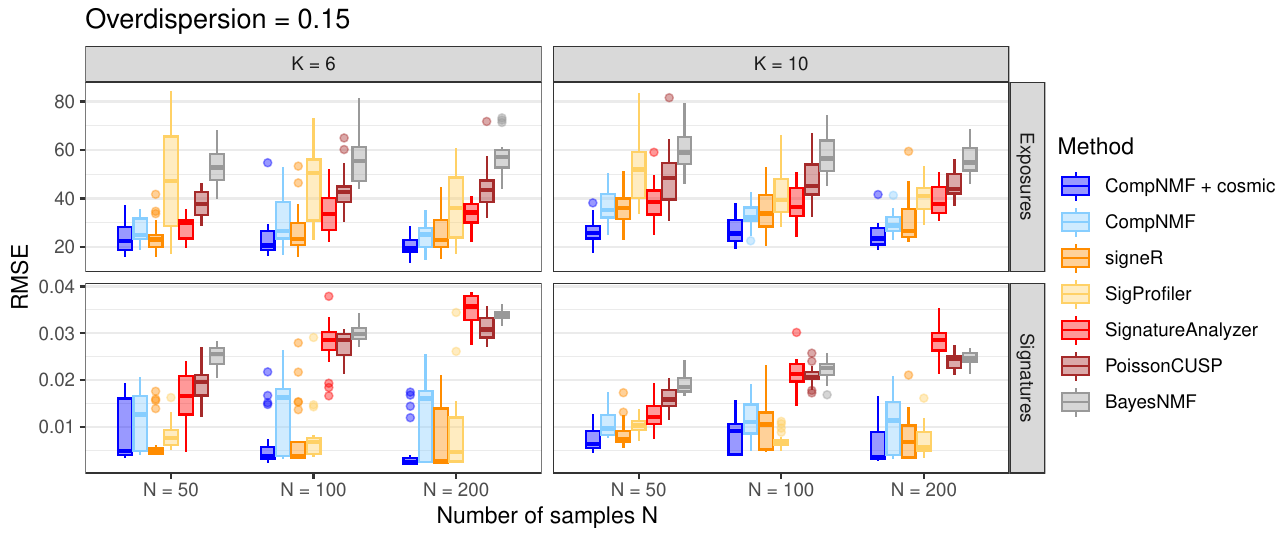}
\caption{RMSE between (top) the estimated exposures $\widehat{\boldsymbol{\Theta}}$ and true exposures $\boldsymbol{\Theta}^0$, and (bottom) the estimated signatures $\widehat{\bm{R}}$ and true signatures $\bm{R}^0$, over 20 replicate datasets, when data are generated with overdispersion $\tau = 0.15$, i.e., negative binomial. }
    \label{fig:SigTheta_015}
\end{figure}

To enable a direct comparison of performance in terms of estimating the true signature and exposures matrices $\bm{R}^0$ and  $\boldsymbol{\Theta}^0$, we first perform a matching step to maximize the total pairwise cosine similarity between the estimated $\widehat{\bm{R}}$ and true $\bm{R}^0$ using the Hungarian algorithm. If the true number of signatures and the estimated number differ, we also pad the smaller matrix with zeros for the non-matched signatures, which penalizes incorrect estimation of $K$. The rows of $\widehat{\boldsymbol{\Theta}}$ and $\boldsymbol{\Theta}^0$ are also permuted accordingly, and are also padded with zeros to make the dimensions the same. We then calculate $\textsc{rmse}(\bm{R}^0, \widehat{\bm{R}}) = \big(\sum_{mk} (r^0_{mk} - \widehat{r}_{mk})^2/MK\big)^{1/2}$ and $\textsc{rmse}(\boldsymbol{\Theta}^0, \widehat{\boldsymbol{\Theta}}) = \big(\sum_{kn} (\theta^0_{kn} - \widehat{\theta}_{kn})^2/KN\big)^{1/2}$, where  $\widehat{r}_{mk}$ and $\widehat{\theta}_{kn}$ are 
the estimated signatures and exposures, obtained by averaging posterior samples in the case of the Bayesian models. 
The results for $\tau = 0$ and $\tau = 0.15$ are shown in \cref{fig:SigTheta_0,fig:SigTheta_015}, respectively. The best performance is attained by CompNMF+cosmic. While this model uses information from the true COSMIC signatures in the prior, it correctly filters out the signatures that are not needed. Moreover, the improved estimation of the COSMIC signatures improves the estimation of the associated exposures. The performance of all the other models, with the exception of SigProfiler, is virtually identical when $\tau =0$. Finally, when $\tau = 0.15$, PoissonCUSP, SignatureAnalyzer and BayesNMF perform poorly due to overfitting, as before.

\subsubsection{\revision{Computation time and effective sample size}}
\label{sec:computational_time}
\begin{figure}
    \centering
    \includegraphics[width = 0.6\linewidth]{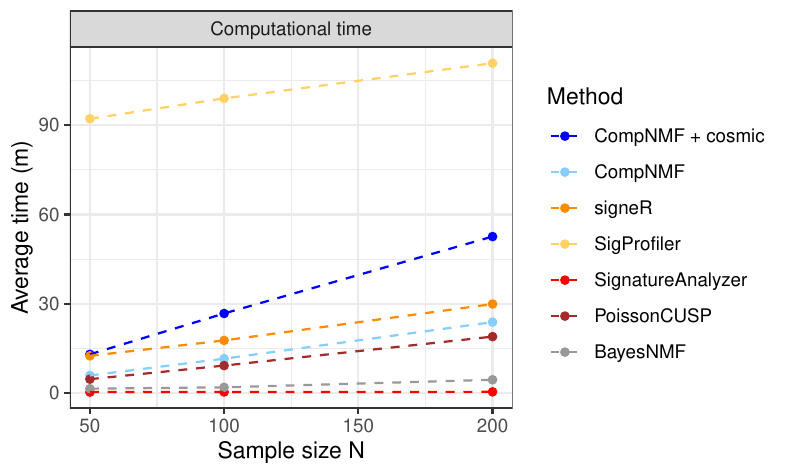}
    \caption{Average computation time (in minutes) for each method. Times shown are the average across all combinations of overdispersion $\tau$ and number of signatures $K^0$ used in the simulation study.}
    \label{fig:simulation_time}
\end{figure}

\begin{figure}
    \centering
    \includegraphics[width = \linewidth]{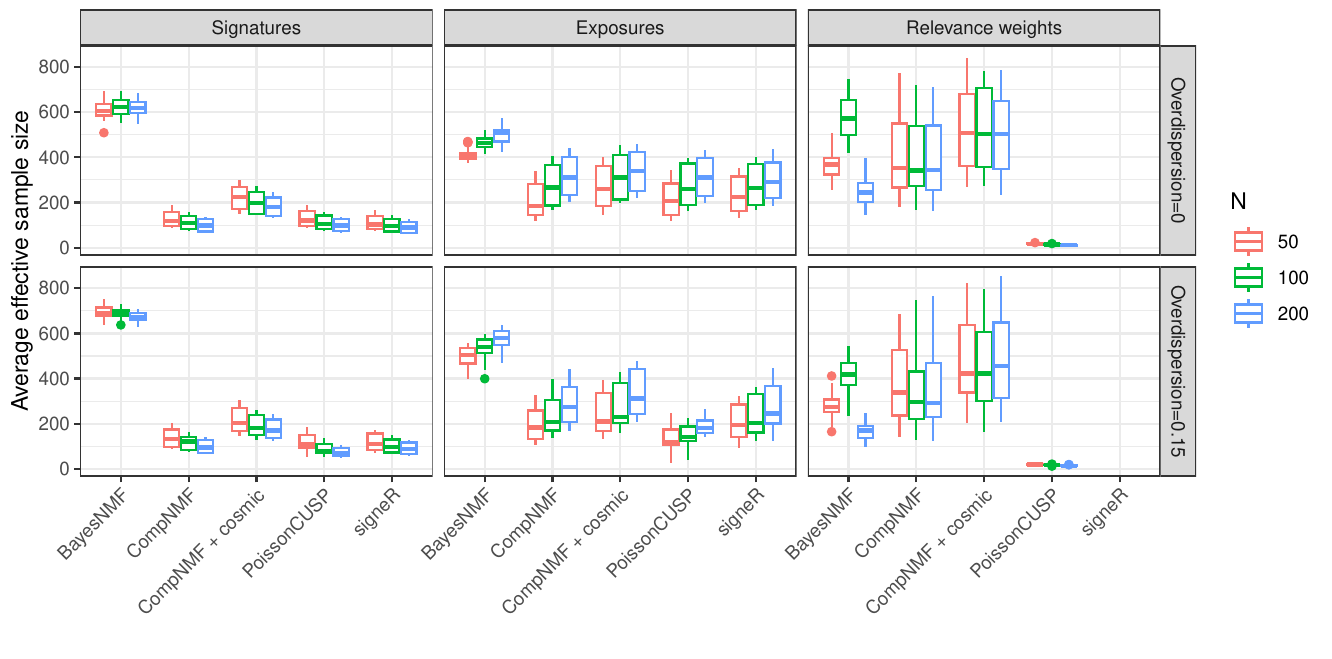}
    \caption{Average effective sample sizes for MCMC algorithms drawing from the posterior of the signatures and the exposures for the Bayesian methods in the simulation.}    \label{fig:simulation_ess}
\end{figure}

\cref{fig:simulation_time} shows the total computation time required by each method, as a function of the number of samples $N$. All computations were performed on an AMD Ryzen 3900-based dedicated server with 128GB of memory, running Ubuntu 20.04, R version 4.3.1 linked to Intel MKL 2019.5-075. Calculations were split across 20 cores via the \texttt{foreach} package, allocating one dataset per core for each combination of $\tau$, $N$, and $K^0$, and running each method sequentially. Not surprisingly, SignatureAnalyzer is the fastest method by an order of magnitude, always taking under one minute to complete. BayesNMF is the second fastest method since inference under the Gaussian NMF model does not require the inclusion of additional latent variables at each iteration of the Gibbs sampler. PoissonCUSP is the third fastest method, because the number of signatures in the model varies adaptively within the Gibbs sampler, preventing unnecessary computations on inactive factors. CompNMF is slightly slower than PoissonCUSP since no adaptation is performed, but it is faster than signeR. 
SigProfiler was the slowest among all methods by a wide margin.
\revision{Thus, in terms of computation time, both CompNMF and CompNMF+cosmic are competitive with the next best-performing existing method, signeR. Furthermore, signeR is written in \texttt{C++}, whereas CompNMF and CompNMF+cosmic are written in \texttt{R} code, so they could be significantly faster if implemented in \texttt{C++} or similar.  Additionally, they could be sped up by updating only the non-compressed factors when running the Gibbs sampler, following the adaptive approach of PoissonCUSP. }
\revision{It is also interesting to compare the performance across models in terms of effective sample size (ESS). These are displayed in \cref{fig:simulation_ess}. ESS values were calculated using the last 1000 posterior samples for each model to ensure a fair comparison. Each boxplot reports the average univariate ESS for $\bm{R}$ and $\boldsymbol{\Theta}$ across 20 replicates. The ESS for the relevance weights $\mu_k$ (or a comparable quantity) are also reported wherever present. We observe a limited difference in the quality of the samples for the models using a Poisson likelihood. This is because the underlying sampler, which is based on multinomial data augmentation, is very similar for all methods. Interestingly, PoissonCUSP performs considerably worse when looking at the relevance weights. BayesNMF, instead, relies on a Gaussian likelihood and has the advantage of not requiring further data augmentation; this leads to considerably better ESS for the signature matrix, though no normalization constraint is imposed in the model. Overall, our CompNMF models exhibit comparable performance in terms of ESS relative to the other Poisson NMF models. }

\begin{figure}[h]
    \centering
    \includegraphics[width=0.9\linewidth]{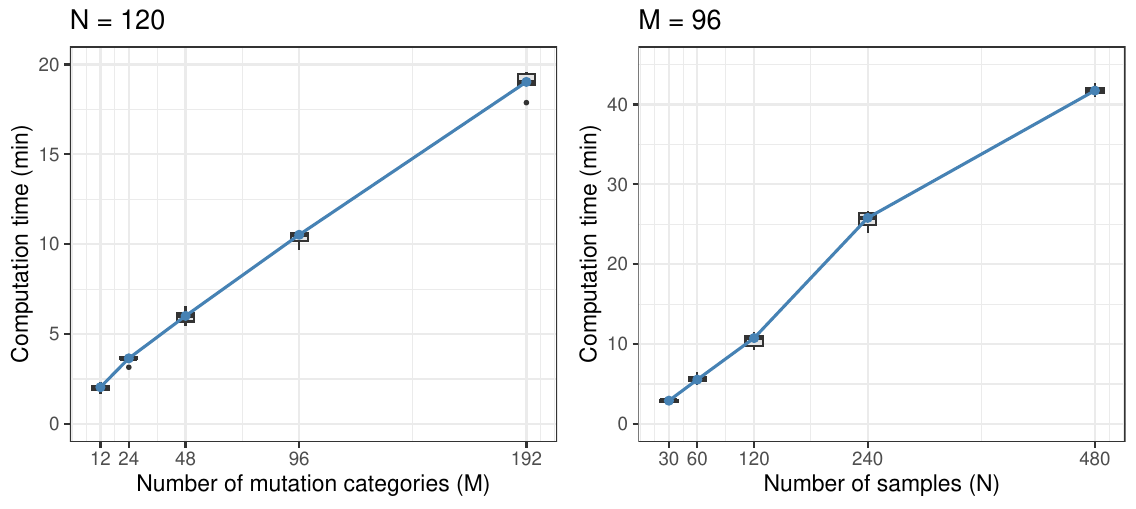}
    \caption{Left: elapsed time in minutes for 3000 MCMC iterations of CompNMF on simulated data, using $N = 120$ and varying $M$. Right: elapsed time keeping $M = 96$ and varying sample size $N$. Boxplots indicate values across 10 replicates.}
    \label{fig:Computation_M_N}
\end{figure}

Finally, we evaluate how the computation time required to run CompNMF grows with increasing $M$ and $N$. 
We simulate 10 data matrices for a range of $M$ and $N$ values, using the same simulation procedure as in Section~\ref{sec:simulation}, keeping $K_{\mathrm{new}} = 6$ and $K_{\mathrm{pre}} = 0$ (since COSMIC signatures only assume $M = 96$). For each data matrix, we run 3000 Gibbs sampling iterations using the CompNMF model in Equations~\eqref{eq:prior_R} to \eqref{eq:hierarchical-model}, initializing at a random starting point.
Figure~\ref{fig:Computation_M_N} shows the elapsed time as a function of $M$ and $N$. We see that the computational time grows approximately linearly with each of $M$ and $N$, separately. 

\subsection{Performance comparison with Bayesian infinite factorization models}\label{subsec:InfiniteFactors}
This section extends the simulation study to compare our CompNMF model versus established Bayesian infinite factorization models from the literature. These approaches posit an infinite number of factors, indexed by $k = 1,2,\ldots$, and use a prior on the exposures that applies an increasing degree of shrinkage for larger values of $k$.
The structure of each model is as follows:

\begin{equation}
X_{mn} \sim \mathrm{Poisson}\left(\sum_{k = 1}^\infty r_{mk}\theta_{kn}\right), \quad 
(r_{1 k}, \ldots, r_{M k})\sim \mathrm{Dir}(\alpha, \ldots, \alpha), \quad (\theta_{k n}) \sim p_{\mathrm{\infty}},
\end{equation}

where $p_{\mathrm{\infty}}$ denotes a prior on the exposures for $k=1,2,\ldots$ and $n=1,\ldots,N$, jointly. We consider two ways of constructing $p_{\mathrm{\infty}}$: the Cumulative Shrinkage Process (CUSP) prior of \citetSupp{Legramanti_2020} and the Multiplicative Gamma Process (MGP) prior of \citetSupp{Bhattacharya_Dunson_2011}. The first one relies on an infinite spike-and-slab process, while the second uses multiplicative shrinkage.  Extensive comparisons of these priors exist for Gaussian factor models, but not for Poisson NMF, to the best of our knowledge. Therefore, we consider the following five alternatives:
\begin{itemize}[leftmargin=*, align=left]
    \item \underline{\textsc{cusp-gamma}}: the CUSP prior of \citetSupp{Legramanti_2020}, with $$\theta_{kn} = \vartheta_{kn}\mu_k, \quad \vartheta_{kn}\sim \mathrm{Ga}(a,a), \quad \mu_k \sim (1-\pi_{k})\mathrm{Ga}(a_0,b_0) + \pi_k \delta_{\mu_\infty}, \qquad (\pi_1, \pi_2, \ldots) \sim \textsc{cusp}(\alpha_\pi)$$ 
    where $\delta_{\mu_\infty}$ is a spike at $\mu_\infty = 0.01$ near zero and $\textsc{cusp}(\alpha_\pi)$ is defined as
    \begin{equation}\label{eq:CUSP_Structure}
    \pi_k = \sum_{\ell = 1}^k  \phi_\ell, \quad  \phi_\ell =  v_\ell \prod_{j = 1}^{\ell-1} (1 - v_j), \quad v_\ell \sim \mathrm{Beta}(1,\alpha_\pi).
    \end{equation}
    Hence, factors with higher $k$ have higher prior probability of setting exposures to near-zero values. This is the version presented in the main manuscript, where we let $\alpha_\pi = 5$, $a = a_0 = b_0 = 1$.
    \item \underline{\textsc{cusp-gammaKC}}: the CUSP prior of \citetSupp{Kowal_Canale2023}. This is an extension of the above model that allows both the spike and the slab to be continuous, and the parameter $\alpha_\pi$ to be given a prior as well. Specifically,
    \begin{align*}
    \theta_{kn} = \vartheta_{kn}\mu_k, \quad \vartheta_{kn}\sim \mathrm{Ga}(a,a), \quad \mu_k &\sim (1-\pi_{k})\mathrm{Ga}(a_0,b_0) + \pi_k \mathrm{Ga}(a_0, b_0/\mu_\infty) \\ (\pi_{1},\pi_2,\ldots) &\sim \textsc{cusp}(\alpha_\pi), \quad \alpha_\pi \sim \mathrm{Ga}(c_0,  d_0).
    \end{align*}
    In our simulations, we set $a = a_0 = b_0 = 1$, $c_0 = 2$, and $d_0 = 1$. Hence, the spike is again centered at $\mu_\infty = 0.01$. Note that Proposition 1 of \citetSupp{Kowal_Canale2023} applies to this case as well.  
    \item \underline{\textsc{cusp-invgamma}}: the CUSP prior of \citetSupp{Legramanti_2020}, this time with $$\theta_{kn} \sim \mathrm{Ga}(a,a/\mu_k), \quad \mu_k \sim (1-\pi_{k})\mathrm{InvGa}(a_0,b_0) + \pi_k \delta_{\mu_\infty}, \quad (\pi_{1}, \pi_2, \ldots) \sim \textsc{cusp}(\alpha_\pi).$$
    Thus, this model uses conjugate inverse gamma priors, like our Compressive NMF model.
    \item \underline{\textsc{cusp-invgammaKC}}: the CUSP prior of \citetSupp{Kowal_Canale2023} applied to the previous case, that is
    \begin{align*}
    \theta_{kn} \sim \mathrm{Ga}(a,a/\mu_k),\quad  \mu_k &\sim (1-\pi_{k})\mathrm{InvGa}(a_0,b_0) + \pi_k \mathrm{InvGa}(a_0,\mu_{\infty}b_0), \\
    (\pi_{1}, \pi_2, \ldots) &\sim \textsc{cusp}(\alpha_\pi), \quad \alpha_\pi\sim \mathrm{Ga}(c_0, d_0).
    \end{align*} We let $c_0 = d_0 = 1$, and $a_0 = 2$, $b_0 = 1$, $a = 1$, so that the spike is again centered at $\mu_\infty$.
    \item \underline{\textsc{PoissonMGP}}: the MGP prior of \citetSupp{Bhattacharya_Dunson_2011}, defined as
    $$\theta_{kn} \sim \mathrm{Ga}(a,a/\mu_k), \quad \mu_k = \prod_{\ell = 1}^k \nu_\ell, \quad \nu_{1} \sim \mathrm{InvGa}(c_0, 1), \quad \nu_{\ell} \sim \mathrm{InvGa}(d_0, 1) \text{ for $\ell\geq 2$}.$$ Hence, a high value of $k$ is linked to a smaller value of the product between inverse gammas, which further shrinks the exposures. We let $a = 1$, and we follow the suggestions proposed by \citetSupp{DURANTE2017198} and set $c_0 = 2.1$ and $d_0 = 3.1$.
\end{itemize}

\subsubsection{Inference details}\label{sec:cusp}
Since each of the infinite factorization models employs conjugate priors, inference can be performed via Gibbs sampling. Specifically, all CUSP models rely on the following steps:
\begin{enumerate}[leftmargin=*, align=left]
    \item \textbf{Mutation counts.} For $m =1, \ldots, M$ and $n =1, \ldots, N$, sample auxiliary variables $\textbf{Y}_{m n} = (Y_{m n 1}, \ldots, Y_{m n K})$ according to 
    $
    \textbf{Y}_{m n}\sim \mathrm{Multinomial}\big(X_{m n}, (q_{m n 1}, \ldots, q_{m n K})\big),
    $
    where $q_{m n k} = r_{m k}\theta_{k n}/\sum_{\kappa = 1}^Kr_{m \kappa}\theta_{\kappa  n}$.
    \item \textbf{Signatures.} Sample the signatures $\bm{r}_k = (r_{1 k}, \ldots, r_{M k})$ from $$(\bm{r}_k\mid -) \sim \mathrm{Dir}\bigg(\alpha + \sum_{n = 1}^{N} Y_{1 n k}, \ldots, \alpha + \sum_{n = 1}^{N} Y_{M n k} \bigg).$$
    \item \textbf{Exposures.} Depending on the choice of model, sample the exposures as:
    \begin{itemize}
        \item \textsc{cusp-gamma} and \textsc{cusp-gammaKC}: sample the exposures as $\vartheta_{k n}$ from $$(\vartheta_{k n}\mid -) \sim \mathrm{Ga}\bigg(a + \sum_{m = 1}^M Y_{m n k},\, a + \mu_k \bigg).$$
        and then set $\theta_{k n} = \vartheta_{k n} \mu_k$
        \item \textsc{cusp-invgamma} and \textsc{cusp-invgammaKC}: sample the exposures as  $$(\theta_{k n}\mid -) \sim \mathrm{Ga}\bigg(a + \sum_{m = 1}^M Y_{m n k},\, 1 + a/\mu_k \bigg).$$
    \end{itemize}
\item \textbf{Spike assignments.}  For each $k = 1, \ldots, K$, sample the categorical auxiliary variables $Z_{k}$ depending on the model:
{\footnotesize
\begin{align*}
\textsc{cusp-gamma}: &\qquad \pr(Z_k = \ell \mid -) \propto 
\begin{cases}
\phi_\ell\, \mu_{\infty}^{\sum_{m n} Y_{m n k}} \exp(-\mu_\infty\sum_n\vartheta_{k n}) &\text{ if } 1 \leq \ell \leq k\\[8pt]
 \displaystyle\phi_\ell\, \frac{b_0^{a_0}}{\Gamma(a_0)} \frac{\Gamma(a_0 + \sum_{m n} Y_{m n k})}{(b_0 + \sum_n \vartheta_{k n})^{a_0 + \sum_{m n} Y_{m n k}}}& \text{ if }  k < \ell \leq K
\end{cases}\\ \\ 
\textsc{cusp-gammaKC}: &\qquad \pr(Z_k = \ell \mid -) \propto 
\begin{cases}
\displaystyle \phi_\ell\, \frac{(b_0/\mu_\infty)^{a_0}}{\Gamma(a_0)} \frac{\Gamma(a_0 + \sum_{m n} Y_{m n k})}{(b_0/\mu_\infty + \sum_n \vartheta_{k n})^{a_0 + \sum_{m n} Y_{m n k}}} &\text{ if } 1 \leq \ell \leq k\\[8pt]
 \displaystyle\phi_\ell\, \frac{b_0^{a_0}}{\Gamma(a_0)} \frac{\Gamma(a_0 + \sum_{m n} Y_{m n k})}{(b_0 + \sum_n \vartheta_{k n})^{a_0 + \sum_{m n} Y_{m n k}}}& \text{ if }  k < \ell \leq K
\end{cases}\\ \\ 
\textsc{cusp-invgamma}: &\qquad \pr(Z_k = \ell \mid -) \propto 
\begin{cases}
\phi_\ell\, \mu_{\infty}^{-aN} \mathrm{exp} (- a \sum_{n = 1}^N \theta_{k n}/\mu_{\infty})&\text{ if } 1 \leq \ell \leq k\\[8pt]
 \displaystyle\phi_\ell\, \frac{b_0^{a_0}}{\Gamma(a_0)} \frac{\Gamma(a_0 + aN)}{(b_0 + a\sum_n \theta_{k n})^{a_0 + \sum_{m n} Y_{m n k}}}& \text{ if }  k < \ell \leq K
\end{cases}\\ \\ 
\textsc{cusp-invgammaKC}:  &\qquad \pr(Z_k = \ell \mid -) \propto 
\begin{cases}
\displaystyle\phi_\ell\, \frac{(b_0\mu_\infty)^{a_0}}{\Gamma(a_0)} \frac{\Gamma(a_0 + aN)}{(b_0\mu_\infty + a\sum_n \theta_{k n})^{a_0 + \sum_{m n} Y_{m n k}}}&\text{ if } 1 \leq \ell \leq k\\[8pt]
 \displaystyle\phi_\ell\, \frac{b_0^{a_0}}{\Gamma(a_0)} \frac{\Gamma(a_0 + aN)}{(b_0 + a\sum_n \theta_{k n})^{a_0 + \sum_{m n} Y_{m n k}}}& \text{ if }  k < \ell \leq K
\end{cases}
\end{align*}
}
where $\phi_\ell = \pr(Z_k = \ell)$ is the prior probability of the auxiliary variables.
\item \textbf{Stick-breaking weights}. For $\ell = 1,\ldots, K-1$, sample 
$$(v_\ell\mid -) \sim \mathrm{Beta}\bigg(1 + \sum_{k = 1}^K\mathds{1}(Z_k =\ell),\; \alpha_\pi + \sum_{k = 1}^K\mathds{1}(Z_k >\ell)\bigg),$$
and fix $v_K = 1$.
\item \textbf{Probabilities}. Calculate $\phi_1, \ldots, \phi_K$ via the stick-breaking construction, $\phi_\ell = v_\ell\prod_{m = 1}^{\ell - 1} (1 - v_m)$.
\item \textbf{Relevance weights}. For $k = 1, \ldots, K$, 
update each $\mu_k$ as follows:
\begin{align*}
\textsc{cusp-gamma}: &\qquad \textrm{if } Z_k \leq k, \ \mu_k = \delta_{\mu_\infty} \textrm{ else } \textstyle \mu_k \sim \mathrm{Ga}\left(a_0 + \sum_{mn} Y_{m n k},\; b_0 + \sum_n \vartheta_{k n}\right).\\
\textsc{cusp-gammaKC}: &\qquad \textstyle \textrm{if } Z_k \leq k, \ (\mu_k \mid -)\sim \mathrm{Ga}\left(a_0 + \sum_{mn} Y_{m n k},\; b_0/\mu_\infty + \sum_n \vartheta_{k n}\right) \\ &\qquad  \textrm{ else } \textstyle (\mu_k\mid -) \sim \mathrm{Ga}\left(a_0 + \sum_{mn} Y_{m n k},\; b_0 + \sum_n \vartheta_{k n}\right).\\
\textsc{cusp-invgamma}: &\qquad \textrm{if } Z_k \leq k, \ \mu_k = \delta_{\mu_\infty} \textrm{ else } \textstyle (\mu_k \mid-)\sim \mathrm{InvGa}\left(a_0 + a N,\; b_0 + \sum_n\theta_{k n}\right).\\
\textsc{cusp-invgammaKC}: &\qquad \textstyle \textrm{if } Z_k \leq k, \ (\mu_k \mid -) \sim \mathrm{InvGa}\left(a_0 + a N,\; b_0\mu_\infty + \sum_n \theta_{k n}\right) \\ &\qquad  \textrm{ else } \textstyle (\mu_k\mid -) \sim \mathrm{InvGa}\left(a_0 + a N,\; b_0 + \sum_n \theta_{k n}\right).
\end{align*}
\item \textbf{Hyperparameter} $\alpha_\pi$. Sample $(\alpha_\pi \mid -) \sim \mathrm{Ga}\big(c_0 + K - 1, d_0 - \sum_{\ell = 1}^{K-1} \log(1 - \nu_\ell)\big)$.
\end{enumerate}

Inference for the number of signatures $\widehat{K}$ in PoissonCUSP can be performed using the same adaptive Metropolis sampler as in Algorithm 2 of \citetSupp{Legramanti_2020}. This automatically updates the number of columns in the factorization during sampling, by eliminating the columns that are assigned to the spike  in a given iteration (that is, the signatures for which $Z_k \leq k$) and potentially adding novel ones. Refer to \citetSupp{Legramanti_2020} for a description. We implement their algorithm with their choice of tuning hyperparameters.

As for the multiplicative gamma process of \citetSupp{Bhattacharya_Dunson_2011}, steps (1) and (2) remain the same as the ones above, with the following additional sampling steps for the exposures and the relevance weights. For this implementation,  we impose an upper bound on the number of signatures, which we refer to as $\overline{K}$. 
\begin{enumerate}[leftmargin=*, align=left]
    \item \textbf{Exposures}. Sample $(\theta_{kn} \mid -) \sim \mathrm{Ga}(a + \sum_{m} Y_{m n k}, 1 + a/\mu_k)$.
    \item \textbf{Relevance weights}  Sample
    \begin{align*}
        \nu_1 &\sim \mathrm{InvGa}\bigg(c_0 + \overline{K}N, 1 + \sum_{k = 1}^{\overline{K}}\sum_{n= 1}^N \frac{\theta_{kn}}{\prod_{\ell = 2}^{\overline{K}} \nu_\ell}\bigg),\\
        \nu_\ell &\sim \mathrm{InvGa}\bigg(d_0 + N(\overline{K} - \ell + 1), 1 + \sum_{k = 1}^{\overline{K}}\sum_{n= 1}^N \frac{\theta_{kn}}{\prod_{t = 1, t\neq \ell}^{\overline{K}} \nu_t}\bigg), \qquad \ell = 2, \ldots, \overline{K}.
    \end{align*}
    and set $\mu_k = \nu_1 \prod_{\ell = 2}^k \nu_\ell$.
\end{enumerate}

\begin{figure}[!t]
    \centering
    \includegraphics[width=0.9\linewidth]{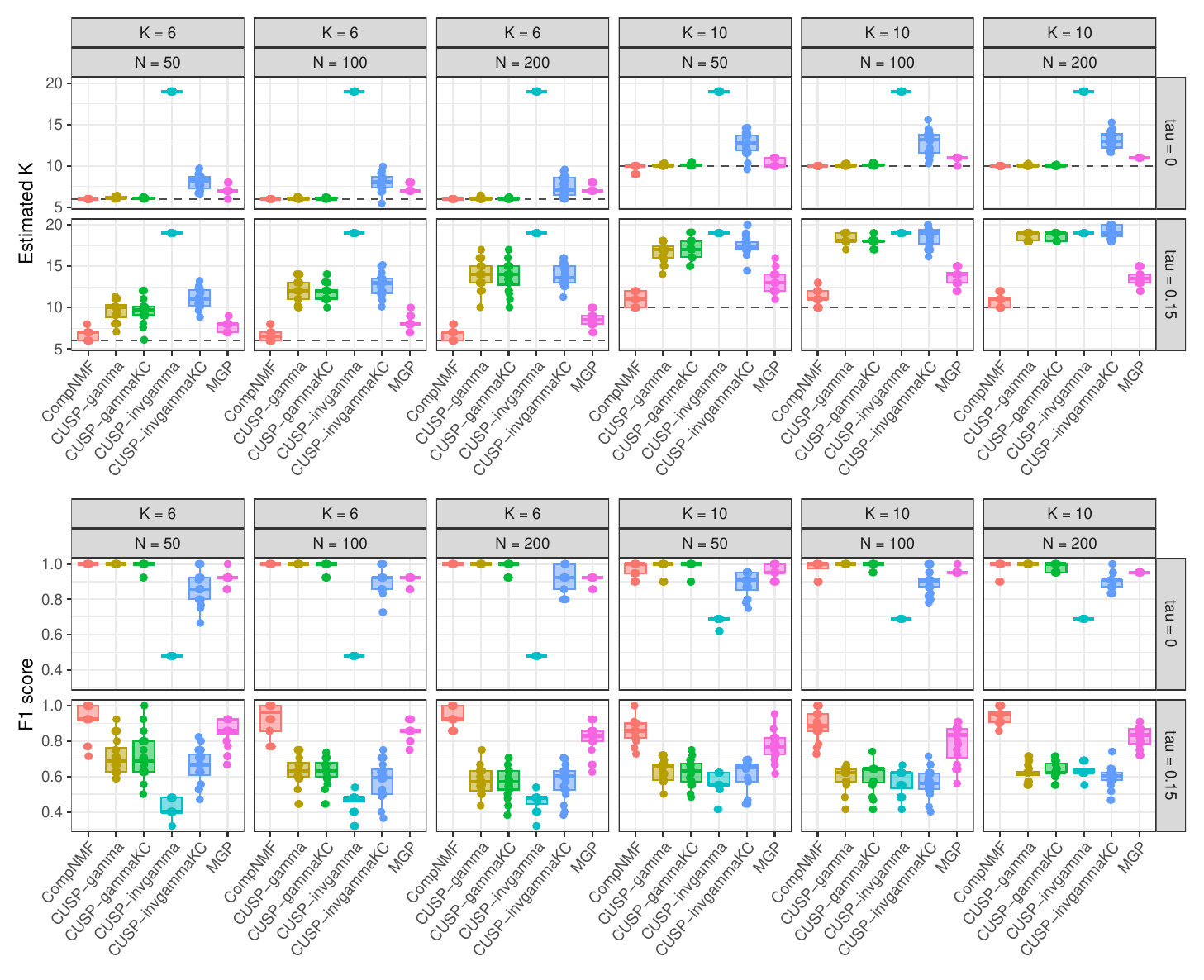}
    \caption{Top panel: estimated number of signatures across all models, simulation scenarios, and overdispersion $\tau$. Points indicate individual replicates. Bottom panel: $F_1$ score calculated in each simulated data, for the same experiment.}
    \label{fig:InfiniteFactors}
\end{figure}

\subsubsection{Simulation results}
Using the same simulation procedures as in Section \ref{sec:simulation} of the main manuscript, we run all models above, using $\mu_\infty = 0.01$. We also re-run CompNMF with $\varepsilon = 0.01$ for comparison. Since infinite factorization models require factor exchangeability, using an informative prior would result in an unnatural asymmetry where the exposures for some known signatures would be penalized more than others. Hence, we only consider the version of CompNMF without informative priors. Each model are estimated by running the MCMC sampler for 5,000 iterations, with the last 1,000 samples used for inference.  For all models, we start with $K = 20$ and then infer the number of signatures according to the specified shrinkage mechanism. Specifically, for CompNMF, we retain all signatures for which the posterior mean satisfies $\widehat{\mu}_k > 5\varepsilon$; for each \textsc{cusp} model, we retain the signatures for which the spike indicator $Z_k$ included the signature in more than 95\% of MCMC samples; and for \textsc{mgp}, we retain the signatures for which $\widehat{\mu}_k > 0.01$ and the cosine similarity between the posterior mean of $\bm{r}_k$ and $(1/96,\ldots,1/96)$ was less than 0.975.

Figure~\ref{fig:InfiniteFactors} displays the results of the simulation. In the misspecified scenario, we observe that CompNMF achieves the best performance overall, both in terms of recovering the correct $K$ (top panel) and attaining the highest $F_1$ score (bottom panel). The second-best performance appears to be from \textsc{PoissonMGP}. This is likely due to the rather conservative behavior of this prior. The worst performance is exhibited by \textsc{cusp-invgamma}, which fails to zero out signatures even in correctly specified settings. However, the variation of \citetSupp{Kowal_Canale2023} leads to improvements. We also detect little difference between \textsc{cusp-gamma} \citepSupp{Legramanti_2020} and \textsc{cusp-gammaKC} \citepSupp{Kowal_Canale2023}. Finally, in the correctly specified scenario, CompNMF and \textsc{cusp-gamma}  are the best models and perform equally well.

\subsection{Simulation with sparse indel mutations}\label{subsec:sparse_data}
In this section, we consider a simulation setting in which the mutation counts are considerably sparser. In practice, sparsity occurs naturally in count data for insertion-deletion mutations (indels), which consist of the addition or removal of one or more nucleotides at a given position in the genome. \citetSupp{Alexandrov_2013} categorize indels into 83 mutational channels, and the resulting indel signatures tend to be lower entropy than SBS signatures.
In COSMIC v3.4, there are 23 recognized indel mutational signatures with a putative etiology, which we can use to define informative priors in our framework. 

We simulate indel counts as $X_{m n} \sim \mathrm{NegBin}\big(1/\tau,\, 1/(1 + \tau\lambda_{m n}^0)\big)$, 
similarly to the SBS simulation study in Section~\ref{sec:simulation},
where  $\lambda^0_{m n} = \sum_{k = 1}^{K_{\mathrm{pre}}^0} \rho^0_{m k}\omega^0_{k n} + \sum_{k = 1}^{K_{\mathrm{new}}^0} r^{0}_{m k}\theta_{k n}^0$. We again set $K_{\mathrm{pre}}^0 = 4$, and for $k = 1,\ldots,4$, we define $\boldsymbol{\rho}_{k}^0 = (\rho_{1 k}^0,\ldots,\rho_{M k}^0)$ to be  COSMIC indel signatures \textsc{id}1, \textsc{id}2, \textsc{id}8, and \textsc{id}9, respectively. \textsc{id}1 and \textsc{id}2 are both related to replication slippage and defective DNA mismatch repair mechanisms. \textsc{id}8 is a clock-like signature and \textsc{id}9 does not have an assigned etiology but has been shown to appear in many cancer types. We also randomly generate $\bm{r}_k^0 = (r_{1 k}^0,\ldots,r_{M k}^0)$ as $\bm{r}_k^0 \sim \mathrm{Dir}(0.05,\ldots,0.05)$, independently for $k = 1,\ldots,K_{\mathrm{new}}^0$, and choose $K_{\mathrm{new}}^0 = 2$ for simplicity, yielding a total of $K = 6$ true signatures.
We generate exposures by setting $\omega^0_{k n} = w_k \xi_{k n}$ where $w_k \sim \mathrm{Ga}(50, 1)$ and $\xi_{k n} \sim \mathrm{Ga}(0.5, 0.5)$ independently, and $\theta^0_{k n}$ in the same way as $\omega^0_{k n}$.
These simulation settings were chosen to produce data similar to observed indel count data.

We generate 20 simulated datasets with $N = 100$ for each $\tau \in\{0,0.15\}$, representing the correctly specified ($\tau = 0$) and misspecified cases ($\tau = 0.15$). The resulting mutation count matrices are sparse, with approximately $65\%$ of entries in the $\bm{X}$ matrix being equal to zero; for comparison, the SBS simulation in Section~\ref{sec:simulation} produces around $30\%$ zeros. This sparsity is a consequence of the sparser signatures and the smaller exposures ($w_k$ has a mean of $50$ instead of $100$). We run the following models:
\begin{enumerate}[leftmargin=*, align=left]
    \item CompNMF with default values $K = 20$, $\varepsilon=0.001$, $\alpha = 0.5$ and $a = 1$.
    \item CompNMF+cosmic with $K_{\mathrm{new}} = 10$ \emph{de novo} signatures and $K_{\mathrm{pre}} = 23$ COSMIC v3.4 \textsc{id} signatures, and with $\varepsilon=0.001$.
    \item SignatureAnalyzer with $K= 25$ and selection method set to \texttt{L1KL} . 
    \item SigProfiler with $K$ ranging from $2$ to $20$, with $10$ NMF replicates and random initialization. 
    \item BayesNMF with $K = 20$ and default hyperparameters.
\end{enumerate}
Note that signeR is not included since the signeR software is only designed to handle SBS mutations.  We also exclude PoissonCUSP from these comparisons due to its lower performance in our other experiments.

\begin{figure}[t]
    \centering
    \includegraphics[width=\linewidth]{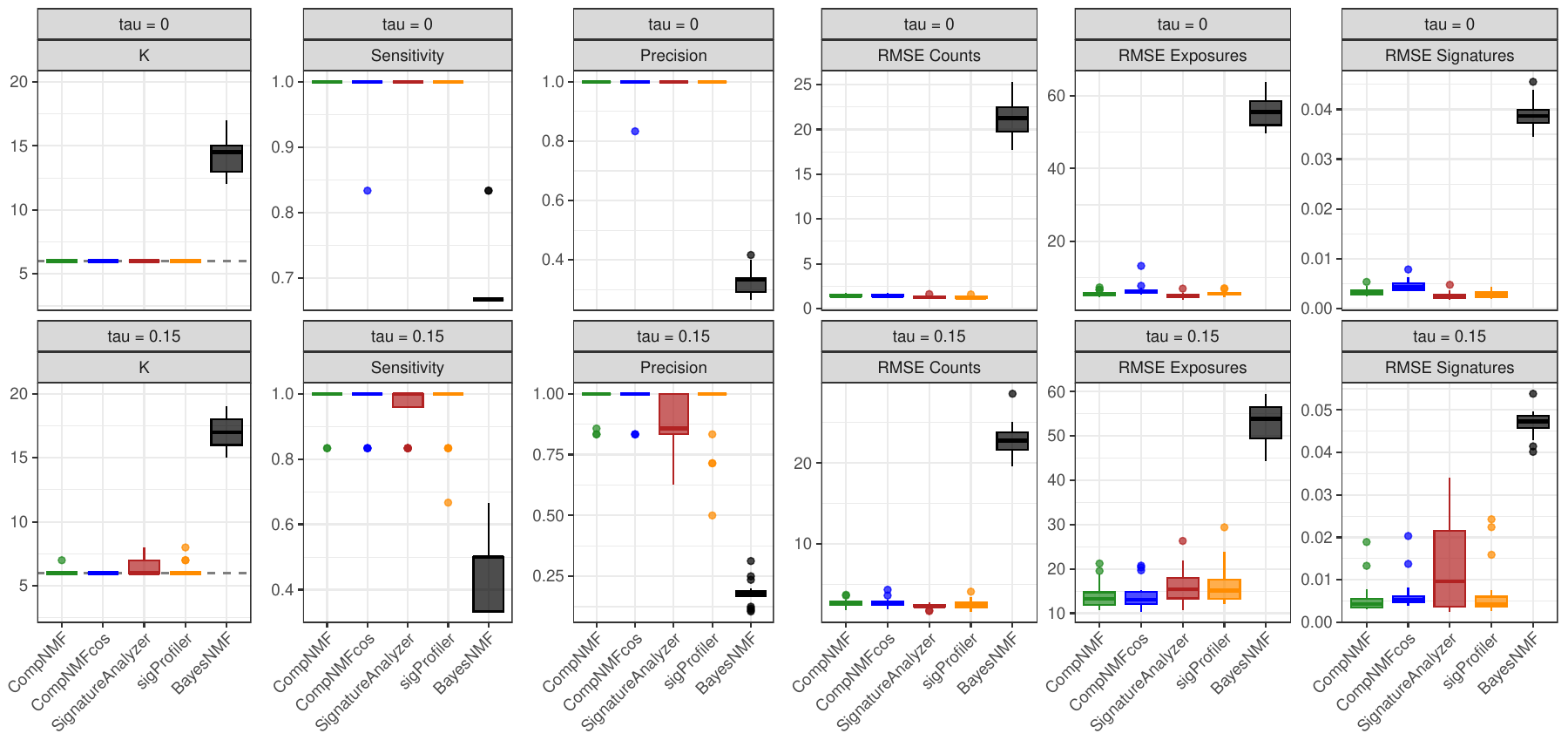}
    \caption{{\revision{Results from the sparse indel simulation. Boxplots summarize the results over 20 replicated datasets in each case. Top row: results from the correctly specified case ($\tau = 0$). Bottom row: results from the misspecified case ($\tau = 0.15$). From left to right, the panels display: (1) number of estimated signatures by each model, (2) sensitivity and (3) precision with respect to the estimated signature profile $\widehat{\bm{R}}$, (4) RMSE between the mutation matrix and the product $\widehat{\bm{R}}\,\widehat{\boldsymbol{\Theta}}$, (5) RMSE between $\bm{R}^0$ and $\widehat{\bm{R}}$, and (6) RMSE between $\boldsymbol{\Theta}^0$ and $\widehat{\boldsymbol{\Theta}}$.}}}
    \label{fig:Indel_simulation_results}
\end{figure}

\cref{fig:Indel_simulation_results} displays the results. When the model is correct ($\tau = 0$), all of the models perform extremely well, with the exception of BayesNMF.  The lower performance of BayesNMF is presumably due to its usage of a Gaussian likelihood, which is not well-suited for sparse count data.  In the misspecified setting ($\tau = 0.15$), CompNMF, CompNMF+cosmic, and SigProfiler perform nearly as well as they do when the model is correct. Meanwhile, SignatureAnalyzer has somewhat reduced performance, and the performance of BayesNMF is degraded even further. These results indicate that the good performance of our compressive NMF models extends to sparse settings as well.

\section{\revision{Sensitivity analyses}}\label{sec:sensitivity_analyses}
\revision{In this section, we conduct a thorough analysis of the sensitivity of our model with respect to the choice of its settings.  
\cref{subsec:comparison_fixed_comp} presents a comparison between our proposed compressive hyperprior and a fixed hyperprior. \cref{subsec:hyperparams_sens} presents sensitivity analyses with respect to the choice of $K$, $\varepsilon$, $a$, and $\alpha$ in the compressive case. }

\subsection{\revision{Fixed versus compressive hyperprior}}\label{subsec:comparison_fixed_comp}
\begin{figure}[h]
    \centering
    \includegraphics[width=0.9\linewidth]{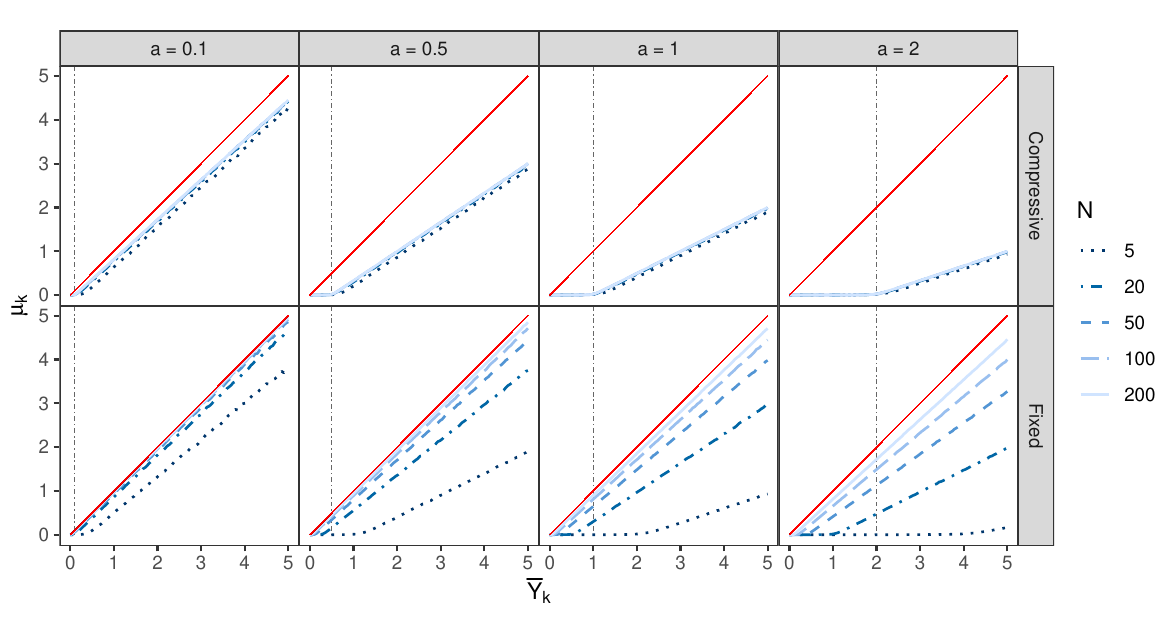}
    \caption{\revision{Comparison between the mean of $\mu_k \mid \bm{Y}$ under our compressive hyperprior $\mu_k^{\textsc{c}} \sim \mathrm{InvGa}(aN + 1, \varepsilon a N)$ and the fixed hyperprior $\mu_k^{\textsc{f}} \sim \mathrm{InvGa}(10 a + 1, 10\varepsilon a)$, i.e., fixing the sample size at $N = 10$, with $\varepsilon = 0.001$, for varying values of $a$ and sample size $N$. The red line indicates where $\overline{Y}_k$ is equal to $\mu_k$, while the vertical dashed line shows the value of $a$.}}
    \label{fig:compressive_vs_fixed_elbow}
\end{figure}

\revision{The form of the hyperprior on the relevance weights $\mu_k$ is essential for automatic relevance determination (ARD) in NMF to work properly. 
Recall that our approach involves letting $\mu_k$ have a small prior mean, like $0.01$ or $0.001$, which makes unneeded factors shrink to small values and effectively removes them from the decomposition \citepSupp{Tan_Fevotte_2013, Brouwer2017b}. However, the strength of this hyperprior on $\mu_k$ can significantly impact the results. Here, we empirically compare our compressive approach, in which $\mu_k \sim \mathrm{InvGa}(aN + 1, \varepsilon a N)$, and the fixed hyperprior $\mu_k \sim \mathrm{InvGa}(a_0, b_0)$ where $a_0$ and $b_0$ are fixed. To make a direct comparison, we set $a_0 = a N_0 + 1$ and $b_0 = \varepsilon a N_0$, where $N_0 = 10$ is fixed. This ensures that $E(\mu_k) = \varepsilon$ in both cases. By the proof of Theorem~\ref{thm:InvKumPost},
\begin{equation}%
(\mu_k\mid \bm{Y}) \sim \mathrm{InvKummer}\big(\mu_k\;\big\vert\; a_0 + N a,\, b_0, \, N a + N\overline{Y}_{k},\, a\big)
\end{equation}
in the fixed case.
We use superscripts to distinguish the relevance weights in the compressive and fixed cases, denoting them $\mu_k^{\textsc{c}}$ and  $\mu_k^{\textsc{f}}$, respectively.} 

\textbf{Effect on relevance weights.} We first compare the distributions of $\mu_k^{\textsc{c}}$ and $\mu_k^{\textsc{f}}$ given the latent counts $\bm{Y}$. \cref{fig:compressive_vs_fixed_elbow} shows the conditional means $E(\mu_k^{\textsc{c}}\mid \bm{Y})$ and $E(\mu_k^{\textsc{f}}\mid \bm{Y})$  for varying values of $\overline{Y}_k$, $a$, and sample size $N$. Suppose $\overline{Y}_k \to y_k$ as $N\to\infty$. From Theorem~\ref{thm:concentration} and \cref{thm:concentration_normality_fixed}, we can fully characterize the concentration point of both distributions as $N\to\infty$, specifically, $\mu_k^{\textsc{c}} \mid \bm{Y}$ concentrates at ${\mu_k^*} = 2 a \varepsilon / \big(\{(y_k-a+\varepsilon)^2 + 8 a\varepsilon\}^{1/2} - (y_k - a + \varepsilon)\big)$ and $\mu_k^{\textsc{f}} \mid \bm{Y}$ concentrates at $y_k$. This is evident from \cref{fig:compressive_vs_fixed_elbow}: In the compressive case, the mean of $\mu_k^{\textsc{c}} \mid \bm{Y}$ is insensitive to the value of $N$, being very close to $\mu_k^*$ even when $N$ is very small. Furthermore, $a$ controls the location of the ``elbow'' of the curves in \cref{fig:compressive_vs_fixed_elbow}, which leads to a sparsity-inducing effect.
Meanwhile, in the fixed case, the mean of $\mu_k^{\textsc{f}}\mid \bm{Y}$ approaches the $45^{\circ}$ line where $\mu_k = \overline{Y}_k$. 
Consequently, the elbow goes away as $N\to\infty$, and $a$ does not play the role of a threshold in the fixed case.
Also see  \cref{subsec:hyperparams_sens} for more on the effect of $a$.

\textbf{Effect on signature recovery performance.} To see the effect on performance for recovering the signatures, we compare the compressive and fixed hyperpriors in the simulation setup of Section~\ref{sec:simulation}.
We set $K^0_{\mathrm{pre}} = 4$ and $K^0_{\mathrm{new}} = 6$, so that there are 10 true signatures in the data. Counts are generated in (i) the correctly specified case (Poisson, $\tau = 0$) and (ii) the misspecified case (negative binomial with overdispersion $\tau = 0.15$).  We generate 20 datasets for each combination of sample size $N \in \{20, 50, 100, 200, 300, 400, 500\}$ and overdispersion  $\tau \in \{0, 0.15\}$, and run the models using $K = 20$, $\varepsilon = 0.01$, $a = 1$, and $\alpha = 0.5$ for $4{,}000$  iterations, using the samples from the last $1{,}000$ iterations for inference. Here, for simplicity, we do not use the informative COSMIC priors. 

\begin{figure}[h]
    \centering
    \includegraphics[width=0.9\linewidth]{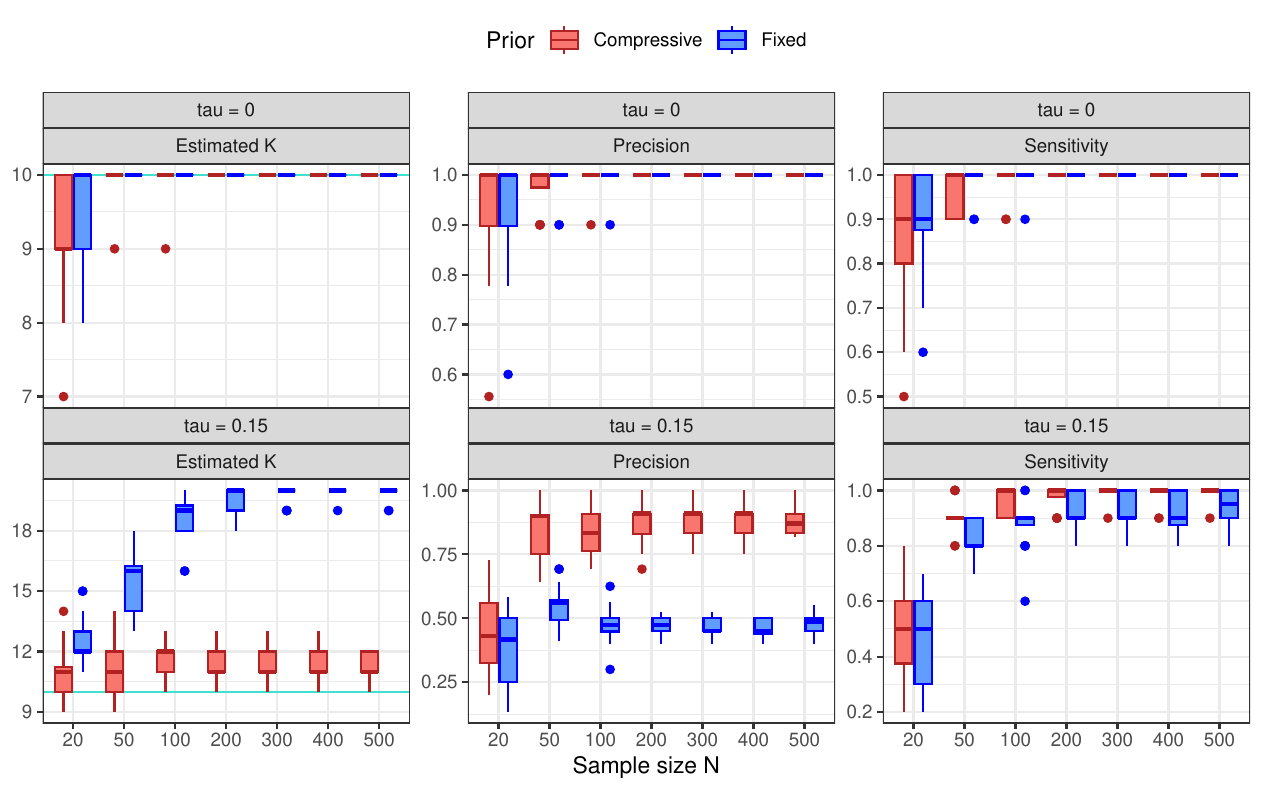}
    \caption{\revision{{Simulation results: Boxplots of the number of active signatures $\widehat{K}$ in the posterior (left column), precision (center column) and sensitivity (right column) across the 20 simulated datasets under compressive and fixed hyperpriors, for $\tau = 0$ (top row) and $\tau = 0.15$ (bottom row).  Signatures are considered active if the average posterior relevance weight is such that $\widehat{\mu}_k > 5\varepsilon$. The horizontal turquoise line indicates the true number of signatures in the data.}}}
    \label{fig:compressive_vs_fixed_K_prec_sens}
\end{figure}

\cref{fig:compressive_vs_fixed_K_prec_sens} shows the estimated number of signatures $\widehat{K}$, precision, and sensitivity.
In the correctly specified case ($\tau = 0$), the compressive and fixed models behave very similarly -- both perform nearly perfectly except for a slight drop in performance when $N$ is small.
However, in the misspecified case ($\tau = 0.15$), there is a stark difference in performance.
We see that the fixed model dramatically overestimates the number of active signatures, exhibits very low precision, and has decreased sensitivity.
The compressive model is much better able to mitigate the effects of misspecification, incurring only a small loss in performance relative to the correctly specified case.

\begin{figure}[!h]
    \centering
    \includegraphics[width=0.85\linewidth]{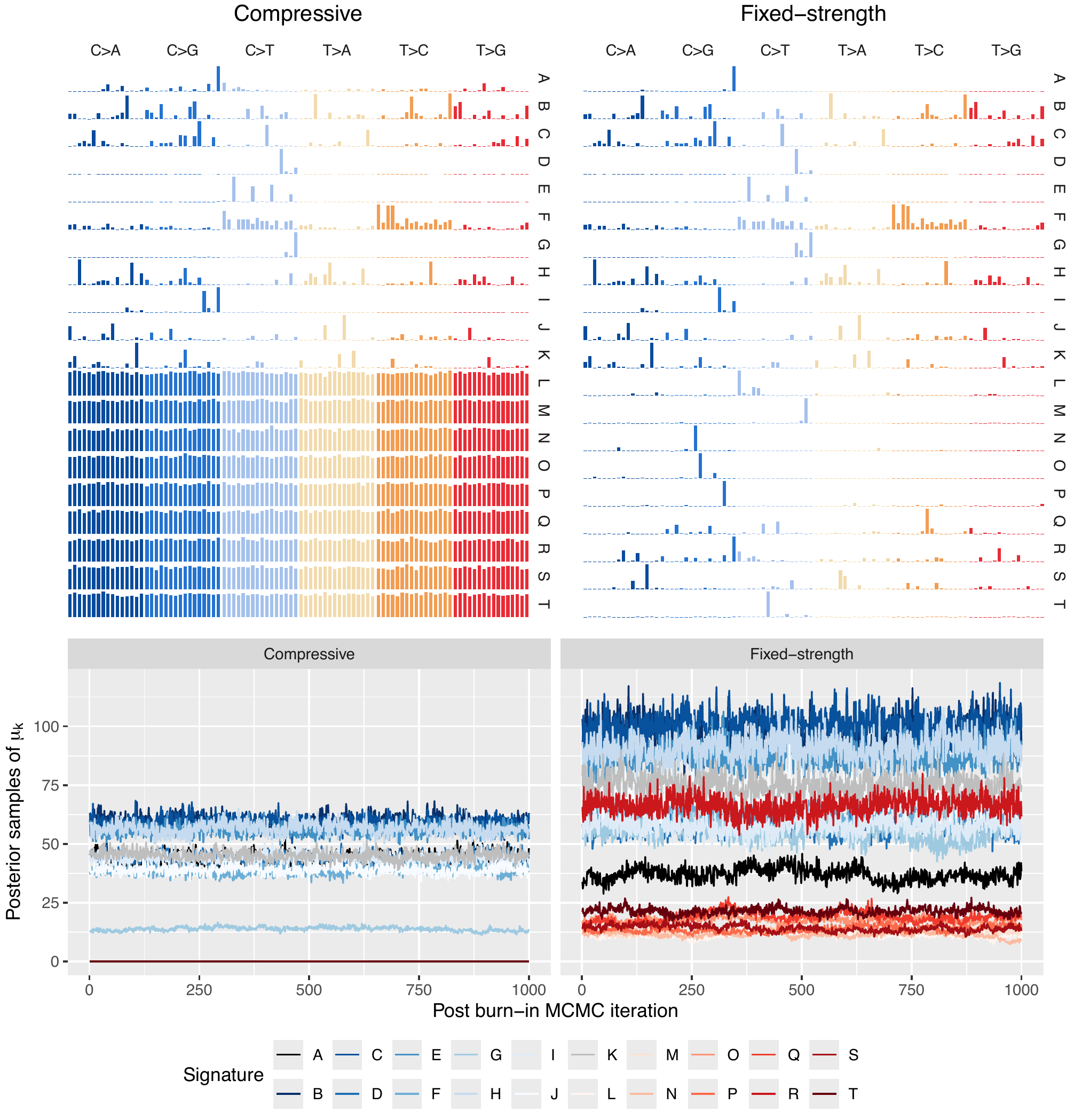}
    \caption{\revision{{Posterior mean of $\bm{R}$ and traceplot for $\boldsymbol{\mu}$ in one of the 20 simulated datasets with $N = 300$ and $\tau = 0.15$. }}}
    \label{fig:compressive_vs_fixed_example}
\end{figure}

\revision{\textbf{Estimated signatures.}
To further examine these differences, \cref{fig:compressive_vs_fixed_example} shows the posterior mean for each of the $K=20$ signatures and their associated relevance weights, for one dataset with $N = 300$ and $\tau = 0.15$. Inferred signatures have been matched and arranged using the Hungarian algorithm to aid visualization so that signatures A, B, C, \dots\ in the compressive case roughly match signatures A, B, C, \dots\ in the fixed counterpart, in terms of cosine similarity.  We see that signatures A through K look very similar in the two cases. On the other hand, signatures L through T are very different.
In the compressive model, signatures L--T are approximately uniform because they correspond to signatures for which $\overline{Y}_k\approx0$ and $\mu_k \approx \varepsilon$; therefore, the full conditional is $(\bm{r}_k\mid \bm{Y}) \sim \mathrm{Dir}(\alpha + \sum_{n = 1}^N Y_{1nk}, \ldots, \alpha + \sum_{n = 1}^N Y_{Mnk}) \approx \mathrm{Dir}(\alpha, \ldots, \alpha)$, for which the mean is $(1/M,\ldots,1/M)$. Meanwhile, in the fixed model, signatures L--T are low-entropy signatures and the corresponding values of $\mu_k$ are far from $\varepsilon$. Note that the traceplots for $\mu_k$ are fairly stable, indicating that the samplers appear to be performing well. }

\begin{figure}[!t]
    \centering
    \includegraphics[width=0.85\linewidth]{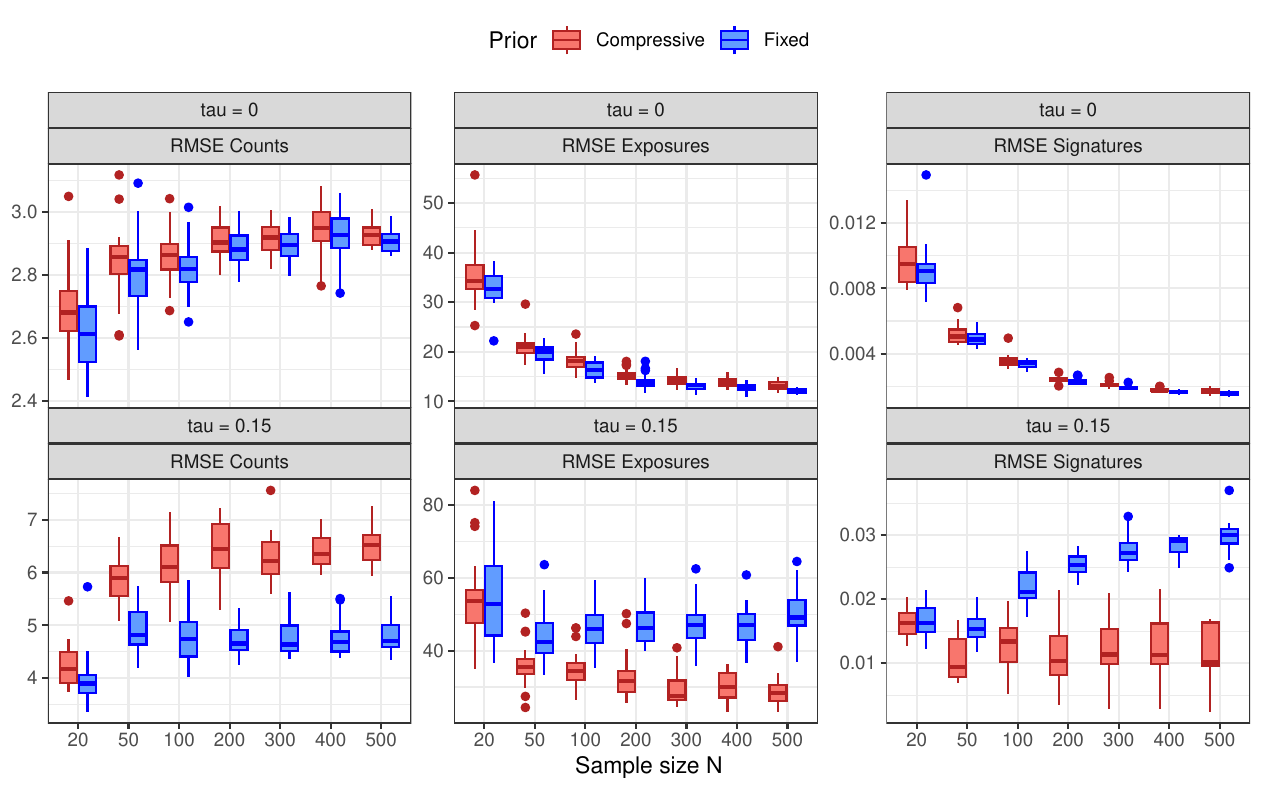}
    \caption{\revision{RMSE between the count matrix $\bm{X}$ and the product $\widehat{\bm{R}}\,\widehat{\boldsymbol{\Theta}}$ (left), RMSE between the true exposures $\boldsymbol{\Theta}^0$ and estimated $\widehat{\boldsymbol{\Theta}}$ (center), and  RMSE between the true signatures $\bm{R}^0$ and $\widehat{\bm{R}}$ (right) for both models under $\tau = 0 $ (top row) and $\tau = 0.15$ (bottom row).}}
    \label{fig:compressive_vs_fixed_rmse}
\end{figure}

\textbf{Effect on estimation performance.} \cref{fig:compressive_vs_fixed_rmse} shows the root mean squared error (RMSE) between the estimated and true values of the signatures, exposures, and count data matrix across the 20 datasets. 
We observe a similar trend as before, with the compressive model outperforming the fixed model in terms of RMSE on the exposures and signatures, especially for larger values of $N$.
While the RMSE of the counts is higher for compressive than fixed, this appears to be due to over-fitting by the fixed model, as evidenced by (i) its numerous spurious signatures in \cref{fig:compressive_vs_fixed_example} and (ii) the fact that the true parameters are more accurately estimated by the compressive model.

\subsection{\revision{Sensitivity to the choice of model settings}} \label{subsec:hyperparams_sens}
\revision{In this section, we evaluate the sensitivity of our proposed model to the choice of settings ($K$, $\varepsilon$, $a$, and $\alpha$).  We find that the model's results are fairly robust to these settings over a wide range of values. The $a$ value has the most important effect, controlling the threshold above which signatures are included in the model. 
We use the same simulation setup as in Section~\ref{sec:simulation}, and generate 40 simulated datasets for each value of overdispersion $\tau \in \{0, 0.15\}$, with $N = 100$, $K_{\mathrm{pre}}^0=4$, and $K_{\mathrm{new}}^{0} = 2$, so that there are 6 true signatures.}

\subsubsection{\revision{Sensitivity to $K$ and $\varepsilon$}}
\begin{figure}[!h]
    \centering
    \includegraphics[width=0.85\linewidth]{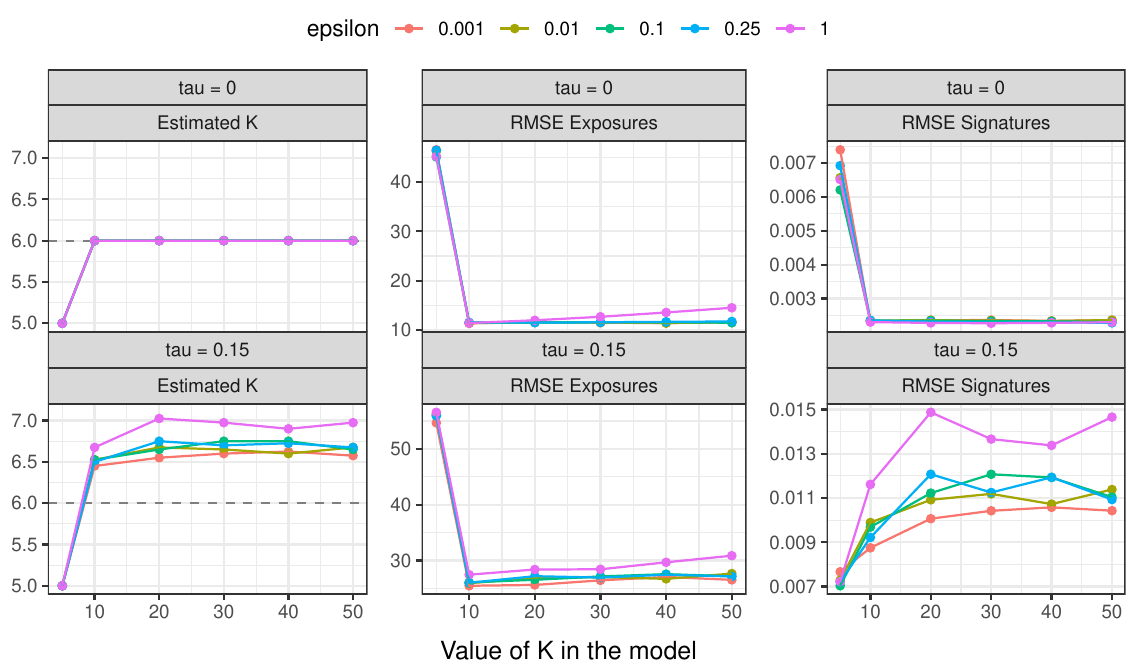}
    \caption{Left:  
    estimated number of signatures. Center:
    RMSE between $\boldsymbol{\Theta}$ and $\widehat{\boldsymbol{\Theta}}$. Right: RMSE between $\bm{R}$ and $\widehat{\bm{R}}$. The dashed horizontal gray line on the left panel indicates the true number of signatures in the data. Each point represents the average over 40 simulated datasets, for each combination of $K$,  $\varepsilon$, and overdispersion $\tau$, holding $a = 1$ and $\alpha = 0.5$ fixed. Each dataset was simulated with $N =100$.
    }
    \label{fig:sensitvity_epsilon_K}
\end{figure}

\revision{We first evaluate sensitivity with respect to the maximum number of signatures in the model, $K$, and the prior mean of the relevance weights, $\varepsilon = E(\mu_k)$. For each dataset, for each combination of $K \in \{5, 10, 20, 30, 40, 50\}$ and $\varepsilon \in \{0.001, 0.01, 0.1, 0.25, 1\}$, we run our compressive model with $a = 1$ and $\alpha = 0.5$ for $4{,}000$  iterations, retaining the samples from the last $1{,}000$ iterations for inference. Notice that when $K = 5$,  the model necessarily has fewer signatures than the true number. }

\cref{fig:sensitvity_epsilon_K} displays the results, quantified in terms of (i) estimated number of signatures, (ii) RMSE between the true exposures matrix $\boldsymbol{\Theta}$ and the model estimate $\widehat{\boldsymbol{\Theta}}$, and (iii) RMSE between the true signature matrix $\bm{R}$ and the model estimate $\widehat{\bm{R}}$.
When $K = 5$, the performance suffers since there are $6$ true signatures. However, the performance is relatively constant for all values of $K$ greater than the true number.
Similarly, the performance is relatively constant as a function of $\varepsilon$, for small values of $\varepsilon$.  Performance degrades for values of $\varepsilon \geq 1$, which is well outside the range of recommended values. Overall, the posterior is robust to the choice of $K$ and $\varepsilon$, as long as $K$ is larger than the true number and $\varepsilon$ is small, such as $0.01$ or $0.001$.

\subsubsection{\revision{Sensitivity to $a$ and $\alpha$}}

\begin{figure}[!h]
    \centering
    \includegraphics[width=0.85\linewidth]{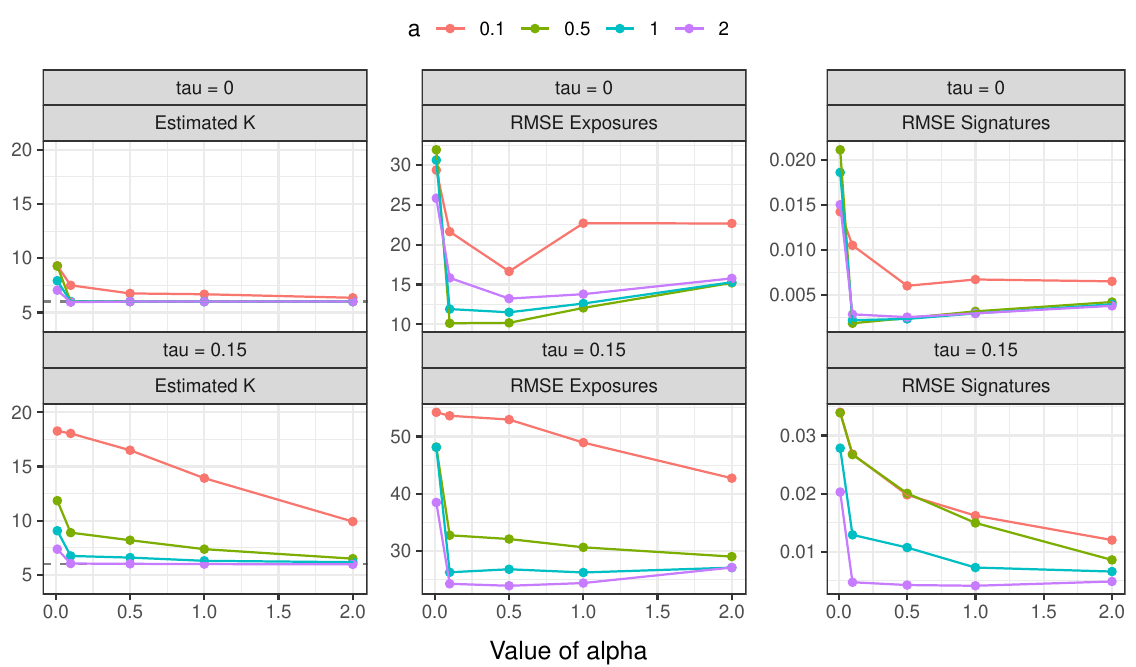}
    \caption{\revision{Left:  
    estimated number of signatures. Center:
    RMSE between $\boldsymbol{\Theta}$ and $\widehat{\boldsymbol{\Theta}}$. Right: RMSE between $\bm{R}$ and $\widehat{\bm{R}}$. The dashed horizontal gray line on the left panel indicates the true number of signatures in the data ($K^0 = 6$). Each point represents the average over 40 simulated datasets, for each combination of $a$, $\alpha$, and overdispersion $\tau$, holding  $\varepsilon = 0.001$ and $K = 20$ fixed. Each dataset was simulated with $N =100$.}}
    \label{fig:sensitvity_alpha_a}
\end{figure}

For each dataset, for each combination of $a \in \{0.1, 0.5, 1, 2\}$ and $\alpha \in \{0.01, 0.1, 0.5, 1, 2\}$, we run our model with $\varepsilon = 0.001$ and $K = 20$ for $4{,}000$  iterations and retain the samples from last $1{,}000$ iterations for inference. 
\cref{fig:sensitvity_alpha_a} displays the results. 
We observe that smaller values of $a$ produce a larger number of active signatures.  This is consistent with the interpretation of $a$: it serves as a threshold such that signature $k$ is included if $\overline{Y}_{k} > a$, where $\overline{Y}_k$ is the average number of mutations due to signature $k$.
In turn, smaller $a$ produces larger RMSE for both $\bm{R}$ and $\boldsymbol{\Theta}$. This is likely due to overfitting in the cases where $a < 1$, as a result of introducing spurious signatures (\cref{fig:sensitvity_alpha_a}). Estimation of signatures and exposures appears more favorable when $a = 2$ in the misspecified case ($\tau = 0.15$), though $a=1$ performs better than $a = 2$ in the correctly specified setting ($\tau = 0$). 
Regarding $\alpha$, we observe that when $a$ is not too small, the choice of $\alpha$ has little effect on the estimated number of signatures.  One exception is that when $\alpha$ is very small ($\alpha = 0.01$), the model introduces a few extra signatures, presumably because this creates a strong prior preference for sparse signatures.

\subsubsection{Effect of putting a hyperprior on $a$}\label{subsec:Random_a_section}
Given that the value of $a$ influences the estimated rank  (\cref{fig:sensitvity_alpha_a}), it might seem natural to put a hyperprior on $a$, but it turns out that this is ill advised.
The issue is that the compressive hyperprior on $\mu_k$ is not intended to accurately represent prior uncertainty in $\mu_k$; instead, its primary role is to shrink $\mu_k$ to zero for unneeded factors. Consequently, putting a hyperprior on $a$ does not improve performance in the way that one would normally expect. 

To illustrate, we consider modifying the model to use a hyperprior $a \sim \pi(a)$. It is straightforward to show that the full conditional for $a$ is then
$$\pi(a\mid \boldsymbol{\Theta}, \boldsymbol{\mu}) \propto\pi(a)\prod_{nk}\mathrm{Ga}(\theta_{kn}; a, a/\mu_k)\prod_{k}\mathrm{InvGa}(\mu_k; aN + 1, \varepsilon a N).$$
For demonstration, we take $\pi(a)$ to be $\mathrm{Ga}(1, 1)$ and we augment the Gibbs sampler to update $a$ from this full conditional using the ratio-of-uniforms method \citepSupp{Devroye86}. We compare this to our proposed model (with fixed $a=1$) on simulated data as in Section~\ref{sec:simulation}. We simulate 20x replicate data sets of $N = 50$ samples each, both in the correctly specified setting ($\tau = 0$) and the misspecified setting ($\tau = 0.15$). For each simulated dataset, we run an MCMC chain for 5000 iterations and use the last 1000 for inference. In each case, the estimated number of signatures is determined using the threshold $\widehat{\mu}_k > 5\varepsilon$, where $\widehat{\mu}_k$ is the posterior mean of $\mu_k$.

\begin{figure}
    \centering
    \includegraphics[width=\linewidth]{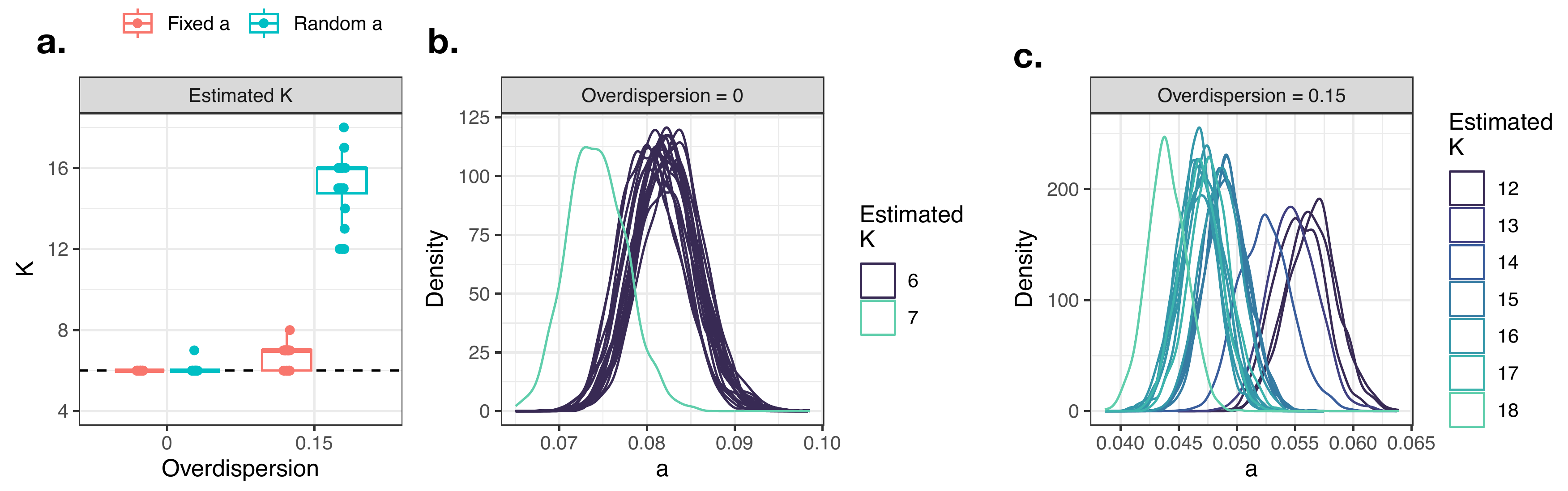}
    \caption{Panel \textbf{a}: Estimated number of signatures $K$ across 20 simulated data sets with $N = 50$ samples, when using fixed $a$ or a hyperprior on $a$ (``Random $a$''). The horizontal dashed line indicated the ground truth. Panel \textbf{b, c}: Posterior densities of $a$ in each simulated dataset  when $\tau = 0$ and $\tau = 0.15$, respectively. Colors denote the estimated number of signatures $K$ in the posterior. }
    \label{fig:random_a_version}
\end{figure}

\cref{fig:random_a_version} displays the results. In the correctly specified case ($\tau = 0$), the hyperprior on $a$ has little effect on the number of estimated signatures (panel a). However,  in the misspecified case ($\tau = 0.15$), the difference becomes substantial. The posterior on $a$ concentrates on small values (panels b and c), which in turn leads to many spurious signatures being included.  Meanwhile, fixing $a = 1$ yields good performance in both settings. Hence, we recommend keeping $a$ fixed.

\section{\revision{Additional results for the 21 breast cancer data}}\label{sec:additional_application_results}
This section contains additional details and results for the applications to the 21 breast cancer data in Section~\ref{sec:application}. We provide further details on the inputs and outputs of the methods, and perform a sensitivity analysis for our CompNMF model. 

\subsection{Details of each method}\label{subsec:details_21breast}

The methods we run are the same as in Section~\ref{sec:simulation}, with only the following changes:
\begin{enumerate}[itemsep=0em,label=(\roman*), leftmargin=*, align=left]
    \item CompNMF: $K = 15$, $\varepsilon = 0.01$, $a=1$, and $\alpha = 0.5$.
    \item CompNMF+cosmic: $K_{\mathrm{new}} = 10$ \emph{de novo} signatures and $K_{\mathrm{pre}} = 67$ COSMIC v3.4 signatures, with $\varepsilon=0.01$.
    \item PoissonCUSP: starting at $K = 15$, with spike at $\mu_\infty = 0.01$ and $a_0 = b_0=1$.
    \item signeR: we set \texttt{estimate\_hyper = TRUE} as in \citetSupp{Rosales_2016}, and let $K$ range from $2$ to $15$; however, unlike  \citetSupp{Rosales_2016}, we do not include the opportunity counts in the model.
    \item SignatureAnalyzer: $K= 15$, selection method set to \texttt{L1W.L2H} (same as \citealpSupp{Alexandrov_2020}). 
    \item SigProfiler: $K$ ranging from $2$ to $15$.
\end{enumerate}
For both compressive methods, we randomly initialize by sampling from the prior, and run the Gibbs sampler for $12{,}000$ iterations, discarding the first $10{,}000$ as burn-in.  This is repeated four times, and we select the run yielding the highest average log-posterior. PoissonCUSP is run for 12,000 iterations for a single chain, discarding the first 10,000 as burn-in.

The effective sample sizes (ESSs) of each Bayesian method are displayed in \cref{tab:effectiveSize_21breast}. We see that CompNMF achieves the best performance and has good mixing overall.  Adding the informative prior leads to lower effective sample sizes, but still higher than both signeR and PoissonCUSP.

\begin{table}[h]
    \centering
    \caption{Effective sample sizes of the Bayesian methods on the 21 breast cancer data. Numbers denote the average univariate effective sample sizes across all entries of $R$, $\Theta$, and $\mu$, with standard deviations in parentheses, and are calculated using $2{,}000$ posterior samples for each method.}
    \begin{tabular}{lcccc}
        \toprule
        & $\bm{R}$  & $\boldsymbol{\Theta}$ &  $\boldsymbol{\mu}$ &  \\
        \midrule
        CompNMF         & $1{,}229 \ (\pm 953)$ & $1{,}065 \ (\pm 866)$ & $670 \ (\pm 531)$ \\
        CompNMF+cosmic  & $219 \ (\pm  161)$ & $65 \ (\pm 27)$  & $697 \ (\pm 867)$ \\
        signeR          & $98 \ (\pm 70)$  & $36 \ (\pm 23)$  &     &     \\
        PoissonCUSP     & $123 \ (\pm 94)$  & $57 \ (\pm 12)$  & $20 \ (\pm 8)$   \\
        \bottomrule
    \end{tabular}
    \label{tab:effectiveSize_21breast}
\end{table}

\subsection{Effect of pruning inactive factors on reconstruction error}\label{subsec:RMSE_differences}

It might seem beneficial to remove inactive factors from the model, either during MCMC sampling or in postprocessing, to improve model fit. 
However, it turns out that the reconstruction error between $\bm{X}$ and $\bm{R}\boldsymbol{\Theta}$ is barely affected by dropping inactive factors. This occurs because the compressive hyperprior shrinks the exposures for inactive factors to $\approx \varepsilon$.

To make this precise, we define the reconstruction error as $\mathrm{RMSE}(\bm{X},\widehat{\bm{R}}\widehat{\boldsymbol{\Theta}})$, where $\mathrm{RMSE}(A,B) = \big(\sum_{m n} (A_{m n} - B_{m n})^2 / M N \big)^{1/2}$.
On the 21 breast cancer application, when using CompNMF, $\mathrm{RMSE}(\bm{X},\widehat{\bm{R}}\widehat{\boldsymbol{\Theta}})$ is equal to $9.511752$ if we drop all but the $\widehat{K} = 6$ active signatures (defined as those with $\widehat{\mu}_k > 5\varepsilon$), and is equal to $9.511745$ if we keep the full set of $K = 15$ signatures. Similarly, for CompNMF+cosmic (that is, using the informative prior), $\mathrm{RMSE}(\bm{X},\widehat{\bm{R}}\widehat{\boldsymbol{\Theta}})$ equals $9.572395$ if we drop all but the $\widehat{K} = 8$ active signatures ($\widehat{\mu}_k > 5\varepsilon$), and is $9.572271$ if we keep all $K = 77$ signatures. Hence, the difference in reconstruction error is negligible. This occurs because the entries of $\widehat{\bm{R}}$ are relatively small ($1/96$ on average, and never greater than $1$), and $\theta_{kn}$ is centered at $\varepsilon$ if the signature is redundant. We set $\varepsilon = 0.001$ by default, making the product $r_{m k} \theta_{k n}$ negligible for all inactive factors. Hence, the difference in reconstruction error is negligible when keeping only the active signatures, compared to using all of them.

\subsection{\revision{Signatures and exposures for each method}}

\begin{figure}%
    \centering
    \includegraphics[width = 0.5\linewidth]{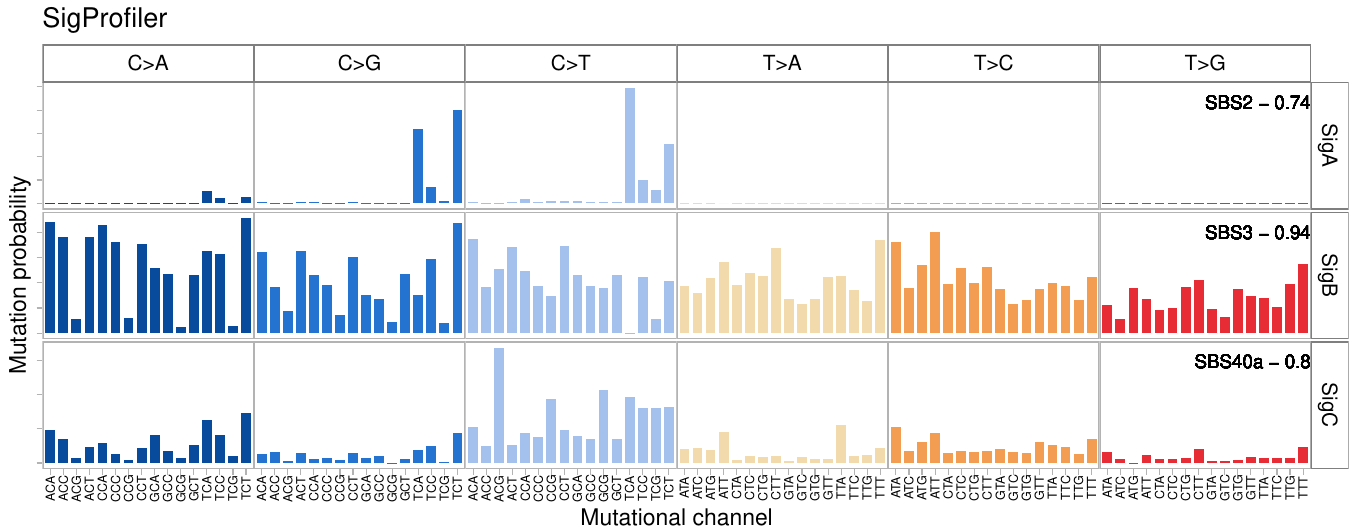}
    \caption{Mutational signatures inferred by SigProfiler on the 21 breast cancer dataset.}
    \label{fig:21br_Sigprofiler}
\end{figure}

\begin{figure}%
    \centering
    \includegraphics[width = 0.5\linewidth]{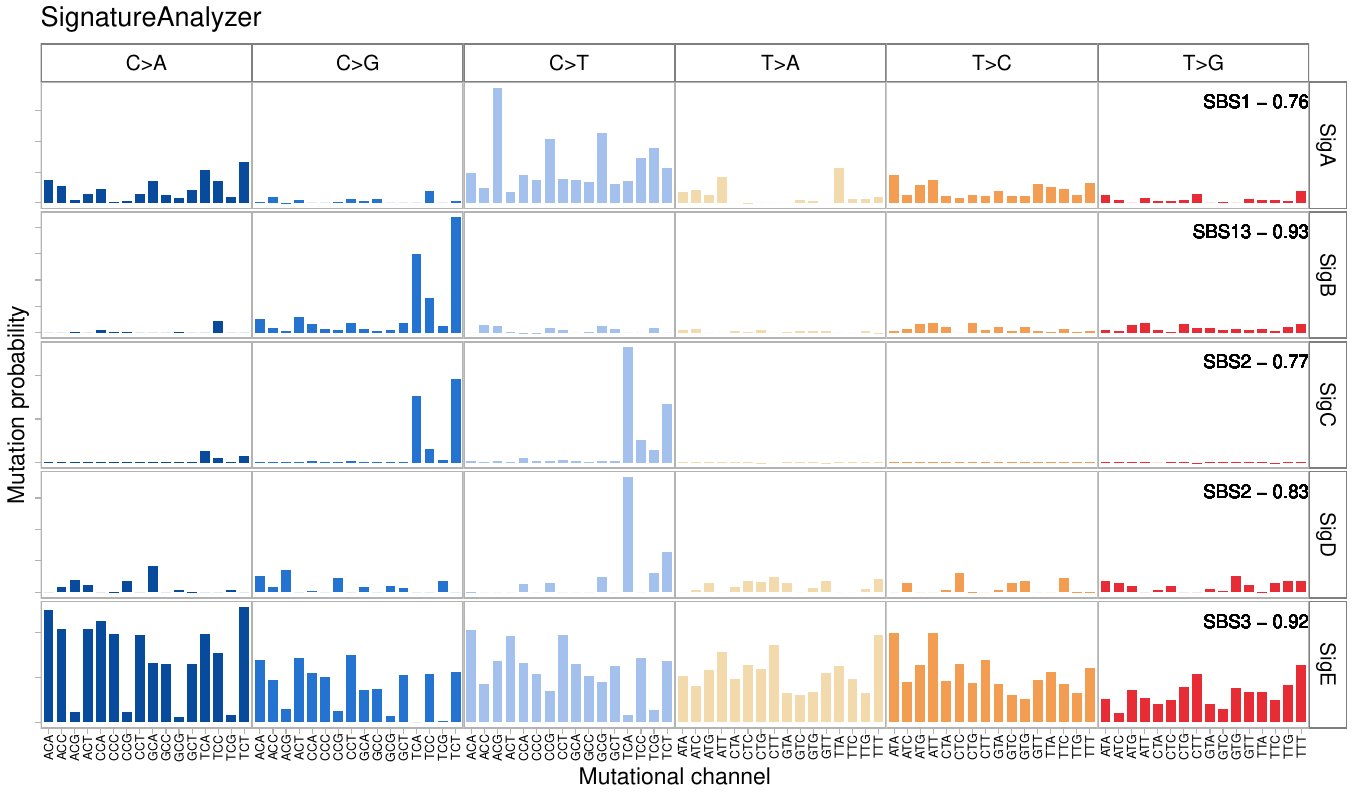}
    \caption{Mutational signatures inferred by SignatureAnalyzer on the 21 breast cancer dataset.}
    \label{fig:21br_SigAnalizer}
\end{figure}

\begin{figure}%
    \centering
    \includegraphics[width = 0.5\linewidth]{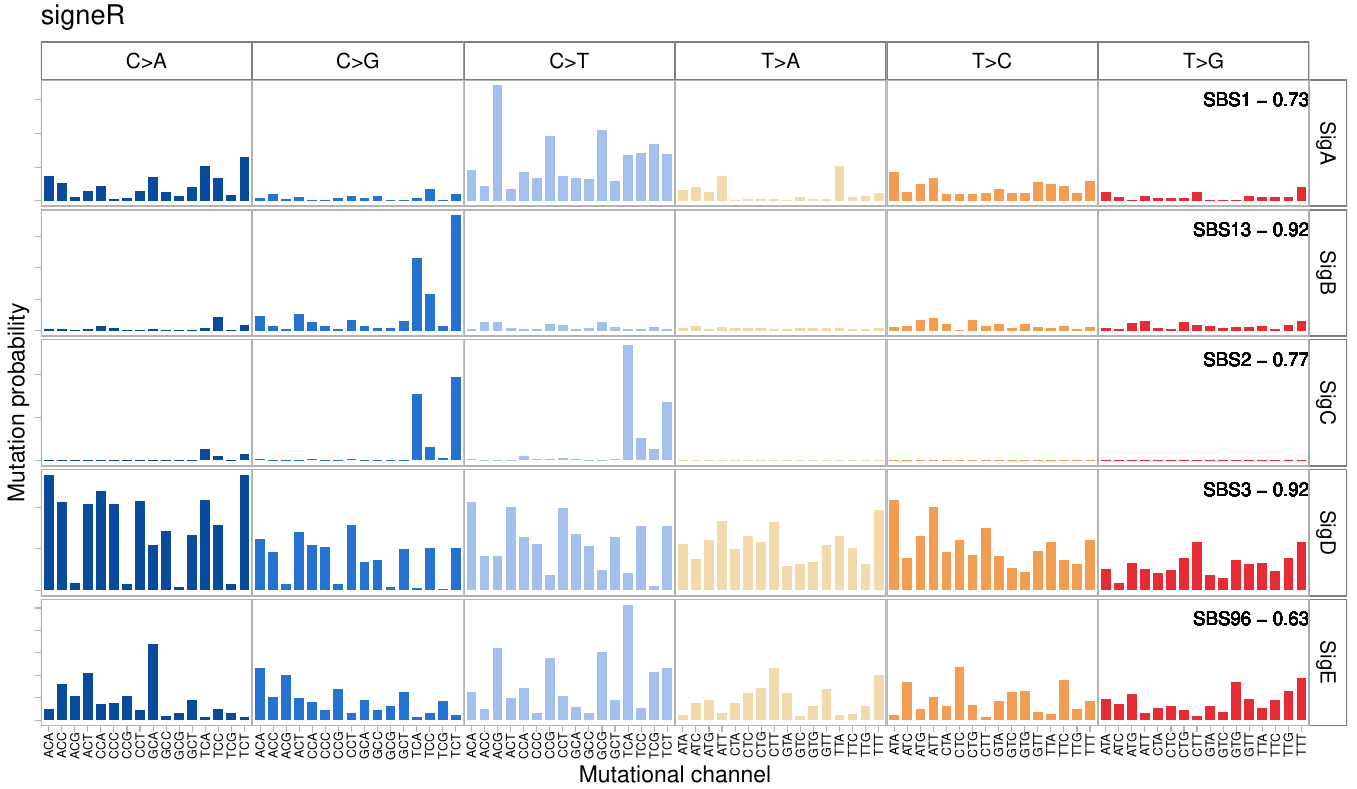}
    \caption{Mutational signatures inferred by signeR on the 21 breast cancer dataset.}
    \label{fig:21br_SigneR}
\end{figure}

\begin{figure}%
    \centering
    \includegraphics[width = 0.5\linewidth]{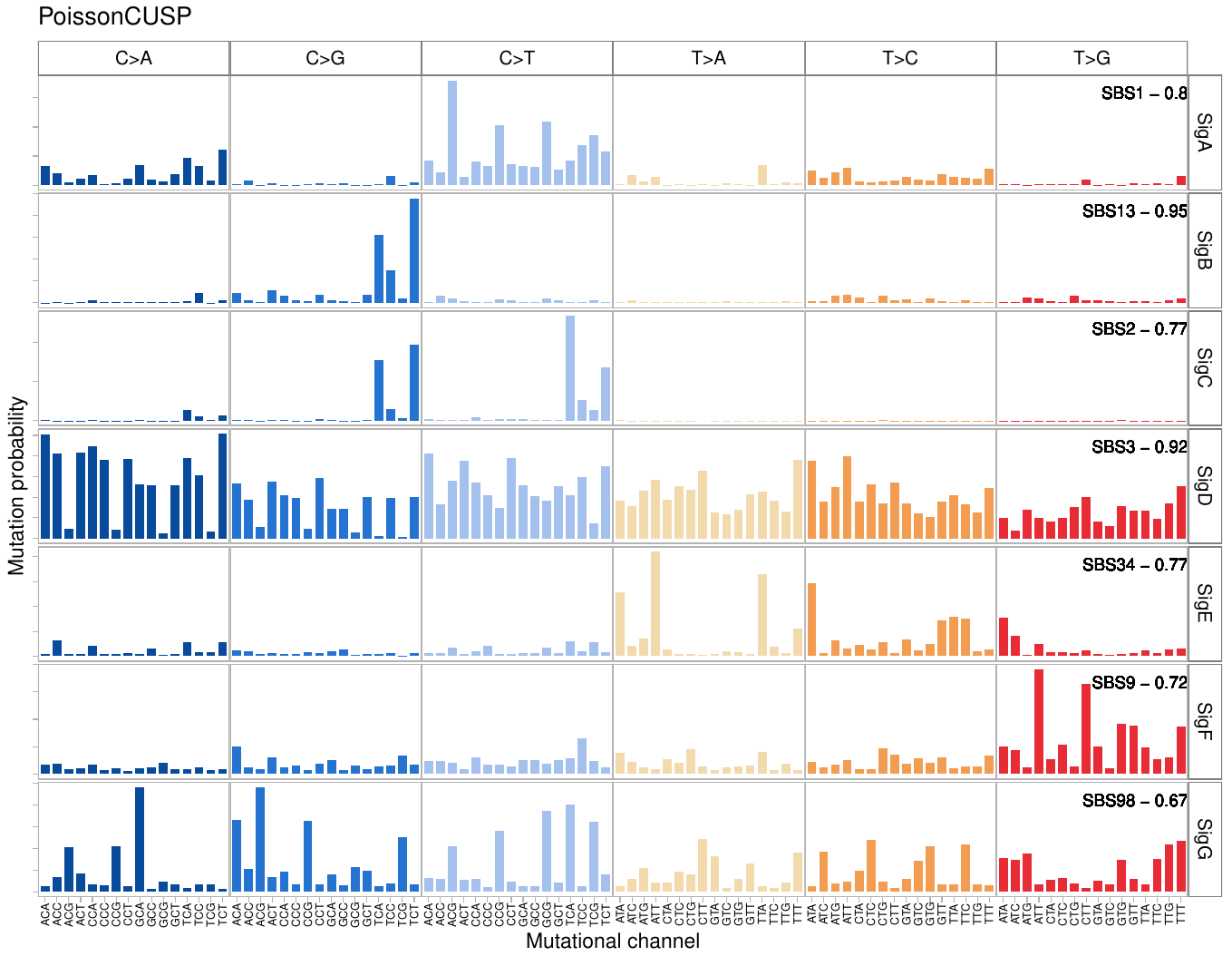}
    \caption{Mutational signatures inferred by PoissonCUSP on the 21 breast cancer dataset.}
    \label{fig:21br_PoissonCUSP}
\end{figure}

\begin{figure}%
    \centering
    \includegraphics[width = 0.5\linewidth]{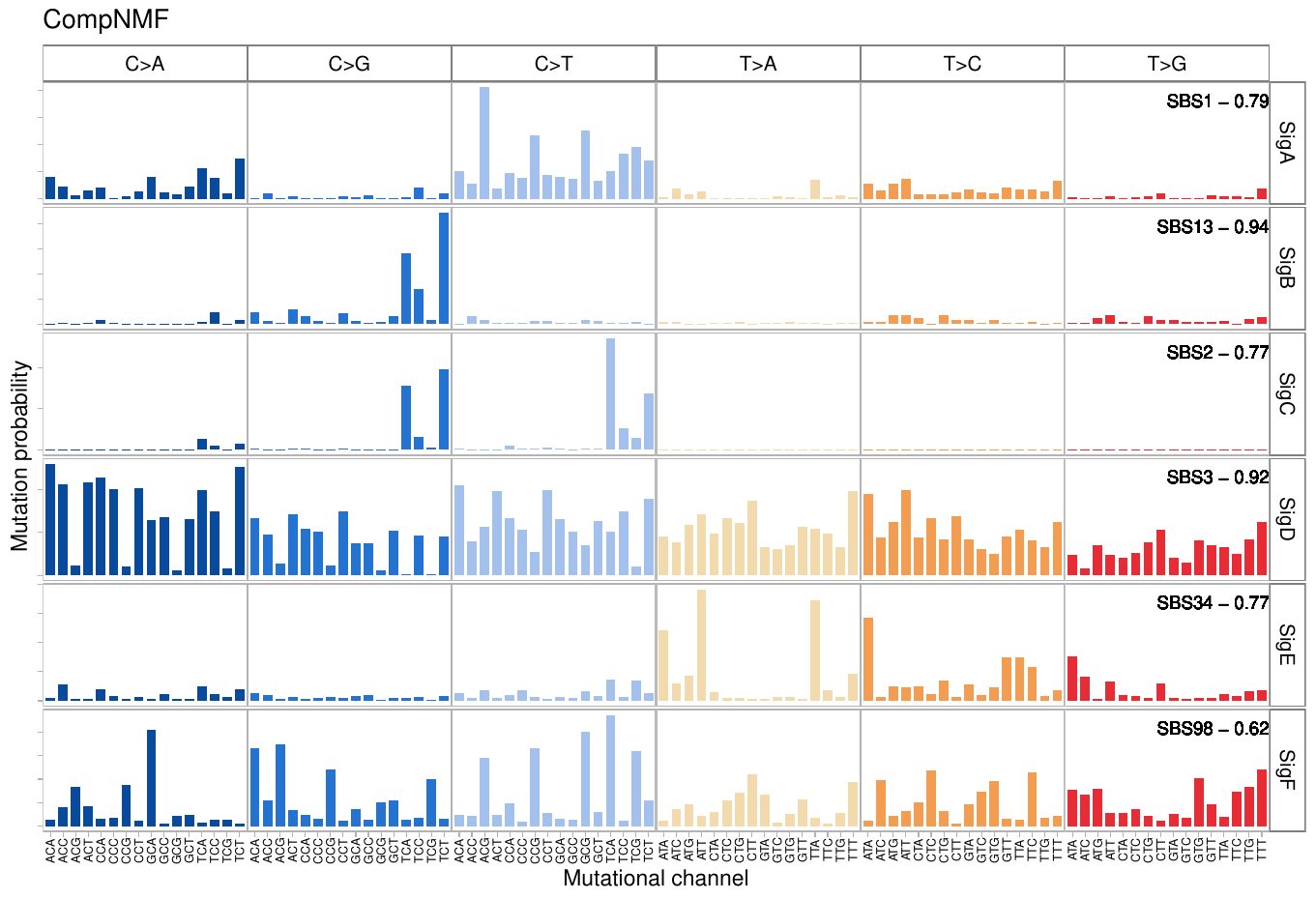}
    \caption{Mutational signatures inferred by CompNMF on the 21 breast cancer dataset.}
    \label{fig:fig:21br_CompNMF}
\end{figure}

\begin{figure}%
    \centering
    \includegraphics[width = 0.5\linewidth]{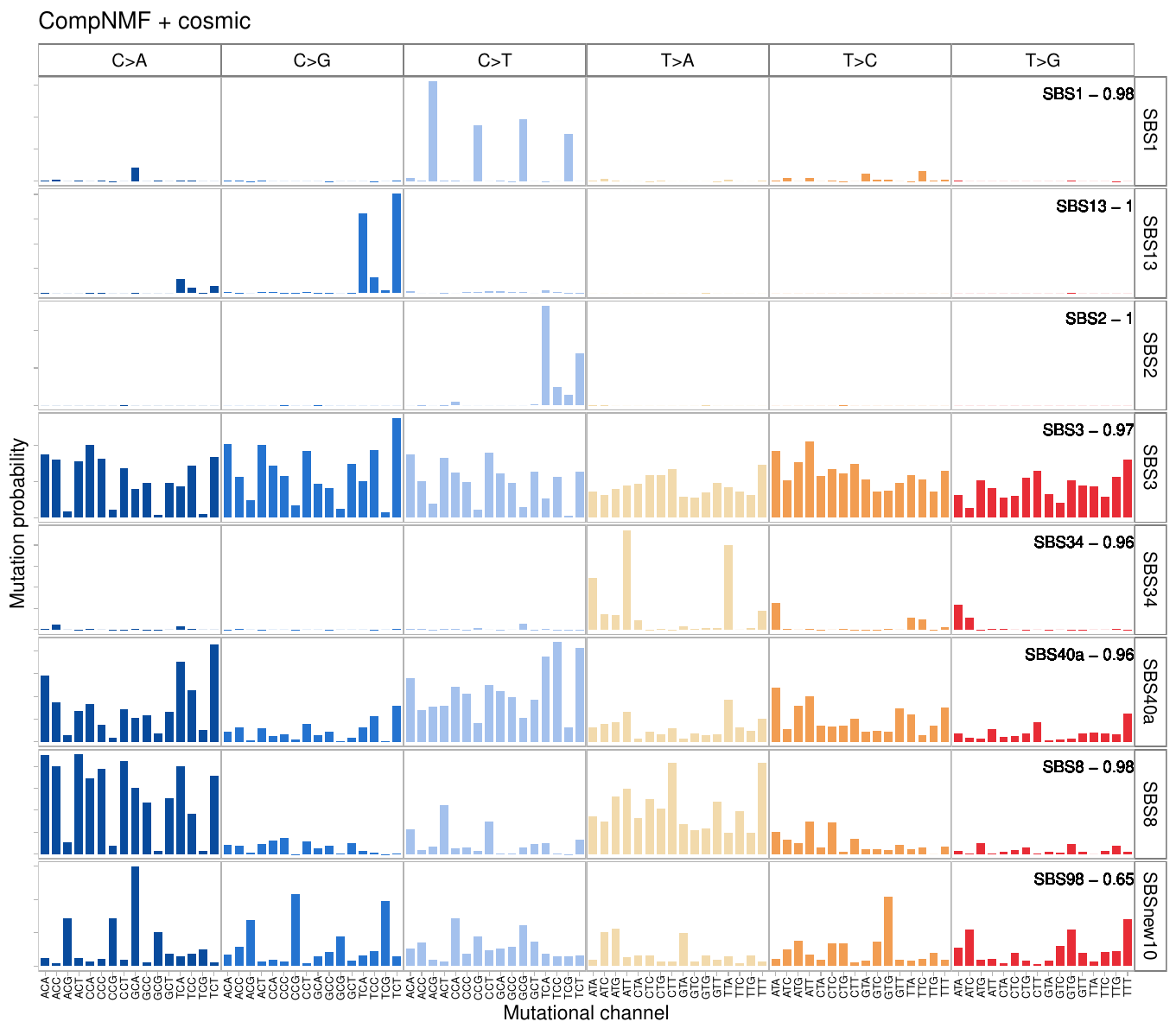}
    \caption{Mutational signatures inferred by CompNMF+cosmic on the 21 breast cancer dataset.}
    \label{fig:21br_CompNMFcosm}
\end{figure}

The complete sets of signatures inferred by each method on the 21 breast cancer data (Section~\ref{sec:application}) are shown in Figures~\ref{fig:21br_Sigprofiler}, \ref{fig:21br_SigAnalizer},  \ref{fig:21br_SigneR}, \ref{fig:21br_PoissonCUSP}, \ref{fig:fig:21br_CompNMF}, \ref{fig:21br_CompNMFcosm} and are discussed below.

SigProfiler estimates only three signatures (matched to SBS2, SBS3, and SBS40a), and the cosine similarity is high only for SBS3; see \cref{fig:21br_Sigprofiler}. This is likely a  consequence of the small sample size, which can lead to the merging of two or more signatures due to insufficient signal to distinguish them. Interestingly, SignatureAnalyzer estimates five signatures (matched to SBS1, SBS2, SBS2, SBS3, and SBS13), but there is duplication since one of them appears to be a combination of SBS2 and SBS13, and ends up being matched to SBS2; see \cref{fig:21br_SigAnalizer}.

We find that signeR infers five signatures (matched to SBS1, SBS2, SBS3, SBS13, and SBS96), similar to the results in the signeR paper \citepSupp{Rosales_2016} but with slight differences since we do not account for the opportunity count matrix, in order to provide a consistent comparison between methods; see \cref{fig:21br_SigneR}.  In the signeR results, SBS1 (which is sparse) appears to have been merged with a flatter signature and thus is retrieved with lower cosine similarity. The method also infers a signature that is matched to SBS96, but with low cosine similarity.

PoissonCUSP and CompNMF estimate a larger number of signatures (seven and six, respectively); see \cref{fig:21br_PoissonCUSP,fig:fig:21br_CompNMF}. In both cases, SBS1, SBS2, SBS3, and SBS13 are inferred with cosine similarities comparable to the other methods. They also both infer SBS34 and SBS98, but these have larger credible intervals, indicating greater uncertainty.  SBS98 also has particularly low cosine similarity in both cases, suggesting that it may be spurious. PoissonCUSP also estimates a signature matched to SBS9, but again with relatively high uncertainty and low cosine similarity. SBS9 has been found in other breast cancer types \citepSupp{Alexandrov_2020}, but its current hypothesized etiology (polymerase eta somatic hypermutation in lymphoid cells) has not been validated.

Finally, our CompNMF+cosmic model estimates eight signatures, all with cosine similarity near $1$, except for SBS98; see \cref{fig:21br_CompNMFcosm}.  The estimated signature matched to SBS98 has very high uncertainty and low cosine similarity, suggesting it is probably a spurious match.  
As in the simulations, the informative prior appears to provide significantly improved sensitivity to detect the presence of signatures, while still allowing for departures from the COSMIC signatures.

\begin{figure}[h]%
    \centering
    \includegraphics[width = \linewidth]{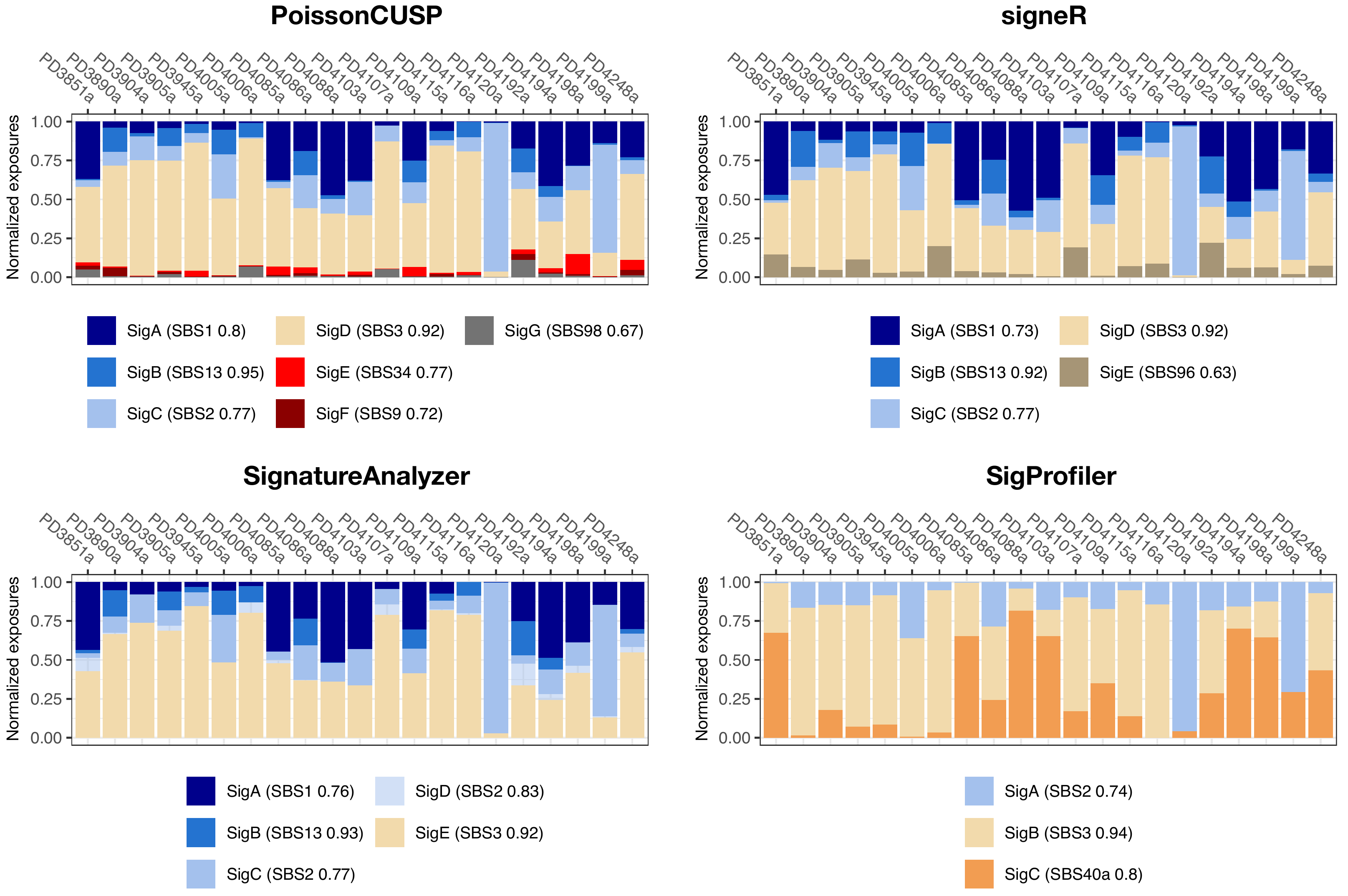}
    \caption{Exposures of each signature (as a percentage of the total exposure) for each patient, for PoissonCUSP, signeR, SignatureAnalyzer, and SigProfiler on the 21 breast cancer dataset.}
    \label{fig:weights_suppl}
\end{figure}

The exposures estimated by PoissonCUSP, signeR, SignatureAnalyzer, and SigProfiler are displayed in \cref{fig:weights_suppl} as percentages.

\newpage
\subsection{Sensitivity analysis for CompNMF on the 21 breast cancer dataset}

\revision{We now assess the effect of the choice of $K$, $\varepsilon$, $a$, and  $\alpha$ on the posterior inferred by our CompNMF model (without the informative prior) applied to the 21 breast cancer dataset. We consider the following ranges of values:
\begin{itemize}
\itemsep0em 
    \item $K = \{15, 20, 25, 40\}$,
    \item $\varepsilon = \{0.01, 0.1, 0.25\}$,
    \item $a = \{1, 2, 0.5\}$,
    \item $\alpha = \{0.5, 1, 2\}$.
\end{itemize}
For each combination of $K$, $\varepsilon$, $a$, and $\alpha$, we run the compressive NMF model for one randomly initialized chain for $12{,}000$ iterations and keep the last $2{,}000$ for inference.}

\begin{figure}%
    \centering
    \includegraphics[width=0.65\linewidth]{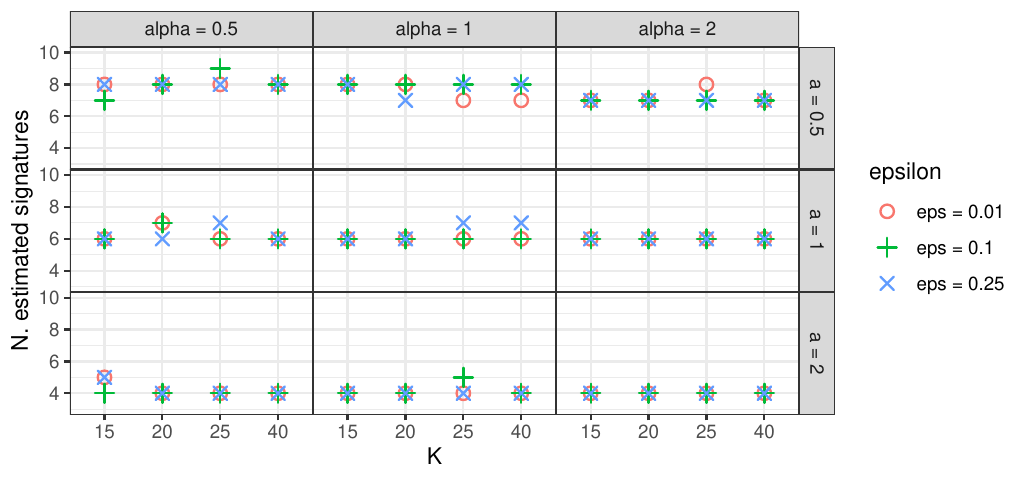}
    \caption{Number of estimated signatures in the 21 breast cancer data for varying $K$, $\varepsilon$, $a$, $\alpha$.}
    \label{fig:21breast_sensitivity_k}
\end{figure}

\cref{fig:21breast_sensitivity_k} displays the estimated value of $\widehat{K}$ across all combinations of model settings. The number of inferred signatures is only slightly affected by the values of $K$, $\varepsilon$, and $\alpha$. This is consistent with the sensitivity analysis in \cref{sec:sensitivity_analyses}: these model settings have minimal impact as long as $K$ is large enough, $\varepsilon$ is small, and $\alpha$ is not too extreme. 
The most impactful setting is the choice of $a$: on average, $a = 0.5$ yields $8$ signatures, $a=1$ yields $6$ signatures, and $a = 2$ yields $4$ signatures. This is expected, since $a$ determines the location of the elbow of the compression curve, as described by Theorem~\ref{thm:concentration} and \cref{fig:compressive_vs_fixed_elbow}. In other words, the larger $a$ is, the more  mutations need to be caused by a signature in order for it to be included.

\begin{figure}%
    \centering
    \includegraphics[width=0.65\linewidth]{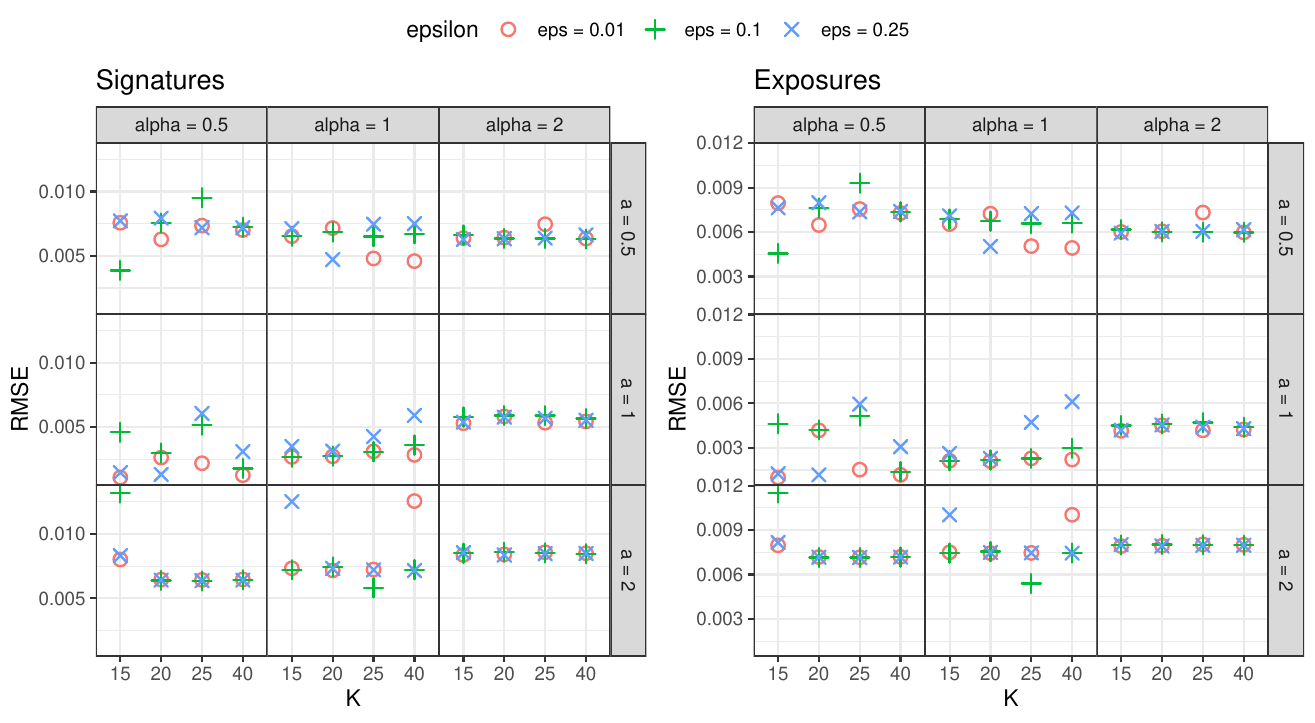}
    \caption{RMSE of $R$ and $\Theta$ for varying values of the model settings $K$, $\varepsilon$, $a$, and $\alpha$.  Here, RMSE is defined as the root mean squared distance from the estimates obtained when using the settings from the main text: $K = 15$, $\varepsilon = 0.01$, $a = 1$, and $\alpha = 0.5$.}
    \label{fig:21breast_sensitivity_rmse}
\end{figure}

\revision{To evaluate the sensitivity of the signatures $R$ and exposures $\Theta$ to the choice of model settings, since the ground truth is not known for this real data, we define $R^\circ$ and $\Theta^\circ$ to be the estimated values when using the model settings presented in \cref{subsec:details_21breast}: $K = 15$, $\varepsilon = 0.01$, $a=1$, and $\alpha = 0.5$.  Then for each combination of $K$, $\varepsilon$, $a$, and $\alpha$, we compute the RMSE between $R^\circ$ and $\Theta^\circ$ and the inferred values of $R$ and $\Theta$, respectively; see \cref{fig:21breast_sensitivity_rmse}. 
As before, there is not much variation in the results across values of $K$, $\varepsilon$, and $\alpha$.  We observe some variation in the RMSE results as $a$ varies, with our recommended value of $a = 1$ performing best overall in these experiments.}


\newpage
\section{\revision{Indels in pancreatic adenocarcinoma}}\label{sec:Indels}

We further test our compressive NMF approach on indel data, which tends to exhibit greater sparsity than SBS data. Indels are the second most common type of mutation and are typically around 10x less frequent than SBS mutations \citepSupp{Alexandrov_2020}. We consider a dataset from a group of pancreatic cancer patients in the ICGC cohort (\texttt{Panc-AdenoCA}, Synapse repository syn7364923, \url{finalconsensuspassonly.snvmnvindel.icgc.public.maf.gz}).
The data comprise $N = 241$ individuals with a total of $233{,}175$ indel mutations across the $M = 83$ channels. The patients are fairly heterogeneous in terms of the number of mutations, with a sample mean of $967$ indels, standard deviation of $3{,}415$, and a maximum of $48{,}137$. The mutation count matrix is sparse, with 41\% of the entries being equal to zero.

We compare CompNMF and CompNMF+cosmic with SigProfiler and SignatureAnalyzer. We exclude signeR and PoissonCUSP, since signeR is not designed for indels while PoissonCUSP showed somewhat worse performance in our simulations.
The 23 indel signatures in COSMICv3.4 are used for the informative prior in CompNMF+cosmic and for interpretation of the results.
For both compressive NMF methods, we run four chains of the sampling algorithm, randomly initializing from the prior and running for $12{,}000$ iterations, keeping the last $2{,}000$ iterations for posterior inference. We use only the chain with the highest log-posterior. For $\bm{R}$, $\boldsymbol{\Theta}$, and $\boldsymbol{\mu}$, respectively, the average effective sample sizes are $1{,}438$, $1{,}549$, and $548$ for CompNMF and $1{,}425$, $1{,}698$ and $578$ for CompNMF+cosmic, indicating good convergence.%

\begin{figure*}%
    \centering
    \includegraphics[width=0.95\linewidth]{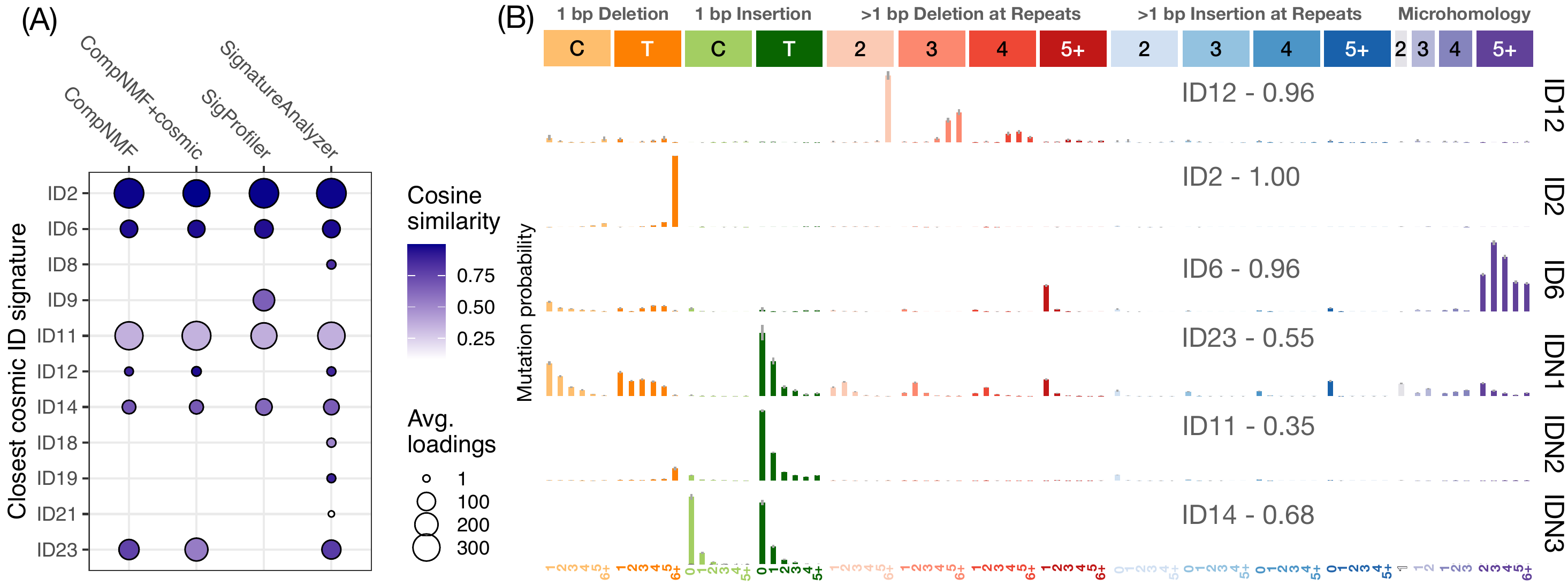}
    \caption{\revision{(A) NMF solution from each model on the Panc-AdenoCA dataset. Points indicate the closest indel signature from COSMIC v3.4 and the color intensity denotes the cosine similarity. Size of points depicts the average exposure associated to the signature across patients. (B) Indel signatures found by CompNMF+cosmic. Grey bars indicate 90\% posterior credible intervals.}} 
    \label{fig:Panc-AdenoCA}
\end{figure*}

Figure~\ref{fig:Panc-AdenoCA}(A) summarizes the results in terms of cosine similarity to the nearest COSMIC signature and estimated exposure. CompNMF and CompNMF+cosmic yield similar results, both estimating six signatures and assigning them similar exposures; in particular, \textsc{id}6, \textsc{id}12, and \textsc{id}2 have high cosine similarity with their matching estimates. SigProfiler finds five signatures, with a match to \textsc{id}9 instead of \textsc{id}12 and \textsc{id}23, although the cosine similarity is not high. SignatureAnalyzer finds 10 signatures, but the ones matching to \textsc{id}8, \textsc{id}18, \textsc{id}19, and \textsc{id}21 have very small exposures and relatively low cosine similarity. This corroborates the finding from our simulations that SignatureAnalyzer tends to return a larger number of signatures, but not necessarily with high quality. The root-mean-squared error between $\widehat{\bm{R}}\, \widehat{\boldsymbol{\Theta}}$ and $\bm{X}$ is $5.24$ for CompNMF, $5.28$ for CompNMF+cosmic, $5.35$ for SignatureAnalyzer, and a much larger $20.57$ for SigProfiler, following the same trend as on the breast cancer data.

Figure~\ref{fig:Panc-AdenoCA}(B) shows the set of signatures inferred by CompNMF+cosmic. The posterior credible intervals are tight, suggesting high confidence about the solution. Based on the cosine similarities, it is clear that the model has recovered COSMIC signatures \textsc{id}2, \textsc{id}12, and \textsc{id}6, while the other three estimated signatures---labelled \textsc{idn}1, 2, and 3---appear to be novel.

\subsection{\revision{Details on the methods and further results}}
\begin{figure}
    \centering
    \includegraphics[width=0.9\linewidth]{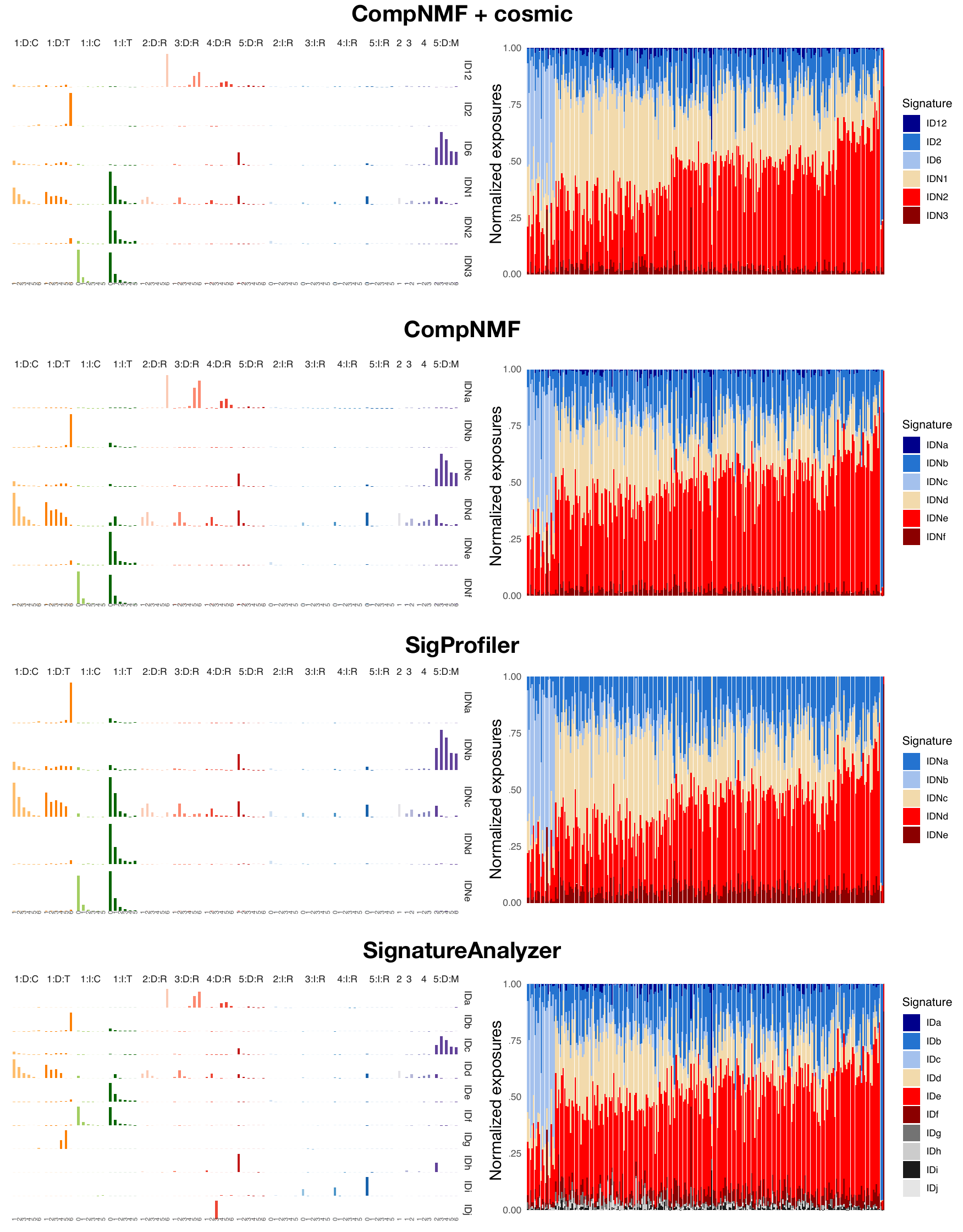}
    \caption{\revision{Signatures and normalized exposures in the four methods. }}
    \label{fig:Plots_AdenoCA}
\end{figure}

\revision{We now report the details for the ICGC \texttt{Panc-AdenoCA} data analysis. We run the following methods:
\begin{enumerate}[itemsep=0em,label=(\roman*)]
    \item CompNMF with $K = 20$, $a = 1$, $\varepsilon = 0.001$, and $\alpha=0.5$,
    \item CompNMF+cosmic with $K_{\mathrm{pre}} = 23$ and $K_{\mathrm{new}} = 10$, $\beta_k$ obtained so that prior samples from the Dirichlet have 0.95 cosine similarity with the associated COSMIC signature,
    \item SignatureAnalyzer with $K = 25$ and \texttt{method = L1W.L2H} (same as \citealpSupp{Alexandrov_2020}),
    \item SigProfilerExtractor, with $K$ ranging from 2 to 15 and 25 replicates each. 
\end{enumerate}
\cref{fig:Plots_AdenoCA} shows the inferred signatures and exposures in each case. }

\newpage
\bibliographystyleSupp{chicago}
\bibliographySupp{references}

\end{document}